\documentclass{JFM-FLM_Au}
\usepackage{subcaption}
\usepackage{multirow}
\usepackage{float}
\usepackage{xcolor}
\usepackage{placeins}

\lefttitle{R. Dushe and Arun K. Saha}
\righttitle{Journal of Fluid Mechanics}

\title{Buoyancy-induced transitions in mixed convective flow past a heated square cylinder}

\author{Rudram Dushe\aff{1}{\footnote{Present address: 
Department of Mechanical Engineering, University of Minnesota Twin Cities, Minneapolis, MN}} ~\and Arun K Saha\aff{1}}

\affiliation{\aff{1}Indian Institute of Technology Kanpur, Kanpur, UP 208016, India}

\corresau{Arun K Saha, \email{aksaha@iitk.ac.in}}

\def \mycolorb {black}
\def \mycolorc {black}
\def \mycolord {black}
\colorlet{mycolor}{black}
\begin{document}
\maketitle

\begin{abstract}
\textcolor{\mycolorc}{The effect of aiding buoyancy on flow past a square cylinder has previously been shown to suppress vortex shedding above a critical Richardson number (Ri), transitioning from unsteady to steady flow. Building on this, the present study uses two-dimensional direct numerical simulations (DNS) of vertically upward flow past a heated square cylinder to examine two buoyancy-induced transitions for a range of Ri and Reynolds number (Re). The secondary flow induced by buoyant force enhances positive entrainment in the momentum-deficit region aft of the cylinder, thereby suppressing vortex shedding. At higher Ri, this buoyant flow continues to gain downstream momentum, leading to a far-field plume-like instability. Extended domain simulations show that the far-field transition depends on the downstream distance available for buoyancy-driven disturbances to grow. Increasing Re delays near-field suppression but advances far-field instability. A local-flow analysis shows that a Reynolds number based on local mean velocity and plume width substantially improves the collapse of the far-field instability onset, while the corresponding plume-based Strouhal number also exhibits an approximate collapse across the investigated cases. Correlations for drag, surface-averaged Nusselt number, and far-field frequency are proposed using a mixed-convective velocity scale. Linear stability analysis (LSA) about the two-dimensional base flow captures both transitions as Hopf bifurcations and accurately predicts the dominant frequencies and finite-domain transition thresholds observed in DNS. The corresponding eigenmodes reveal localized oscillatory structures that shift downstream with increasing Ri. These results demonstrate the dual role of buoyancy in stabilizing the near wake while destabilizing the far field.}
\end{abstract}


\section{Introduction}
\label{Intro}

The temperature difference between a heated body and its surrounding quiescent fluid generates thermal buoyancy, resulting in a free convective flow characterised by a non-dimensional number termed the Grashof number. Similarly, a forced convective flow characterised by the non-dimensional number called the Reynolds number is evident if the relative contribution of the thermal buoyancy is far less compared to the inertial effect in the flow. However, a mixed convective flow occurs when the two forces are comparable in magnitude, with their relative orientation (aligned or opposing) playing a key role in determining flow behavior. Depending on the combination of critical parametric values, the flow past a heated bluff body exhibits either steady flow or periodic vortex shedding. By maintaining the flow in the active/suppressed vortex shedding regime, various factors, such as heat transfer from the bluff body and the forces acting on it, can be controlled. Mixed convective flows past heated bluff bodies are of practical importance in engineering because of their occurrence and applications in heat exchangers, the cooling of machine equipment, etc. The study of vortex shedding in flows past bluff bodies is also important to avoid resonance, in which the structure's natural frequency matches the vortex-shedding frequency, leading to sustained, large-scale vibrations. For these reasons, many previous studies have focused on mixed convective flows past bluff bodies.

Several experimental and theoretical studies on free-convection flows have shown that buoyant plumes can be unsteady beyond the critical values of a few dimensionless parameters. The heated horizontal cylinder experiment of \citet{kuehner2015experimental} revealed the plume's oscillation at high Rayleigh and Prandtl numbers. 
~\citet{sasmal2011laminar} calculated laminar natural convection heat transfer from a heated long square cylinder in static fluids over Grashof, Prandtl, and power-law index ranges. The planar-plume study of \citet{hunt2022two} showed that buoyancy in an entraining plume contained in a uniform co-flow can increase downstream momentum flux locally. Although their configuration lacks the separated shear layers and recirculating near-wake, the physical implication that buoyancy can progressively amplify vertical momentum with distance may help explain the buoyant flow observed in a mixed convective flow past a bluff body.

From previous studies involving purely forced convective flows past bluff bodies, we know that for a given fluid, the critical Reynolds number (Re) for the transition of flow from steady (characterized by a symmetric pair of steady vortices) to unsteady (characterized by active vortex shedding leading to a K\'arm\'an vortex street) occurs at a critical Reynolds number of 45 \citep{yoon2010flow, sohankar1998low, saha2000transition}. A critical Reynolds number lying between 150 and 175 was identified \citep{saha2003three, saha2009effect} as the point of transition of the flow from two-dimensional (2D) unsteady flow, which exhibits a K\'arm\'an vortex street, to a three-dimensional (3D) flow past a square cylinder. Within the laminar regime considered in the present study, variations in Reynolds number primarily influence the strength and spatial extent of wake structures rather than introducing fundamentally new instability mechanisms. Recent numerical investigations have shown that increasing Reynolds number \textcolor{\mycolord}{to} low to moderate values enhances the wake velocity deficit, elongates recirculation regions, and amplifies vortex strength, thereby intensifying momentum and heat transport while preserving the underlying flow topology \citep{mashhadi2022two, sohankar1998low}. Consequently, Reynolds number acts as a controlling parameter for the intensity and spatial development of flow and thermal fields, motivating its systematic consideration alongside buoyancy effects in mixed convection. 

The literature on mixed convective flow past bluff bodies builds upon the foundations laid by the above-mentioned studies. Studies in the open literature examine the influence of the Richardson number on steady mixed convective flow past bluff bodies of various shapes~\citep{sharma2012mixed, patel2015effect}. For an unsteady mixed convective flow, through a numerical investigation performed by \citet{sharma2004effect} which focused on the characteristics of mixed convection flow around a square cylinder at a Reynolds number of 100 and a Prandtl number (Pr) of 0.7, it was observed that the degeneration of the K\'arm\'an vortex street in the aiding buoyancy regime \textcolor{\mycolorb}{(i.e., buoyancy acting in the same direction as the incoming flow)} resulted in a transition from unsteady periodic flow to steady flow at a critical Richardson number (Ri) of 0.15. However, subsequent transitions beyond this point were not investigated. Some studies have focused on identifying the mechanism underlying vortex shedding suppression. It was identified that the baroclinic vorticity production is the primary factor responsible for the suppression of vortex shedding, because of the symmetry-preserving nature of the baroclinic vorticity production term in the transport equation for vorticity \citep{arif2019vortex, hasan2013vortex}. In a study on mixed convective flow past a circular cylinder conducted by \citet{chang1990effect}, an increase in Grashof number was observed to lead to the acceleration of the wall boundary layer and an increase in velocity in the wake region \textcolor{\mycolorb}{in the aiding buoyancy configuration}. This resulted in a thinner boundary layer and an increase in shedding frequency. However, the weakened roll-up process hindered the entrainment of ambient flow into the wake cavity, eventually leading to the cessation of vortex shedding from the cylinder surface. Interestingly, a critical Ri of 0.15 for vortex shedding suppression was found, matching the earlier findings for a square cylinder by \citet{sharma2004effect}. Thus, a striking similarity in vortex shedding suppression can be observed for square and circular cylinders. For flow past a circular cylinder at a fixed Reynolds number, the Nusselt number was found to increase with the Richardson number for the aiding buoyancy \citep{singh1998cir}. Also, it is worth noting that the phenomenon of vortex shedding was found to initiate beyond a critical Richardson number for a fluid flowing past a square cylinder in a cross-flow configuration by \citet{chatterjee2011effect} (where the direction of incoming flow is perpendicular to the direction of the gravity vector). However, the phenomenon of vortex shedding suppression was found to be absent in the cross-flow configuration \citep{bhattacharyya2005vortex}, which reveals an interesting difference between aiding-buoyancy and cross-flow configurations. For the 3D counterpart to the present problem (flow past a sphere), a similar phenomenon as that of the onset of vortex shedding was observed by \citet{kotouvc2008loss}- the onset of axisymmetry breaking. In this case as well, the addition of buoyancy was found to have a stabilizing effect and pushes the Reynolds number for the onset of instabilities from Re = 212 for an unheated sphere to Re = 1470 at Pr = 0.72 and Ri = 0.7.

A few experimental studies have also focused on studying mixed convective flow past a cylinder. The existence of a secondary flow due to the buoyant forces at higher Richardson numbers was identified in the experimental study on near-wake characteristics of a heated circular cylinder \textcolor{\mycolorb}{in a buoyancy-aided configuration} conducted by \citet{michaux1997near}. This secondary flow was found to add sufficient momentum to offset the momentum deficit caused by the cylinder drag (blockage). This causes the wake to shrink, become short, and ultimately suppress vortex shedding. In an experimental study to investigate the combined effects of buoyancy and orientation on vortex shedding from a heated square cylinder conducted by \citet{kakade2010schlieren}, the shear layers originating from the two upstream corners of the square cylinder were found to merge into a single plumelike structure beyond a critical Richardson number. This phenomenon was attributed to the buoyancy-induced motion \textcolor{\mycolorb}{aligned with the incoming flow direction} that counteracts the vorticity generated by flow separation and impedes the transport of circulation across the opposing shear layers.

While direct numerical simulations (DNS) and experimental studies have been the primary tools for studying transitions in mixed convective flows, linear stability analysis (LSA) offers a complementary framework for assessing the onset and nature of instabilities. Several studies have employed linear stability analysis (LSA) as an aiding tool to understand the dynamics of flow past bluff bodies. \citet{zebib1987stability} investigated the stability characteristics of viscous flow around a circular cylinder and demonstrated that while symmetric perturbations remained stable, asymmetric perturbations became unstable beyond \(Re \approx 40\), resulting in global flow instability. \citet{kelkar1992numerical} performed LSA for steady flow past a square cylinder and identified the onset of unsteadiness at \(Re = 53\), with the associated perturbation modes exhibiting characteristic inverted heart-shaped structures \textcolor{\mycolord}{in the streamfunction contours}. These results provided physical insight into the origin and structure of instabilities in bluff-body wakes. Further, \citet{camarri2007inversion} found a correlation between vorticity inversion \textcolor{\mycolord}{in the wake} and wake momentum deficit through LSA. However, most investigations have applied LSA to purely forced convective flows past bluff bodies, and applications to mixed convective flows are scarce. \citet{zhang2016numerical} conducted LSA in the aiding buoyancy regime for \(0 \leq Ri \leq 0.10\), showing that buoyancy dampens oscillations of flow variables, but without investigating the transition types and mechanisms. Thus, despite its potential to reveal instability structures, growth rates, and dominant frequencies, LSA remains underutilized in the analysis of buoyancy-aided flows.

Despite these insights, a major limitation of the existing literature is that prior studies have primarily focused on how buoyancy modifies vortex shedding and flow transitions in the near-field, while its influence on the far-field regime remains essentially unexplored. This is particularly important because of the competing effect between buoyancy and forced convection \textcolor{\mycolorb}{(in the aiding configuration considered here)}, which creates the possibility of other flow transitions at higher Richardson numbers, especially in regions further downstream where buoyancy effects intensify. Furthermore, while LSA has been widely applied to forced-convective flows to identify instability mechanisms and bifurcation behavior, its application in mixed convective regimes remains limited. The present work aims to bridge these gaps by providing a combined DNS and LSA framework to gain insights into a unified mechanism governing the effect of buoyancy on dynamics in both the near- and far-field regimes \textcolor{\mycolorb}{under aiding buoyancy conditions}. It is to be noted here that the entire shear flow region behind the cylinder is divided into two regions, namely, near-field ($y/B \leq 15$) and far-field ($y/B > 15$). 

It is well established that the isothermal flow past a square cylinder becomes three-dimensional for Re$\geq$163.5~\citep{saha2009effect}. The present study considers the maximum Reynolds and Richardson numbers to be 150 and 0.9, respectively. The two-dimensionality of the flow field is checked by conducting a three-dimensional simulation with a cylinder length (spanwise length) of 3$B$ for Re=150 and Ri=1.0, which exceeds the chosen maximum Ri=0.9. The extracted iso-surface of vortical structures from the instantaneous flow field shows no variation in the spanwise direction, suggesting the two-dimensionality of the flow field. The time signal stored at the far-field also reveals that the instantaneous spanwise velocity component is negligible at all times, indicating the two-dimensional nature of the flow.

In the present work, two-dimensional direct numerical simulations are performed for vertically upward mixed convective flow past a heated square-section cylinder, \textcolor{\mycolorb}{with gravity aligned vertically with the inflow (an aiding buoyancy configuration)}. A constant \(Re = 100\) is used with air as the working fluid, for which \(Pr = 0.7\) for the combined DNS and LSA study. The assisted buoyancy regime is investigated by varying \(Ri\) in the range \(0 \leq Ri \leq 0.8\). In addition, a limited set of additional direct numerical simulations at different Reynolds numbers is used to assess the influence of $\mathrm{Re}$ on wake-transition characteristics. Besides analyzing flow transitions, wake dynamics, and heat transfer using DNS, a global linear stability analysis is conducted to examine the origin, type, and spatial structure of instabilities across the Richardson-number range at $\mathrm{Re} = 100$. The analysis identifies two successive transitions: suppression of near-field vortex shedding beyond a critical Richardson number (\(Ri_c \approx 0.145\)) and the emergence of far-field unsteadiness at higher buoyancy levels. These transitions are characterized as Hopf bifurcations through eigenvalue analysis, and the associated mode shapes provide mechanistic insight into how buoyancy modifies the global instability landscape. The LSA predictions for both the critical \(Ri\) values and dominant frequencies closely match DNS results, underscoring the synergy between the two approaches. Unlike prior studies focused on near-field effects, the present work reveals a second transition in the far-field and explains it using both DNS and global stability analysis.

\section{Numerical Methodology}

This section comprehensively presents the numerical methodology employed in this study. Direct Numerical Simulations (DNS) are used to capture the full nonlinear behavior of the mixed convective flow, while Linear Stability Analysis (LSA) is performed to investigate the onset and nature of instabilities.

\subsection{Direct Numerical Simulations}

\subsubsection{Problem formulation}~\label{problemformulation}
\begin{figure}
    \centering
    \includegraphics[width=0.4\linewidth]{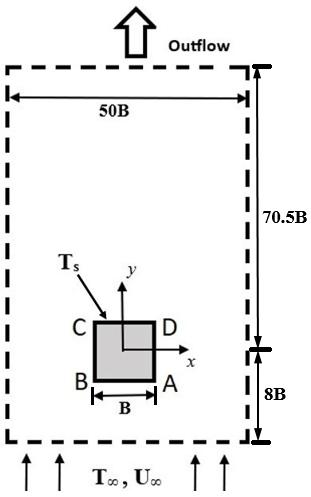}
    \caption{Schematic of the problem set-up and computational domain.}
    \label{domain}
\end{figure}

The computational domain for the problem is shown in Figure \ref{domain} and consists of a vertically aligned channel with a centrally placed heated square cylinder of edge length \( B \). The domain extends \( 8B \) upstream, \( 70.5B \) downstream, and \( 25B \) laterally from the cylinder center, which results in the overall domain size of $78.5B \times 50B$, ensuring negligible boundary effects. The stream-wise length was purposefully chosen to be sufficiently long to capture flow behaviour in the far-field, especially for higher Richardson numbers. Considering the values of the flow parameters used in the present study, the equations are taken to be incompressible, two-dimensional and laminar \citep{saha2003three, kakade2010schlieren, hasan2013vortex}. Hence, the non-dimensional continuity, Navier-Stokes and energy equations (under the Boussinesq approximation) governing the current problem are as follows:
\begin{align}
\nabla \cdot \boldsymbol{u} &= 0 \label{eq:continuity} \\
\frac{\partial \boldsymbol{u}}{\partial \tau} + (\boldsymbol{u} \cdot \nabla)\boldsymbol{u} &= -\nabla p + \frac{1}{Re} \nabla^2 \boldsymbol{u} + Ri\, \theta\, \hat{\boldsymbol{y}} \label{eq:momentum} \\
\frac{\partial \theta}{\partial \tau} +(\boldsymbol{u} \cdot \nabla)\theta &= \frac{1}{Re\,Pr} \nabla^2 \theta \label{eq:energy}
\end{align}

Here, \( \boldsymbol{u} = (u, v) \) denotes the non-dimensional velocity field, \( p \) the pressure, and \( \theta \) the non-dimensional temperature. The momentum equation along the \textit{y} direction has an additional term $Ri\theta$ arising due to the free convection caused by the thermal buoyancy effect. The reference length, velocity, and time scales used are the cylinder edge length \( B \), the freestream velocity \( U_\infty \), and the ratio $B/U_{\infty}$ respectively. The pressure is non-dimensionalized using $\rho U_{\infty}^2$ while the non-dimensional temperature is written as:
\[
\theta = \frac{T - T_\infty}{T_s - T_\infty}
\]
Where $T_{\infty}$ is the free-stream temperature and $T_s$ is the surface temperature of the cylinder. The above non-dimensionalization using the reference values results in the following dimensionless parameters:
\[
Re = \frac{U_\infty B}{\nu}, \quad Pr = \frac{\mu c_p}{k}, \quad Gr = \frac{g\beta (T_s - T_{\infty})B^3}{\nu^2}, \quad Ri = \frac{g \beta (T_s - T_\infty) B}{U_\infty^2} = \frac{Gr}{Re^2}
\]

Here, $Re$ is the Reynolds number, $Pr$ is the Prandtl number, $Gr$ is the Grashof number, and $Ri$ is the Richardson number of the fluid.

In the above equations, $g$ is the acceleration due to gravity, \( \beta \) is the coefficient of volumetric thermal expansion, \( k \) is the thermal conductivity,  $\nu$ is the kinematic viscosity, $\rho$ is the density, $\mu$ is the dynamic viscosity, and $c_{p}$ is the specific heat of the fluid.

For the boundary conditions, at the inlet (bottom boundary), a uniform vertical inflow was imposed with \( u = 0 \), \( v = 1 \), and \( \theta = 0 \). The cylinder surface was treated as no-slip and isothermal, enforcing \( u = v = 0 \) and \( \theta = 1 \). Along the lateral walls, free-slip conditions for velocity: \( u = 0 \), \( \partial v/\partial x = 0 \), and adiabatic conditions for temperature: \( \partial \theta/\partial x = 0 \) were applied. At the outlet (top boundary), determining a suitable outflow boundary condition is difficult due to the absence of a unique prescription for the outflow boundary of external flows. An incorrect choice of the outflow boundary condition can significantly influence the upstream flow, especially under subsonic conditions. Consequently, it should be selected to have minimal effect on upstream flow. Previous studies, including those by~\citet{sohankar1998low}, have demonstrated that the convective boundary conditions shown below are preferable to other alternative boundary conditions, such as the homogeneous Neumann boundary conditions (\( \partial u/\partial y = 0 \) and \( \partial v/\partial y = 0 \)) or vanishing second derivatives (\(\partial^2 u/\partial y^2 = 0 \) and \(\partial^2 v/\partial y^2 = 0 \)).~\citet{sohankar1998low} have demonstrated that the convective boundary conditions decrease computational time, iterations required for convergence, and the duration needed to achieve periodic flow. Furthermore, the convective boundary conditions facilitate the smooth convection of vortices out of the domain at outlet, whereas the homogeneous Neumann boundary conditions and vanishing second derivatives conditions deform and smear vortices as they exit the domain. Accordingly, convective outflow conditions, first prescribed by~\citet{orlanski1976simple}, were specified for all variables using
\[
\frac{\partial q}{\partial \tau} + V_c \frac{\partial q}{\partial y} = 0, \quad \text{where } q \in \{ u, v, \theta \} \text{ and } V_c = 0.8.
\]

\subsubsection{Numerical Method}
The numerical method employed in this study is a modified Marker-And-Cell (MAC) method (\citet{hoffmann1994numerical}). The governing equations have been solved on a staggered grid, which facilitates the coupling of pressure and velocity fields. In a staggered grid arrangement, pressure is defined at the cell center, while velocity components are located on the cell face. The velocity components are interpolated to the cell center, wherever required. A second-order central difference scheme is employed to discretize convective and diffusive terms. Similarly, a second-order time-accurate explicit Adams-Bashforth method is employed to advance the time of both convective and diffusive terms.

The velocity and pressure fields are obtained using a two-step predictor-corrector method. The initial predictor step involves the explicit calculation of provisional velocity values by the time advancement of the results of the previous time step. In the subsequent step, the correction of pressure and velocity fields is sought by iterating the pressure correction equations till a divergence-free velocity field is obtained. The solution for the next time step is initiated once the correct velocity and pressure fields are obtained after the divergence of the velocity field reaches a pre-defined threshold. The details of the numerical methodology can be found elsewhere~\citep{kavin2025diamond}.


A non-uniform Cartesian mesh was used, with clustering near the cylinder surface and in the wake to resolve boundary layers and shear structures accurately. The grid size was finalized based on a grid-independence test (section \ref{validGCT}) to have 268 cells along $x$ and 468 cells along $y$. The minimum grid size along the $x$ and $y$ directions is 0.005. The simulations were run for sufficiently long times so that a statistically steady or dynamic steady state was reached. 
For imposing boundary conditions, linear interpolation of the variables was used as required.

Finally, various quantities of interest were computed from the flow variables. The drag and lift coefficients were obtained by integrating the pressure and viscous stress components over the cylinder surface. The local Nusselt number was evaluated from the wall-normal temperature gradient and subsequently averaged over the cylinder surface. The Strouhal number was extracted from the dominant frequency obtained through Fast Fourier Transform (FFT) of the transverse velocity signal in the shear-layer region. The mathematical definitions used to calculate these global flow parameters are provided in Appendix A.

\subsubsection{Code Validation and Grid Convergence}~\label{validGCT}

\begin{figure}
    \centering
    \includegraphics[width=0.5\linewidth]{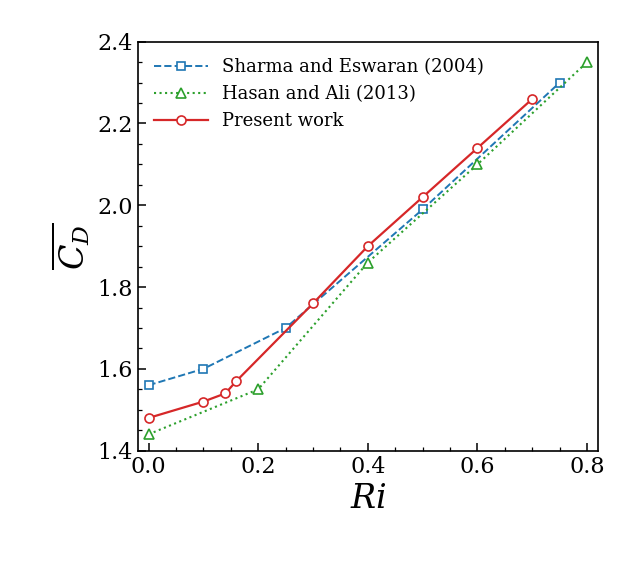}
    \caption{Comparison of the variation of mean drag coefficient at various Ri and Re = 100.}
    \label{drag_validation}
\end{figure}

To validate the DNS solver, results from the present computations were compared with benchmark data for an unbounded flow scenario at \( Ri = 0 \), \( Re = 100 \), and \( Pr = 0.7 \). These include the time-averaged drag coefficient, Strouhal number, and average Nusselt number on the cylinder surface, drawn from previous numerical and experimental studies with minimal blockage. Table~\ref{tab:validation} shows good agreement between the present work and earlier studies, demonstrating the accuracy of the solver. Furthermore, a comparison of the mean drag coefficient obtained at various Richardson numbers in the present work was made with prior studies \citep{sharma2004effect, hasan2013vortex} for Re = 100 and Pr = 0.7. Results are shown in Figure \ref{drag_validation}, and the three studies agree well with each other.

To ensure grid-independence, simulations were conducted on three meshes—coarse (\( 238 \times 410 \)), regular (\( 268 \times 468 \)), and fine (\( 276 \times 486 \)) at \( Ri = 0.13 \), where the flow is strongly nonlinear, and vortex shedding is active. Table~\ref{tab:gridconv} summarizes the results for the drag coefficient, Strouhal number, and Nusselt number. The observed changes in these quantities between the regular and fine grids were minor (less than 1\% in drag, 0.2\% in Nusselt number, and 3.7\% in Strouhal number), indicating that the regular grid offers an optimal balance between accuracy and computational cost. The regular grid was therefore used in all simulations.

\begin{table}
\centering
\small
\setlength{\tabcolsep}{8pt}
\caption{Comparison of time-averaged drag coefficient, Strouhal number, and Nusselt number for \( Re = 100, Ri = 0 \)}
\label{tab:validation}
\begin{tabular}{lccc}
\toprule
Studies & \( \overline{C}_D \) & \( St \) & \( \overline{Nu} \)
\\
\midrule
\cite{sharma2004effect} & 1.56 & 0.1481 & 4.07 \\
\cite{ranjan2008numerical}       & 1.45 & 0.1453 & 4.12 \\
\cite{hasan2013vortex}       & 1.44 & 0.1434 & 4.05 \\
Present work               & 1.48 & 0.1512 & 4.67 \\
\end{tabular}
\end{table}

\begin{table}
\centering
\small
\setlength{\tabcolsep}{8pt}
\caption{Grid-independence test results at \( Re = 100 \) and \( Ri = 0.13 \)}
\label{tab:gridconv}
\begin{tabular}{lcccc}
\toprule
Grids & \( \overline{C}_D \) & \( \overline{Nu} \) & \( St \) & Nearwall grid spacing
\\
\midrule
Coarse (238 × 410)  & 1.518 & 4.47 & 0.1202 & 0.007 \\
Regular (268 × 468) & 1.535 & 4.77 & 0.1641 & 0.005 \\
Fine (276 × 486)    & 1.550 & 4.76 & 0.1704 & 0.003 \\
\bottomrule
\end{tabular}
\end{table}

\subsection{Linear Stability Analysis}

\subsubsection{Mathematical Formulation}
The mathematical formulation is developed by decomposing the flow variables into a steady base component and a time-dependent perturbation. This decomposition forms the basis for deriving the perturbation equations governing the linearized dynamics. In the flow field, each flow variable \( \phi(x, y, \tau) \) is decomposed into a steady base component \( \phi_b(x, y) \) and a time-dependent perturbation \( \tilde{\phi}(x, y, \tau) \) as:
\begin{equation} ~\label{decomposition}
    \phi(x, y, \tau) = \phi_b(x, y) + \tilde{\phi}(x, y, \tau)
\end{equation}
where \( \phi \in \{ u, v, p, \theta \} \) is used as a generalization of flow variables. This formulation is then substituted into the governing equations.

The base flow corresponds to the steady, symmetric solution of the governing equations, denoted by \( \boldsymbol{u}_b \), \( p_b \), and \( \theta_b \). It was computed by solving the governing equations in the left half of the domain using the same numerical methodology as the DNS, with a symmetry boundary condition applied along the centerline (\( x = 0 \)). The complete base flow was reconstructed by reflecting the solution about the centerline using the relations:
\begin{gather}
    u_b(-x, y) = -u_b(x, y) \quad \text{and} \quad   v_b(-x, y) = v_b(x, y)  \notag \\
    p_b(-x, y) = p_b(x, y) \notag \\
    \theta_b(-x, y) = \theta_b(x, y)
\end{gather}
The imposition of symmetry boundary conditions along the centerline enforces a steady base flow, regardless of whether the actual flow field is steady or unsteady.

At the inlet, a uniform vertical flow was imposed with \( u_b = 0 \), \( v_b = 1 \), and \( \theta_b = 0 \). The cylinder surface was treated as no-slip and isothermal, enforcing \( u_b = v_b = 0 \) and \( \theta_b = 1 \). Along the lateral boundaries—including both the centerline and the outer wall—symmetry conditions were imposed: \( u_b = 0 \), \( \partial v_b / \partial x = 0 \), and \( \partial \theta_b / \partial x = 0 \). At the outlet, the convective condition reduces to a homogeneous Neumann condition due to the steady-state assumption, yielding \( \partial u_b / \partial y = \partial v_b / \partial y = \partial \theta_b / \partial y = 0 \).

\begin{figure}[!h]
    \centering
    \begin{subfigure}{0.3\textwidth}
        \centering
        \includegraphics[width=\textwidth]{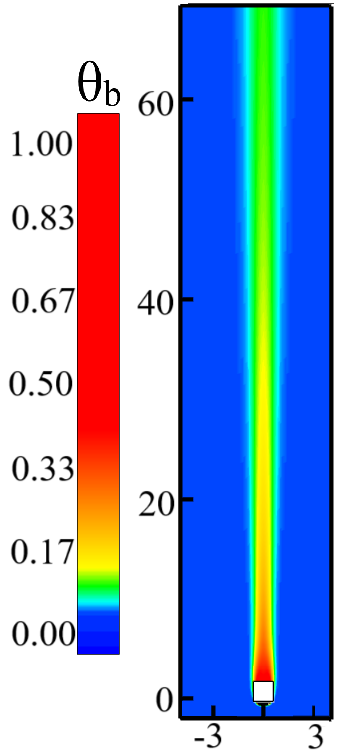}
        \subcaption{Temperature}
    \end{subfigure}
    \hfill
    \begin{subfigure}{0.3\textwidth}
        \centering
        \includegraphics[width=\textwidth]{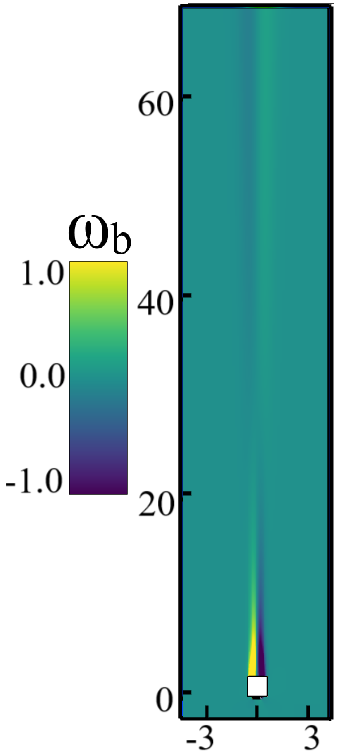}
        \subcaption{Vorticity}
    \end{subfigure}
    \hfill
    \begin{subfigure}{0.33\textwidth}
        \centering
        \includegraphics[width=\textwidth]{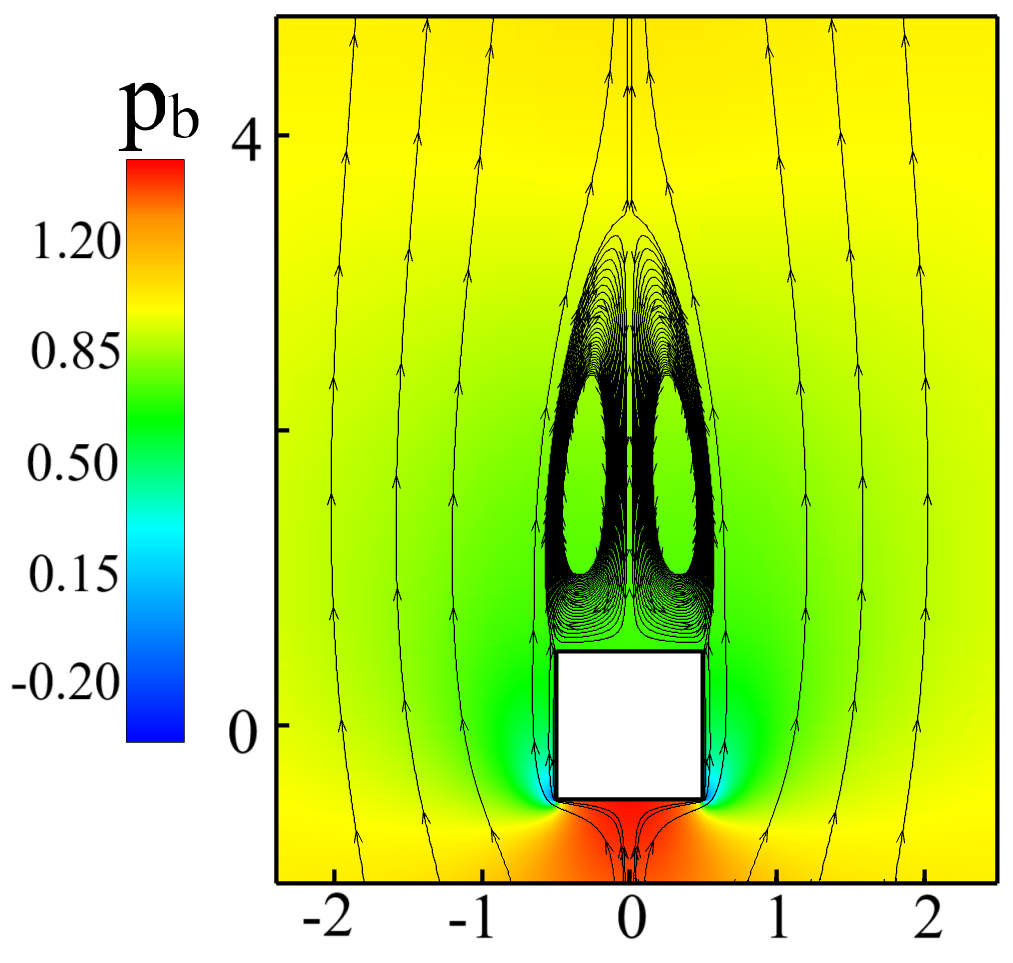}
        \subcaption{Near-field streamlines}
    \end{subfigure}

    \caption{Base flow at \( Ri = 0.10 \) and $Re=100$. (a, b) full-domain temperature and vorticity contours. (c) streamlines and pressure field in the near-field region showing separation and recirculation structures.}
    \label{fig:baseflow}
\end{figure}

An illustration of the base flow for a representative case at \( Ri = 0.10 \), including temperature ($\theta_b$) contours, vorticity ($\omega_b$) contours and pressure distribution superimposed with streamlines, is shown in Figure~\ref{fig:baseflow}.

The base flow variables act like flow field variables with no temporal variation, while still satisfying the governing equations. Hence, when the flow field decomposition relation involving base and perturbation components given by Eq.~\ref{decomposition} is substituted in the governing equations, and the base flow is subtracted, only the base flow terms that appear in product with the perturbation variables are left. Finally, the nonlinear terms arising from products of perturbation components are neglected to obtain the linear perturbation equations. These equations are written component-wise in conservative form as:

\begin{gather}
\frac{\partial \tilde{u}}{\partial x} + \frac{\partial \tilde{v}}{\partial y} = 0 \notag \\
\frac{\partial \tilde{u}}{\partial \tau} 
+ 2 \frac{\partial (u_b \tilde{u})}{\partial x} 
+ \frac{\partial (v_b \tilde{u})}{\partial y} 
+ \frac{\partial (u_b \tilde{v})}{\partial y}
+ \frac{\partial \tilde{p}}{\partial x}
- \frac{1}{Re} \left( \frac{\partial^2 \tilde{u}}{\partial x^2} + \frac{\partial^2 \tilde{u}}{\partial y^2} \right) = 0 \notag \\
\frac{\partial \tilde{v}}{\partial \tau} 
+ \frac{\partial (u_b \tilde{v})}{\partial x} 
+ \frac{\partial (v_b \tilde{u})}{\partial x} 
+ 2 \frac{\partial (v_b \tilde{v})}{\partial y}
+ \frac{\partial \tilde{p}}{\partial y}
- Ri\, \tilde{\theta}
- \frac{1}{Re} \left( \frac{\partial^2 \tilde{v}}{\partial x^2} + \frac{\partial^2 \tilde{v}}{\partial y^2} \right) = 0 \notag \\
\frac{\partial \tilde{\theta}}{\partial \tau}
+ \frac{\partial (u_b \tilde{\theta})}{\partial x}
+ \frac{\partial (\theta_b \tilde{u})}{\partial x}
+ \frac{\partial (v_b \tilde{\theta})}{\partial y}
+ \frac{\partial (\theta_b \tilde{v})}{\partial y}
- \frac{1}{Re\,Pr} \left( \frac{\partial^2 \tilde{\theta}}{\partial x^2} + \frac{\partial^2 \tilde{\theta}}{\partial y^2} \right) = 0 \label{eq:pert_all}
\end{gather}


Assuming normal mode form \( \tilde{\phi}(x,y,\tau) = \hat{\phi}(x,y)e^{\lambda \tau} \), which assumes an exponentially varying temporal variation, it is possible to morph the perturbation equations into a generalized eigenvalue problem (GEVP) formulation, where the time dependence in the equations is represented implicitly through the value of $\lambda$. The GEVP is given as:
\begin{equation}
    \mathcal{A}\hat{\phi} = \lambda \mathcal{B}\hat{\phi}
\end{equation}


where $\mathcal{A}$ and $\mathcal{B}$ matrix operators corresponding to the linearized perturbation equations, \( \hat{\phi} = [\hat{u}, \hat{v}, \hat{p}, \hat{\theta}]^T \) represent the spatial variation of the perturbation modes, and \( \lambda \in \mathbb{C} \) is the eigenvalue. The matrix operator $\mathcal{B}$ represents the coefficients of the perturbation variables in the linear perturbation equations arising due to the time-derivative terms after applying the normal-mode approximation, and is solely made up of constants. The matrix operator $\mathcal{A}$ contains the remaining coefficients of the perturbation variables, and is made up of constants and the base flow variables. The boundary conditions for the perturbation fields were as follows. At the inlet, all perturbation variables were set to zero: \( \hat{u} = \hat{v} = \hat{p} = \hat{\theta} = 0 \). Along the left and right confining boundaries, symmetry-compatible conditions were applied: \( \partial \hat{v} / \partial x = 0 \), \( \hat{u} = 0 \), \( \partial \hat{\theta} / \partial x = 0 \), and \( \partial \hat{p} / \partial x = 0 \). At the outlet, a convective boundary condition was imposed of the form \( V_c \, \partial \hat{q} / \partial y = -\lambda \hat{q} \), for each \( \hat{q} \in \{ \hat{u}, \hat{v}, \hat{\theta} \} \), while the pressure field was implicitly accounted for through the continuity equation. On the cylinder surface, homogeneous Dirichlet conditions were imposed: \( \hat{u} = \hat{v} = \hat{p} = \hat{\theta} = 0 \).

In general, the values of eigenvalues to be obtained as a solution to the GEVP in the present work would be complex. It is worth noting that since the linear perturbation equations are being considered, the final perturbation field would be represented as the linear combination of all the perturbation modes obtained corresponding to the found values of $\lambda$:

\begin{align}~\label{field}
    \tilde{\phi}(x, y, \tau) = \Sigma_{k=1}^nC_k\hat{\phi_k}(x,y)e^{\lambda_k \tau}
\end{align}

Where \textit{n} is the total number of eigenvalues of the system, and $\hat{\phi_k}$ is the solution to the GEVP for eigenvalue $\lambda_k$. It can be readily observed that the perturbation field in equation \eqref{field} is oscillatory in nature, having a time-varying oscillation amplitude (which depends on the real parts of $\lambda$) and an oscillation frequency comprised of multiple fundamental frequencies (which depend on the imaginary parts of $\lambda$). If all eigenvalues have negative real parts, perturbations decay with time and only the steady-symmetric components remain. Conversely, the presence of any eigenvalue with a positive real part leads to temporal growth of perturbations, indicating flow unsteadiness. The eigenmodes ($\hat{\phi}$) corresponding to the most unstable eigenvalues can be analyzed for a systematic investigation of the mechanisms responsible for flow transitions. 

\subsubsection{Numerical method}

The linear perturbation equations were discretized using a second-order central difference scheme on a Cartesian grid. All perturbation variables were defined at cell centers for consistency and ease of matrix assembly. Since the base flow was computed on a staggered grid, linear interpolation was used to express base quantities at cell centers. The boundary values were also interpolated to preserve spatial accuracy wherever required.

At the lateral artificial boundaries, where central differencing was not applicable, a first-order one-sided scheme was employed to enforce symmetry-compatible conditions. However, at the outlet boundary, which plays a critical role in determining far-field instability, a second-order backward difference scheme was retained to accurately capture gradients in the streamwise direction.

The discretized system was assembled into a generalized eigenvalue problem of the form
\[
\mathbf{A} \hat{\boldsymbol{\Phi}} = \lambda \mathbf{B} \hat{\boldsymbol{\Phi}}
\]
where \( \mathbf{A} \) and \( \mathbf{B} \) are large, sparse matrices, and \( \hat{\boldsymbol{\Phi}} \) is a column vector containing all perturbation variables. For each interior cell, four discretized equations were constructed, corresponding to the four governing equations—yielding a total of \( 4mn \) equations and unknowns where $m$ and $n$ denote the number of grid cells along \textit{x} and \textit{y} respectively.

Matrix assembly was carried out in a structured manner, taking care of the discretized terms for each variable as well as the eigenvalue terms. The resulting generalized eigenvalue problem was solved using MATLAB’s built-in \texttt{eigs()} function. The 50 eigenvalues with the largest real parts were extracted using the implicitly restarted Arnoldi method to investigate the most unstable eigenmodes. A convergence tolerance of \( 10^{-8} \) was imposed to ensure spectral accuracy of the computed modes.

\subsubsection{Validation of LSA results}~\label{sec:LSAvalid}

To evaluate the accuracy of LSA, comparisons were made between eigenvalue-based predictions and both DNS results and previous LSA studies. The imaginary part of the most unstable eigenvalue, denoted by \( \lambda_i \), provides the non-dimensional angular frequency of the corresponding perturbation mode. The Strouhal number based on the most unstable eigenvalue, denoted by \( St_\lambda \), can be expressed as
\[
St_\lambda = \frac{fB}{U_{\infty}} = \frac{\lambda_i}{2\pi}
\]
The unstable eigenmode with the largest positive real part dominates the perturbation dynamics and thus determines the global oscillation frequency of the system. Table~\ref{table:st_validation} compares the Strouhal numbers obtained from DNS with those predicted using LSA, demonstrating excellent agreement between the two approaches.

\begin{table}
\centering
\caption{Comparison of Strouhal numbers obtained from DNS and LSA at $Re=100$.}
\label{table:st_validation}
\begin{tabular}{cccc}
\toprule
Ri & $St$ (DNS) & $St_\lambda$ (LSA) & \% Error \\
\midrule
0.10 & 0.1520 & 0.1519 & 0.07\% \\
0.12 & 0.1569 & 0.1576 & 0.45\% \\
0.14 & 0.1621 & 0.1627 & 0.37\% \\
0.65 & 0.2153 & 0.2148 & 0.23\% \\
\bottomrule
\end{tabular}
\end{table}

To further validate the LSA methodology, results for the case \( \mathrm{Ri} = 0 \) (pure forced convection) were compared with previous LSA studies on flow past square cylinders. As shown later in Sec.~\ref{modes_analysis} and Figure~\ref{REM}, the contours of the stream function $\psi$ of the most unstable eigenmode exhibit the characteristic inverted heart-shaped structures associated with shear-layer instability (vortex shedding), as observed by~\citet{kelkar1992numerical} and~\citet{camarri2007inversion}, confirming the fidelity of the current implementation.

Additionally, the eigenvalue spectrum near critical Richardson numbers indicates that the transition from unsteady to steady flow, and vice versa, corresponds to the emergence of a complex-conjugate pair of eigenvalues crossing the imaginary axis, a hallmark of Hopf bifurcation. This behavior is seen clearly near \( \mathrm{Ri} = 0.15 \) and \( \mathrm{Ri} = 0.62 \) for LSA, as shown later in Sec.~\ref{sec:eigenvalue_analysis} and Figure~\ref{EV}, and agrees closely with the DNS results, which indicate the suppression of near-field vortex shedding and the onset of far-field instability, respectively (Figure~\ref{vorticity}). The DNS flow fields display approximately sinusoidal oscillations around the former steady state (Figures ~\ref{signal1near} and \ref{signal1_far}), consistent with oscillatory behaviour associated with a Hopf bifurcation, as discussed by~\citet{strogatz2018nonlinear}. In summary, the Strouhal number predictions, eigenmode structures, and eigenvalue behavior collectively affirm the validity and reliability of the linear stability analysis employed in this study.

\section{Results and Discussion}
  
In this section, we first examine the instantaneous and time-averaged flow fields obtained from direct numerical simulations (DNS), followed by a thorough investigation of eigenmodes obtained from linear stability analysis, to understand the mechanisms of flow transitions associated with near-field vortex shedding suppression and the onset of far-field instability. 

\subsection{Instantaneous flow characteristics}

We begin by examining the instantaneous flow behaviour across a range of Richardson numbers at a fixed Reynolds number of $\mathrm{Re} = 100$, particularly focusing on the evolution of near- and far-field structures associated with the observed flow transitions.

\subsubsection{Near-field characteristics}

\begin{figure}
    \centering
        \includegraphics[width=0.95\textwidth]{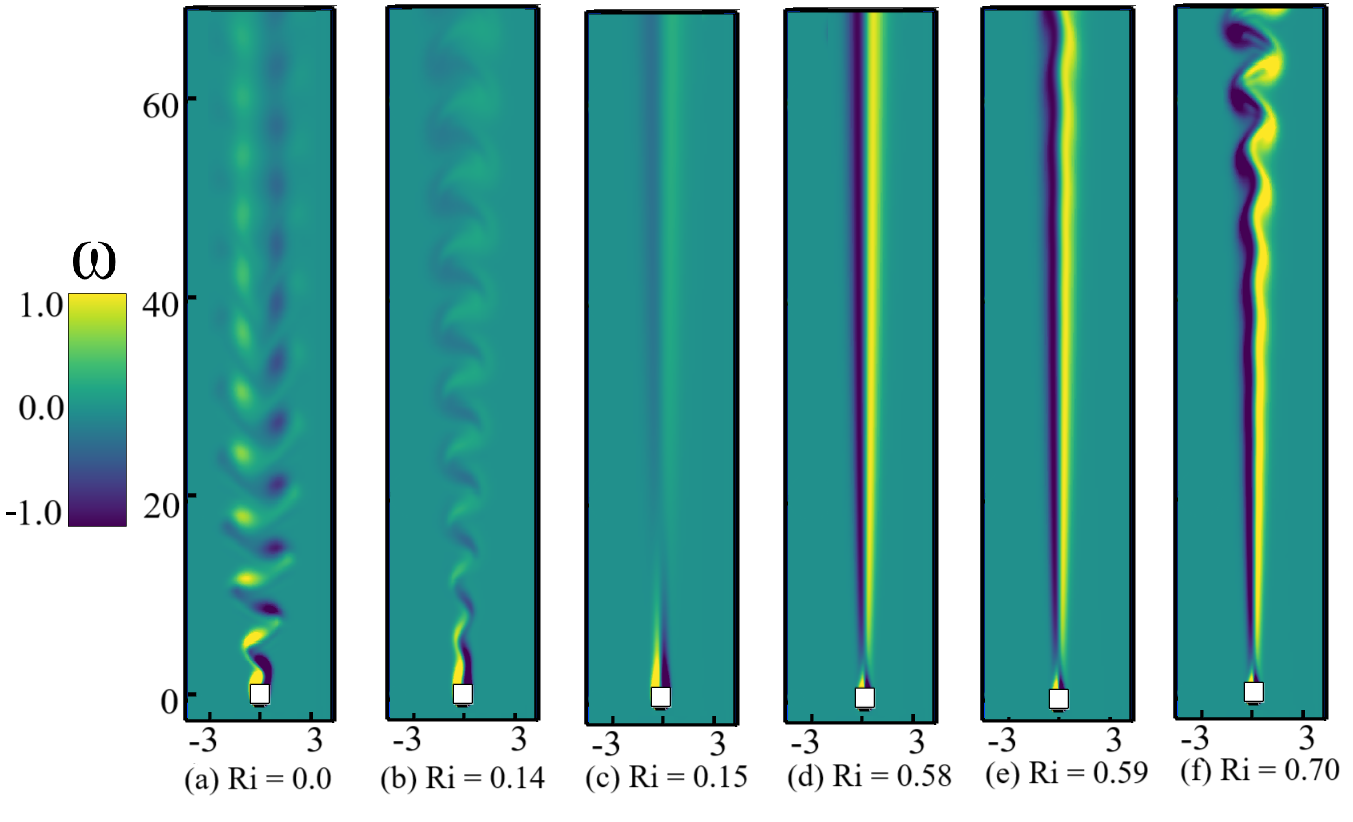}
    \caption{Instantaneous vorticity contours at various Ri and $Re=100$.}
   \label{vorticity}
\end{figure}

\begin{figure}
    \centering
        \includegraphics[width=0.95\textwidth]{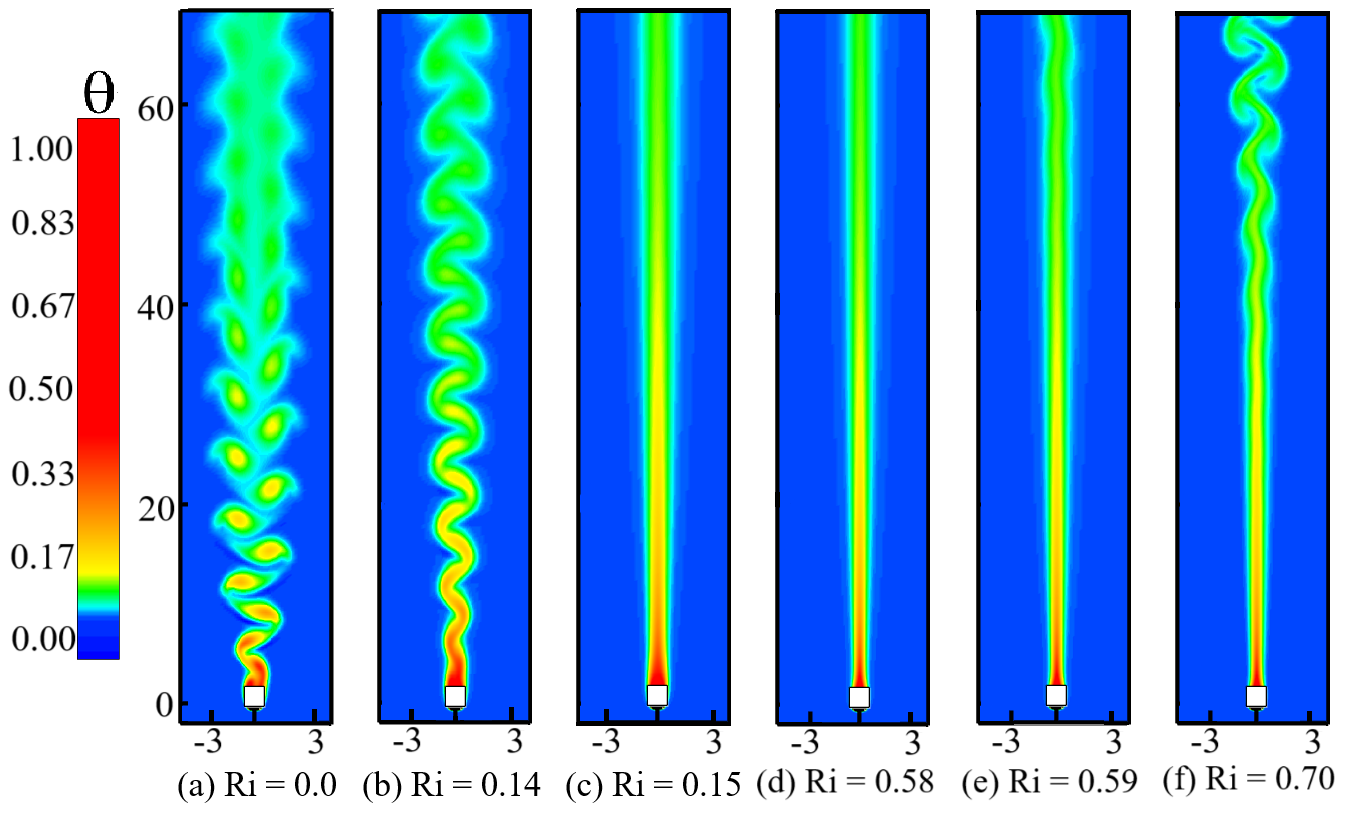}
    \caption{Instantaneous temperature contours at various Ri and $Re=100$.}
   \label{inst_temp}
\end{figure}

\begin{figure}
    \centering
        \includegraphics[width=0.95\textwidth]{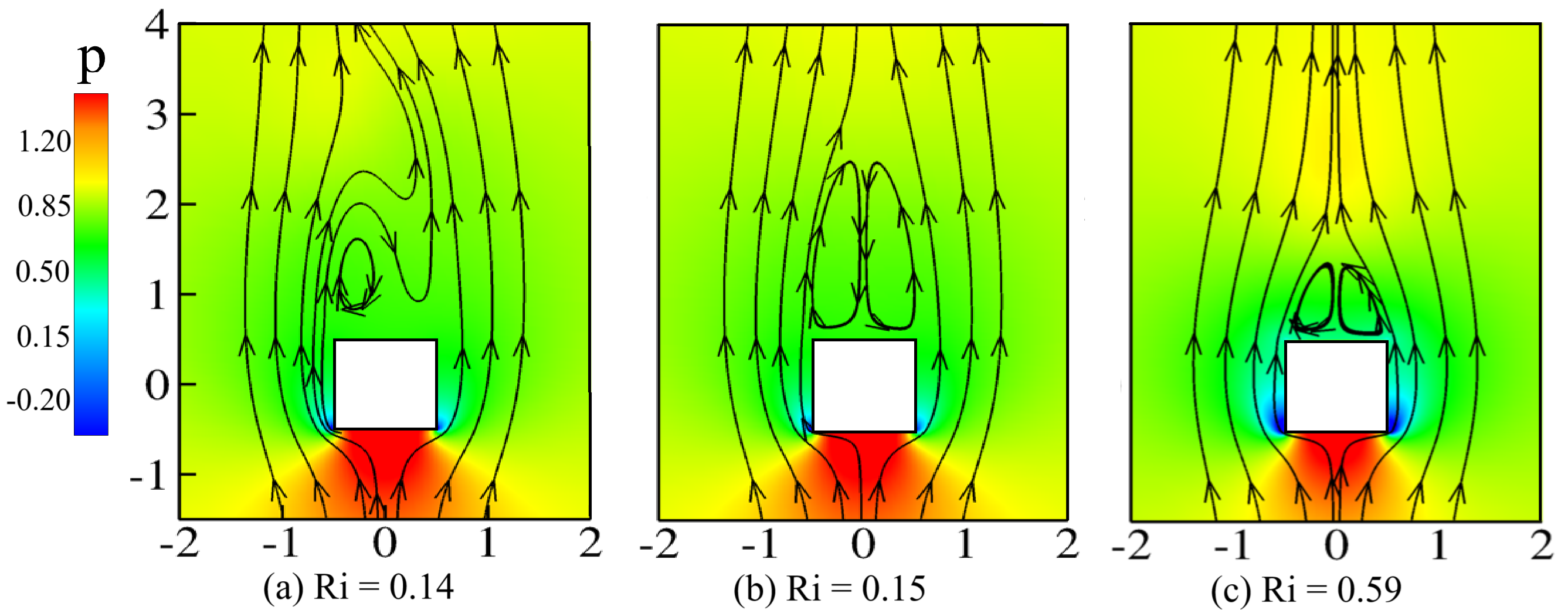}
    \caption{Instantaneous streamline plots and pressure contours in the near-field region for $Re=100$, (a) Ri=0.14 (b) Ri=0.15 and (c) Ri=0.59.}
    \label{stream}
\end{figure}

\begin{figure}
    \centering
    \begin{minipage}[t]{0.5\textwidth}
        \centering
        \includegraphics[width=0.95\textwidth]{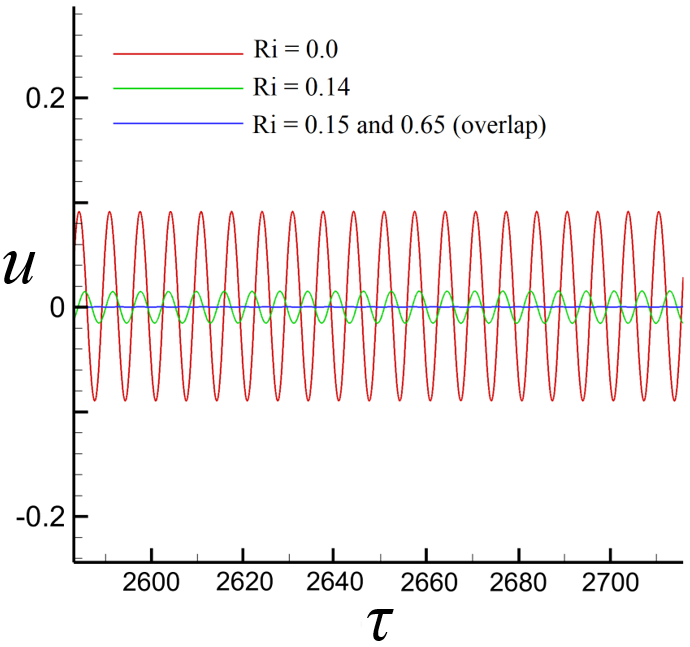}
        \caption{Transverse velocity signals at $x = 0$, \\ $y = 5$ for three different Richardson numbers.}
        \label{signal1near}
    \end{minipage}%
    \hfill
    \begin{minipage}[t]{0.5\textwidth}
        \centering
        \includegraphics[width=0.95\textwidth]{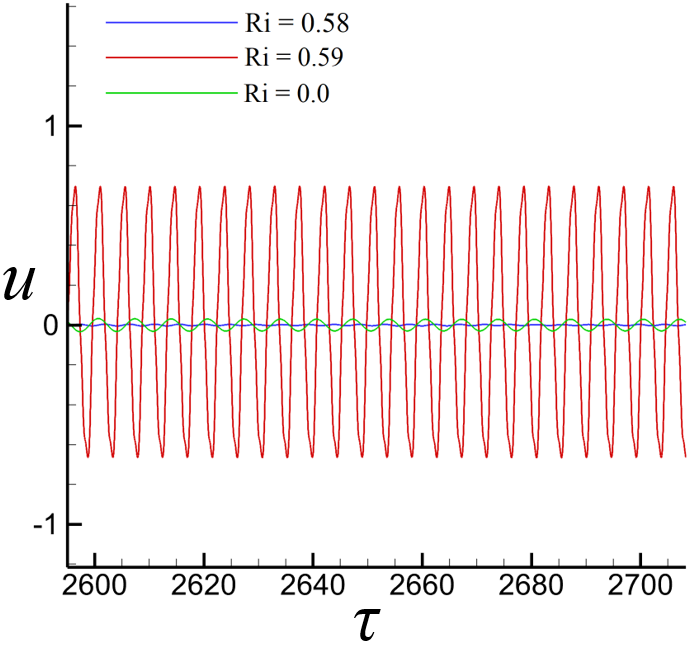}
        \caption{Transverse velocity signal at $x = 0$, \\ $y = 65$ for three different Richardson numbers.}
        \label{signal1_far}
    \end{minipage}
\end{figure}

The instantaneous vorticity ($\omega$) and temperature ($\theta$) contours for different Richardson numbers are shown in Figures~\ref{vorticity} and~\ref{inst_temp}, respectively. For low values of $\mathrm{Ri}$ (cases (a) and (b), i.e., Ri=0 and 0.14), vortex shedding is clearly active. However, at $\mathrm{Ri} = 0.15$ (case (c)), vortex shedding is completely suppressed in the entire computational domain, and the wake transitions to a steady regime. This change is also apparent in the streamline plots (Figures~\ref{stream}(b) and (c)), where a stable pair of symmetric recirculating vortices is observed immediately downstream of the cylinder. The transition is further supported by the temporal signals of flow variables shown in Figure~\ref{signal1near}, which demonstrate a progressive decay in oscillation amplitude with increasing $\mathrm{Ri}$ in the near-field region. At $\mathrm{Ri} = 0.15$, these oscillations vanish entirely, marking the first flow transition by vortex shedding suppression. \citet{sharma2004effect} also reported an identical Ri (Ri=0.15) for the vortex shedding suppression.

\subsubsection{Far-field characteristics}~\label{farwake}
The development of far-field unsteadiness is clearly visible in the instantaneous vorticity and temperature contours (Figures~\ref{vorticity} and~\ref{inst_temp}). At $\mathrm{Ri} = 0.59$, in the original computational domain, a second critical transition occurs beyond which the flow becomes unsteady again in the far-field, resembling a plume-like flow and thermal field. The transition is further illustrated in Figure~\ref{signal1_far}, which shows a sudden increase in oscillation amplitude at $\mathrm{Ri} = 0.59$. None of the earlier studies, including \citet{sharma2004effect}, could show such a far-field unsteadiness, which may be due to their insufficient computational domain length downstream of the cylinder. The vorticity and temperature contours further suggest that the onset point of the far-field instability moves progressively closer to the cylinder as $\mathrm{Ri}$ increases. At any fixed location along the domain centreline in the far field, the amplitude of oscillations increases with $\mathrm{Ri}$ for $\mathrm{Ri} \geq 0.59$ (Figure~\ref{signal1_far}), indicating a strengthening of the instability. This raises an important question: is $\mathrm{Ri} = 0.59$ the absolute critical value for the onset of the far-field instability, or would the instability emerge at lower Richardson numbers if the domain were extended?

\begin{figure}[htbp]
    \centering
    \begin{subfigure}[t]{0.48\textwidth}
        \centering
        \includegraphics[width=\textwidth]{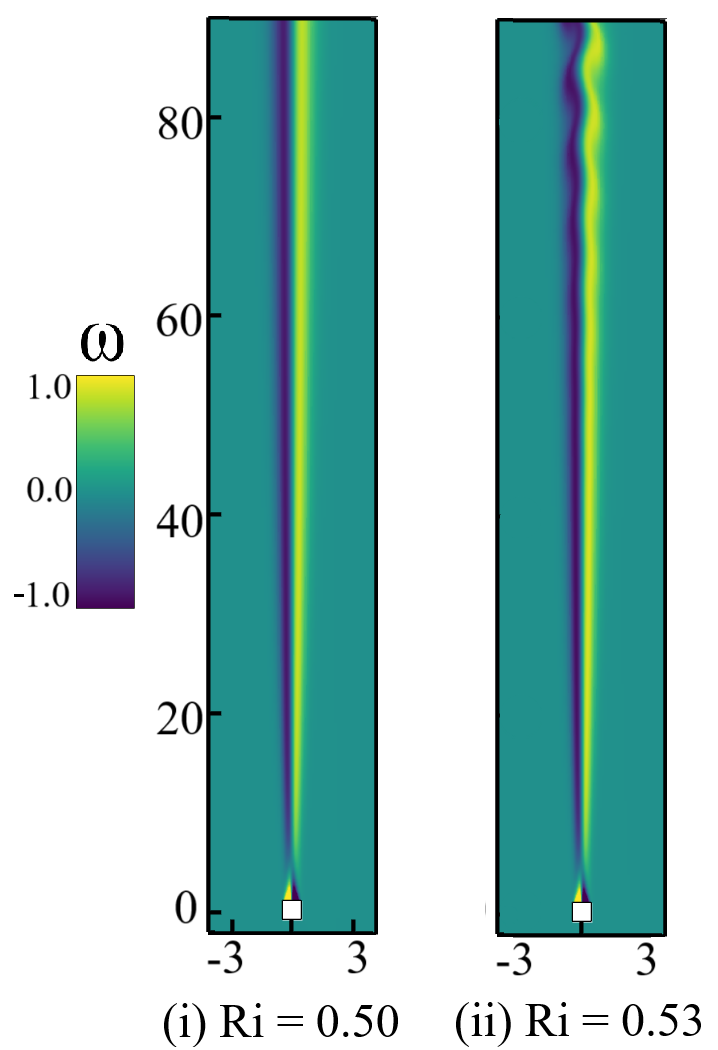}
        \caption{Vorticity contours in the extended domain}
            \label{for_critical3}
    \end{subfigure}
        \hfill
    \begin{subfigure}[t]{0.5\textwidth}
        \centering
        \includegraphics[width=\textwidth]{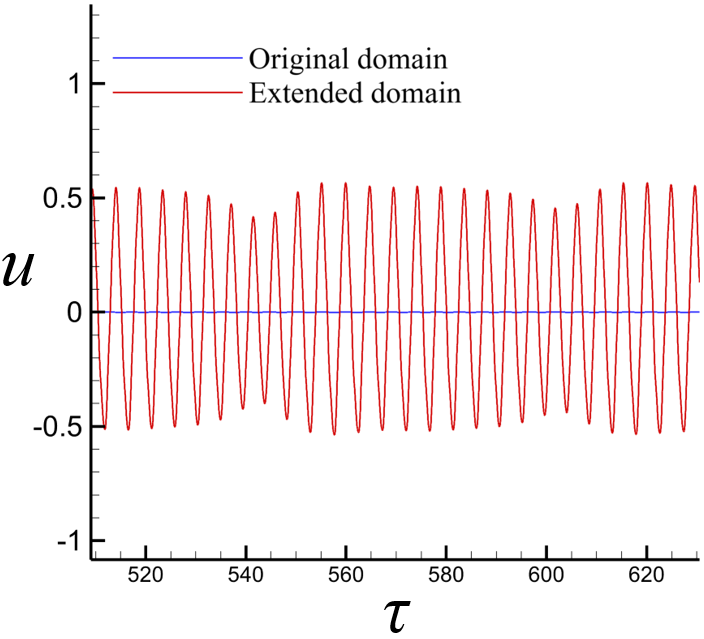}
        \caption{Far-field signals at $x = 0$, $y = 65$ for $\mathrm{Ri} = 0.55$.}
        \label{for_critical2}
    \end{subfigure}%
    \caption{Domain dependence test results of the far-field instability for Re=100.}
\end{figure}

Figure~\ref{for_critical3}, which presents DNS results for a vertically extended domain at $\mathrm{Ri} = 0.53$, offers insight into this question. This new domain is extended by 20 units more along the \textit{y-}direction compared to the original domain used in this present study (Figure \ref{domain}). In contrast to the original domain, where the flow remains steady at $\mathrm{Ri} = 0.53$, the new extended domain exhibits far-field unsteadiness. This indicates that the steady behaviour observed earlier was a consequence of insufficient streamwise extent, rather than the absence of instability. The far-field instability at any particular Ri happens beyond a specific downstream location. If the domain size is smaller than the downstream distance required for the onset of unsteadiness, then a steady regime is observed in the entire domain of interest, which is necessarily not true to identify the flow as steady. Therefore, the onset of the instability is contingent on the downstream distance over which buoyancy can act.

The spatial dependence is further corroborated by the comparison of transverse velocity signals for $\mathrm{Ri} = 0.55$, at which the flow in the original domain is closer to the transitional Ri, in Figure~\ref{for_critical2}. At a location ($y=65$) in the far-field common to both domains, the extended domain exhibits clear oscillations, while the original domain does not. However, this discrepancy diminishes as one approaches the cylinder, where the signals become nearly identical. For instance, comparing the attenuation of transverse velocity in the near-field for $\mathrm{Ri} = 0.65$ (Figure~\ref{signal1near}) and the strong far-field oscillations at $\mathrm{Ri} = 0.59$ (Figure~\ref{signal1_far}) highlights the spatially confined nature of the instability even in the original domain. Importantly, although the far-field region differs significantly between the two domains, the key integral quantities, such as the mean drag coefficient and the mean cylinder Nusselt number, remain essentially unchanged. This suggests that, for cases where the far-field instability originates sufficiently downstream, its influence does not propagate upstream to alter the near-field dynamics.

The far-field instability, as discussed earlier, is expected to appear naturally beyond a specific streamwise length at any particular combination of Reynolds number and Richardson number. The streamwise length in a computational model is therefore crucial to capture this transition. However, it is quite challenging to know the right downstream length a priori.
The critical Richardson number for far-field instability in the computational model is therefore not unique but domain-dependent, in the sense that the flow may be deemed steady if the chosen downstream length fails to capture the naturally occurring unsteadiness that persists beyond it. For the extended domain, unsteadiness appears between $\mathrm{Ri} = 0.50$ and $0.53$ (Figure~\ref{for_critical3}), whereas in the original domain it emerges between $\mathrm{Ri} = 0.58$ and $0.59$ (Figure~\ref{signal1_far}). This supports the interpretation that far-field instability is not only governed by a fixed $\mathrm{Ri}$, but by a combination of Re, Ri and the critical downstream distance beyond which buoyancy-induced perturbations can grow sufficiently. For clarity and consistency, all references to the critical point for the second transition in this study are made with respect to the original domain.

\subsubsection{Strouhal number behaviour}

Strouhal numbers were determined from Fourier analysis of transverse velocity signals. Near-field signals were used in the vortex shedding regime, while far-field signals were analyzed in cases exhibiting far-field unsteadiness. Representative Strouhal numbers obtained from DNS are reported in Table~\ref{table:st_validation}. In the active vortex shedding regime, the Strouhal number increases with $\mathrm{Ri}$ even though the amplitude of oscillations gradually diminish. This happens until $\mathrm{Ri} = 0.15$, beyond which the Strouhal number drops abruptly to zero due to the suppression of shedding. This trend is consistent with the experimental observations of \citet{chang1990effect}. \textcolor{\mycolorb}{A similar increase in vortex shedding frequency with increasing cylinder heating has also been reported by \citet{noto1987numerical} and \citet{jain1979unsteady}, further supporting the observation that buoyancy can alter the characteristic timescale of wake oscillations prior to the complete suppression of shedding.} 

The far-field unsteadiness is characterized by Strouhal numbers significantly higher than those observed in the near-field, indicating the emergence of a distinct, higher-frequency unsteadiness. This can be inferred by comparing values of $St$ reported for near-field unsteadiness at lower $Ri$ with the values reported for far-field unsteadiness at higher $Ri$ in Table~\ref{table:st_validation} (also see $St$ values for far-field unsteadiness shown later for $\mathrm{Ri} = 0.8$ in Table~\ref{tab:St_vs_Re}). Moreover, the Strouhal number continues to increase with $\mathrm{Ri}$ in the far-field regime as well ($St = 0.2153$ for $\mathrm{Ri} = 0.65$ and $St = 0.2400$ for $\mathrm{Ri} = 0.8$ at $\mathrm{Re} = 100$). The observed increase in the Strouhal number with increasing $\mathrm{Ri}$ in the unsteady regime may also be interpreted in terms of increasing flow inertia associated with buoyancy effects. As buoyancy adds streamwise momentum to the flow, the axial (vertical) velocity increases, thereby modifying the characteristic convective timescale and contributing to higher oscillation frequencies. This motivates examining the frequency using velocity scales that account for buoyancy-induced momentum augmentation. A more detailed discussion of the downstream velocity amplification and the corresponding global and local frequency scalings is provided later (Secs.~\ref{meanflowanalysis} and \ref{sec:Re_effect}).

\subsubsection{Landau model for determining critical points of transition}

\begin{figure}
\centering
\includegraphics[width=0.5\textwidth]{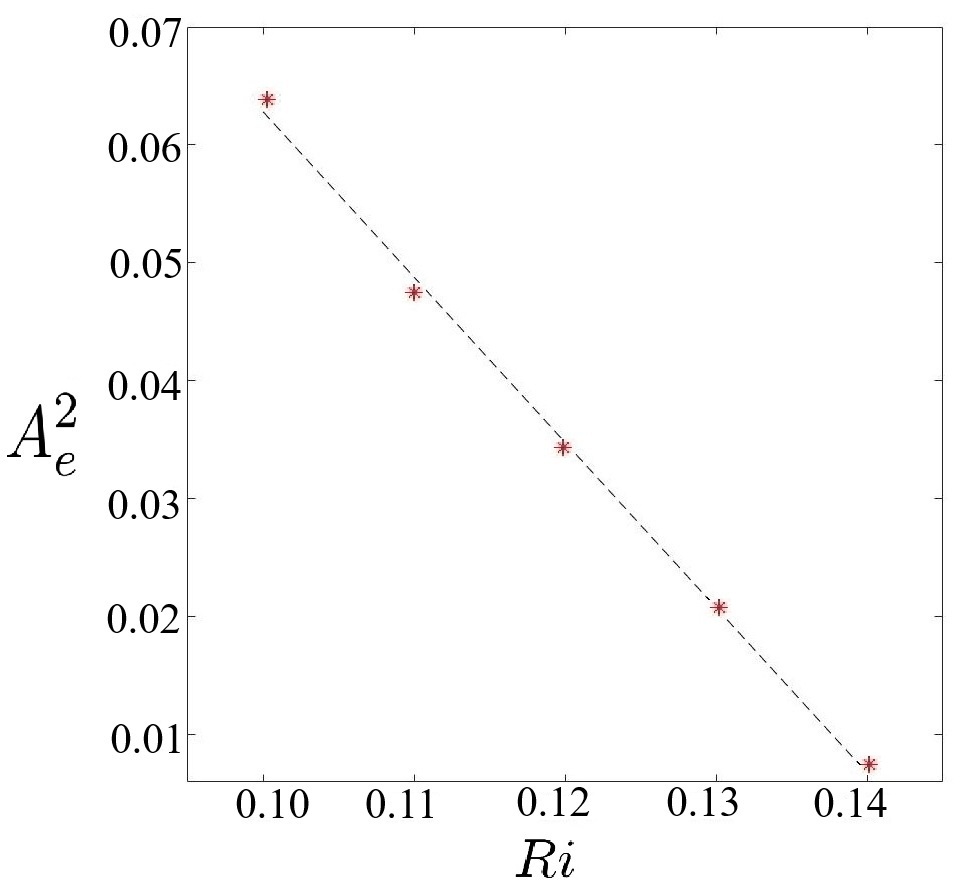}
\caption{Variation of saturation amplitude of $C_L$ with varying Ri at Re=100.}
\label{landau}
\end{figure}

The Landau model is used in the present work to determine the exact critical point of transition in a system showing a Hopf bifurcation (\citet{drazin2004hydrodynamic}). It has been proven appropriate in previous works \citep{stuart1960non,watson1960non} to work on systems for which the unstable state just after achieving instability is a small-amplitude, sinusoidal, limit cycle oscillation about the former steady state, such as the present case (can be verified by referring to Figure \ref{signal1near}). For supercritical instabilities governed by the Landau equation, one can show that the equilibrium amplitude of oscillations near the critical point varies with the control parameter (Ri in the current problem for a constant Re) as: 

\begin{equation}\label{landaueq}
    A_{e}^2 = C(Ri - Ri_c)
\end{equation}

where $A_e$ is the equilibrium amplitude of oscillation of the flow variables, $Ri_c$ is the critical Richardson number, and C is a constant. This equation has been used here to determine the exact value of the critical Ri for the vortex shedding suppression. The constant C in equation~\ref{landaueq} is a negative constant (so that, for $Ri > Ri_c$, equilibrium amplitude does not exist and a stable system without oscillations is obtained). Here, the amplitude of the time signal of the lift coefficient has been used as $A_e$. This has been shown to give excellent results in previous studies \cite{hasan2013vortex, mishra2020passive, meng2021wake} for the fact that the lift coefficient signal captures the flow stability well, as the oscillations in its signal achieve a constant equilibrium amplitude for unstable cases, while they decay to zero for stable cases. Also, the rms values of the lift coefficient were shown to be much more sensitive to flow unsteadiness compared to other flow parameters. Thus, the square of the amplitude of oscillation of the lift coefficient has been plotted against the Richardson number in Figure \ref{landau}. Based on the extrapolation of the line produced by the least-square approach (shown in Figure \ref{landau}), the exact critical Richardson number for the first transition corresponding to vortex shedding suppression is calculated to be $Ri_{c} = 0.1448$, confirming the values computed using DNS that claim that the exact critical point lies between Ri = 0.14 and Ri = 0.15. The linear trend in the present study confirms that the first transition is a supercritical Hopf bifurcation. 

The Stuart–Landau model is not employed to analyze the far-field transition, as this transition does not occur at a fixed critical Richardson number. Instead, it is governed by a critical downstream distance beyond which the instability emerges, making it inherently domain-dependent. As a result, the fundamental assumption of a well-defined bifurcation parameter (such as Ri) controlling the onset of global unsteadiness does not hold, rendering the Landau model inapplicable in this regime.

\subsection{Time-averaged flow}~\label{meanflowanalysis}

\begin{figure}
    \centering
        \includegraphics[width=0.8\textwidth]{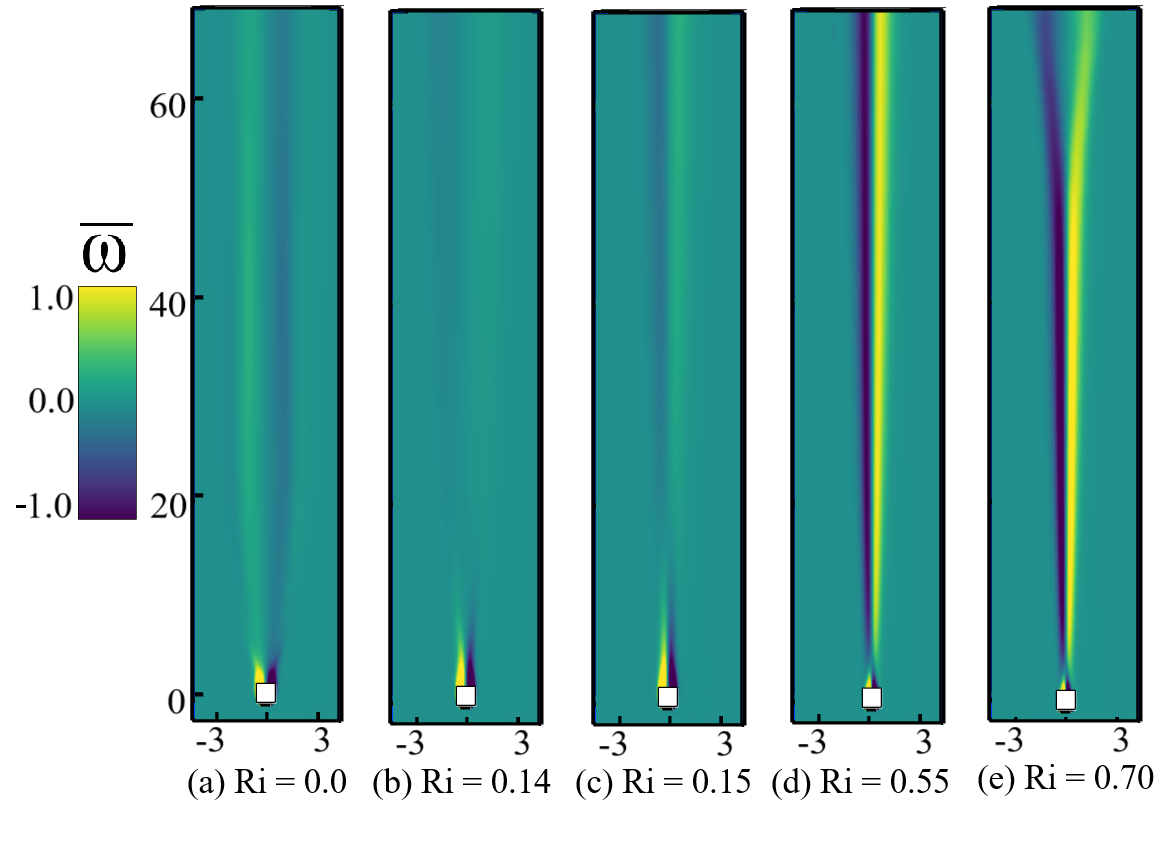}
    \caption{Time-averaged vorticity contours at various Ri for Re=100.}
    \label{mean vorticity}
\end{figure}

\begin{figure}
    \centering
        \includegraphics[width=0.8\textwidth]{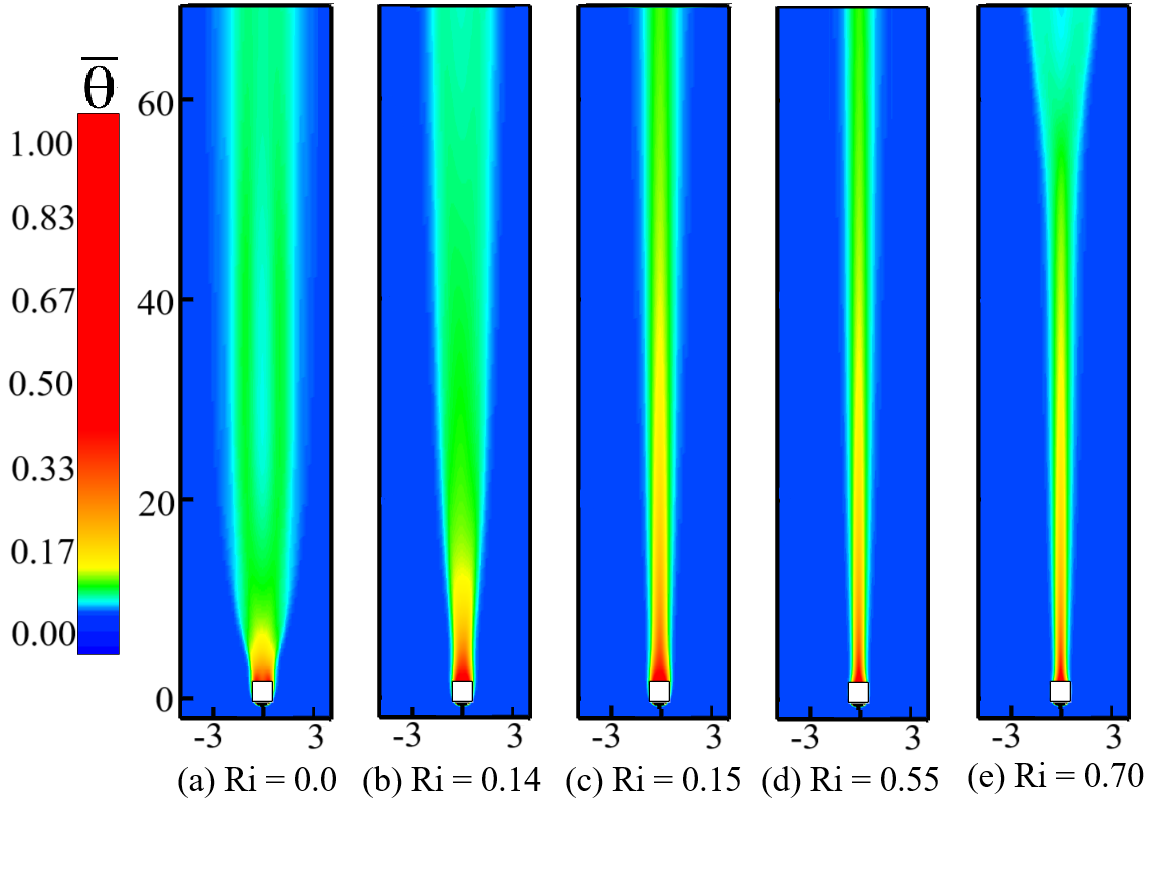}
    \caption{Time-averaged temperature contours at various Ri for Re=100.}
    \label{mean tempe}
\end{figure}

The time-dependent flow variables are averaged over large periods of time (covering a minimum of 10 flow-through cycles) and analyzed in this section for a fixed Reynolds number of $\mathrm{Re} = 100$. Time-averaged vorticity and temperature contours are presented in Figures \ref{mean vorticity} and \ref{mean tempe}, respectively. The vortex shedding suppression causes the wake/shear layer in the near field to become thinner, while the onset of far-field instability makes the shear layer diverge transversely in the far-field. The vorticity contours reveal an overall increase in vorticity magnitude with increasing Richardson number. Reasons for such behaviour, along with the characteristics of the drag coefficient, flow structures, and vorticity inversion, are presented in this section.

~\\
\subsubsection{Flow structure and transport characteristics}~\label{sec:transport}

\begin{figure}
    \centering
    \begin{subfigure}{0.33\textwidth}
        \includegraphics[width=1\textwidth]{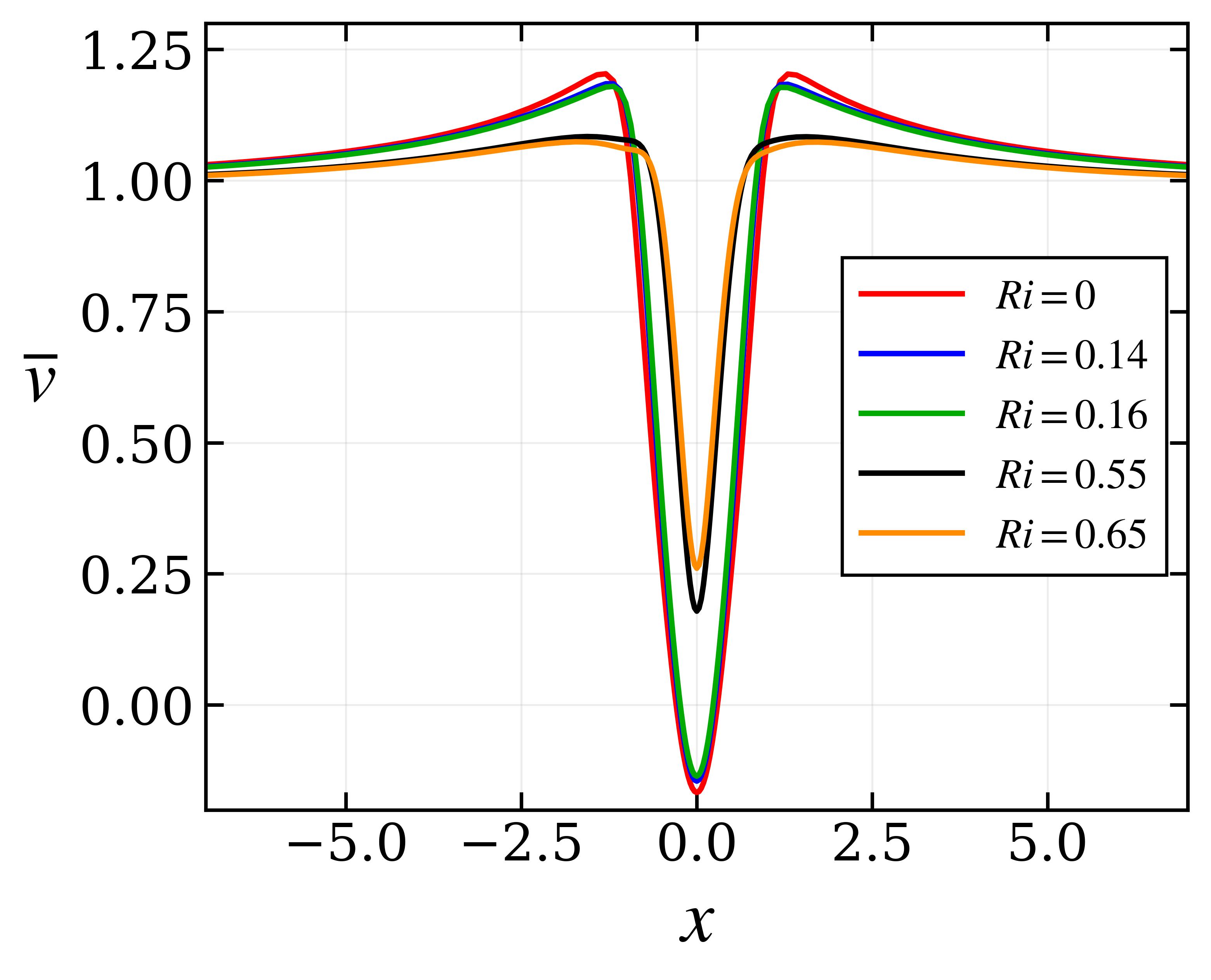}
        \subcaption{$y$ = 1.8}
    \end{subfigure}
    \hfill
    \begin{subfigure}{0.33\textwidth}
        \includegraphics[width=1\textwidth]{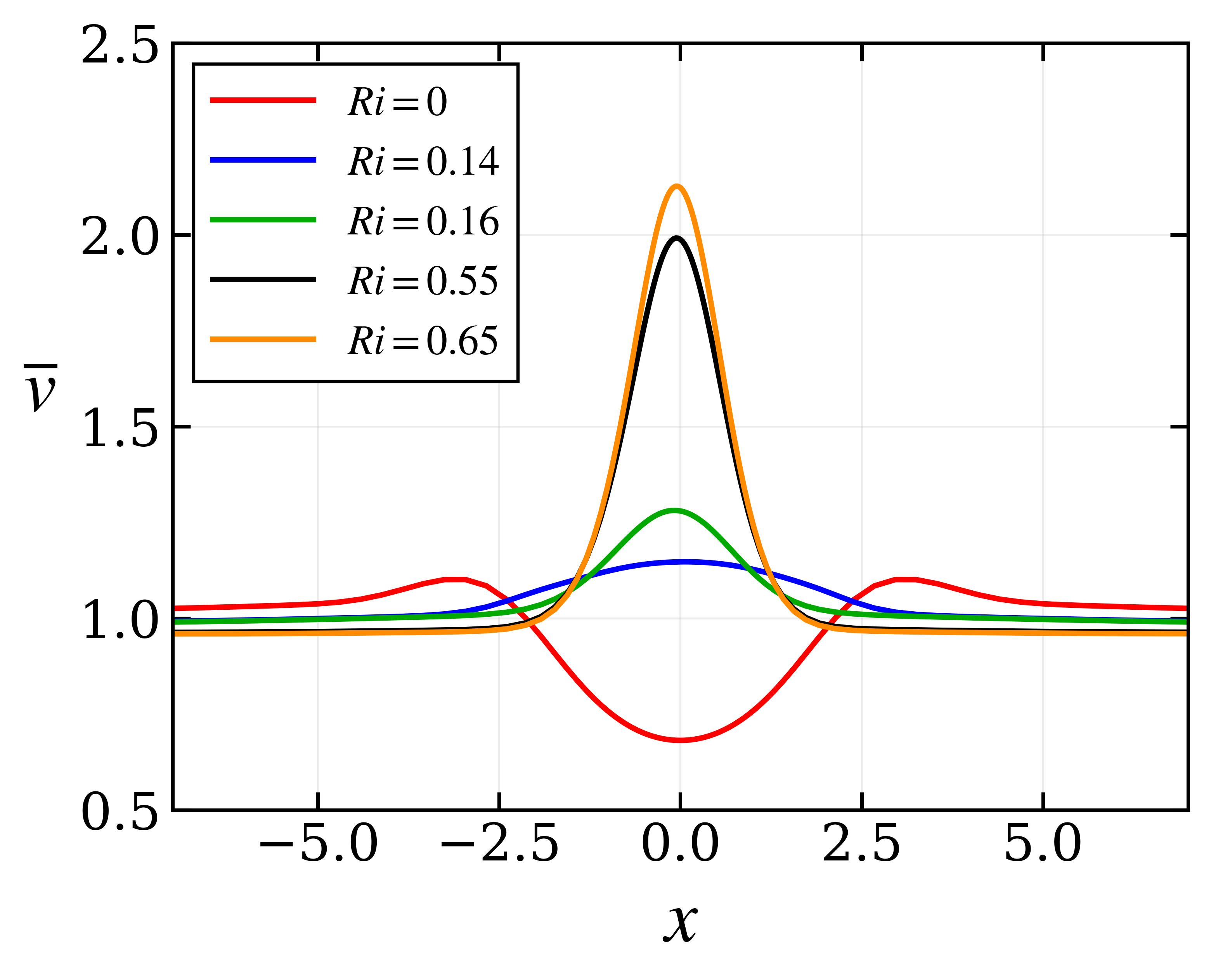}
        \subcaption{$y$ = 30}
    \end{subfigure}
    \hfill
    \begin{subfigure}{0.32\textwidth}
        \includegraphics[width=1\textwidth]{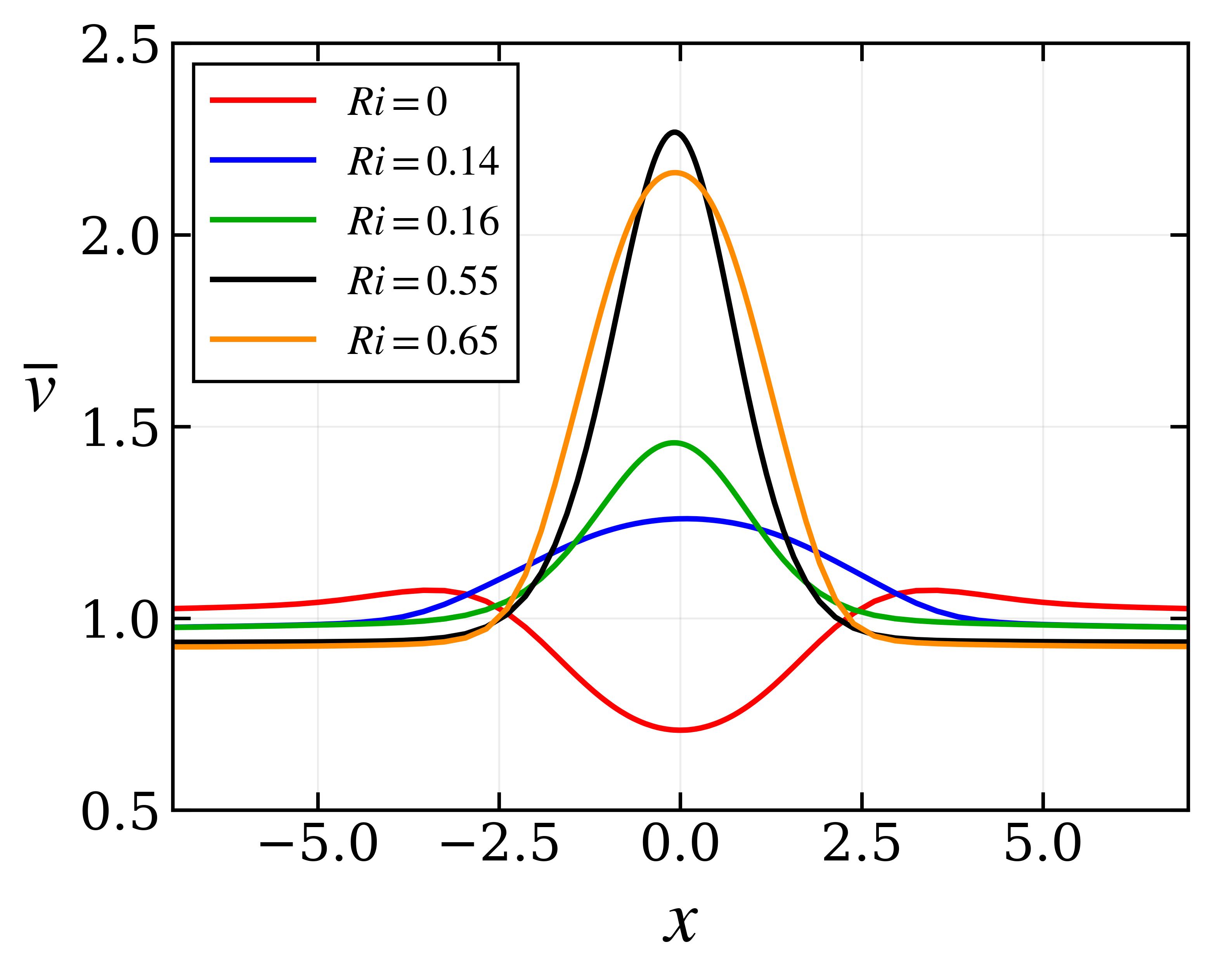}
        \subcaption{$y$ = 60}
    \end{subfigure}
    \caption{Time-averaged axial velocity at various axial locations ($y$-axis) for different Ri at Re=100.}
    \label{deficit1}
\end{figure}

Figure~\ref{deficit1} presents the vertical velocity ($\overline{v}$) profiles at various Richardson numbers and downstream locations. A pronounced momentum deficit is evident in the near-field region for all cases, consistent with the characteristic low-speed recirculation immediately behind bluff bodies. As $\mathrm{Ri}$ increases, this momentum deficit progressively decreases, particularly at downstream locations. For non-zero Richardson numbers ($\mathrm{Ri} > 0$), the profiles reveal a clear enhancement in the mean vertical velocity in the far field relative to the $\mathrm{Ri} = 0$ case. This behaviour indicates the emergence of a buoyancy-driven secondary flow (i.e., natural convective flow), aligned with the primary inlet direction, which adds streamwise momentum to the wake and counteracts the local velocity deficit. \textcolor{\mycolorc}{The far-wake profiles in Figure~\ref{deficit1} also clarify the redistribution of streamwise flow across the wake. For $Ri=0$, the profile retains a centreline velocity deficit together with a weak off-centre velocity excess. In the present incompressible configuration, the impermeable lateral boundaries imply conservation of the cross-sectional volume flux. Consequently, the centreline velocity deficit of the forced wake is accompanied by a compensating velocity excess away from the centreline. The off-centre maxima are associated with the incoming flow being diverted and locally accelerated around the transverse sides of the cylinder, producing higher-speed regions bordering the central wake deficit. As the wake develops downstream, cross-stream momentum redistribution and viscous diffusion progressively smooth these velocity gradients; however, over the streamwise extent shown in Figure~\ref{deficit1}, this relaxation is not complete, and the off-centre maxima remain discernible even at the farthest location shown for $Ri=0$. As $Ri$ increases, buoyancy progressively accelerates the heated fluid near the centreline, first reducing the wake deficit and eventually producing a plume-like excess in centreline velocity. Consistent with conservation of the total cross-sectional volume flux, the outer-region velocity correspondingly decreases and may fall slightly below the freestream value at higher $Ri$.}

\begin{figure}
\begin{minipage}{0.48\textwidth}
\centering
\captionsetup{width=0.95\textwidth}
\includegraphics[width=1\textwidth]{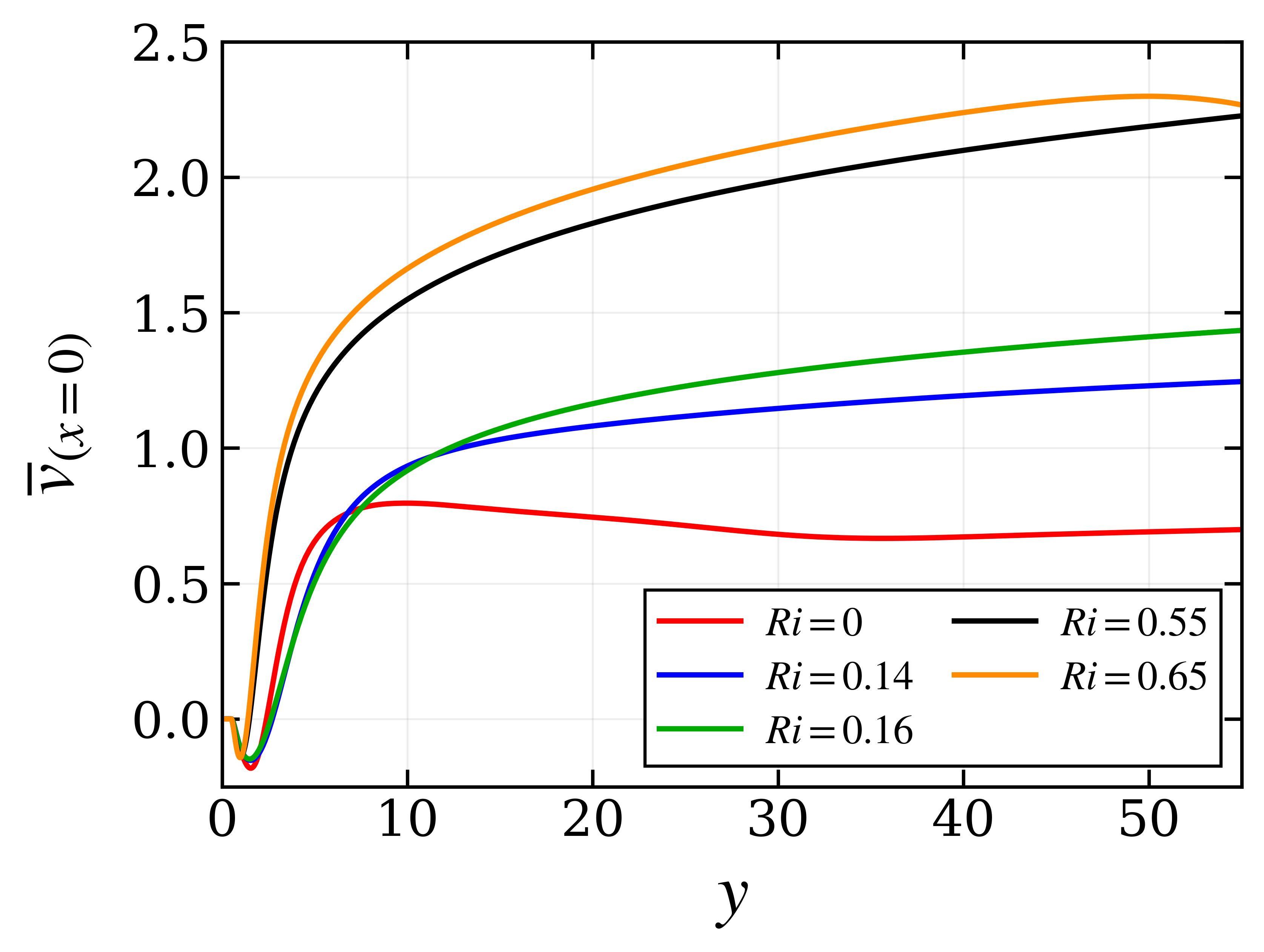}
\caption{Variation of $\overline{v}$ along the domain centerline for different Ri at Re=100.}
\label{profile}
\end{minipage}
\hfill
\begin{minipage}{0.44\textwidth}
\centering
\captionsetup{width=0.95\textwidth}
\includegraphics[width=1\textwidth]{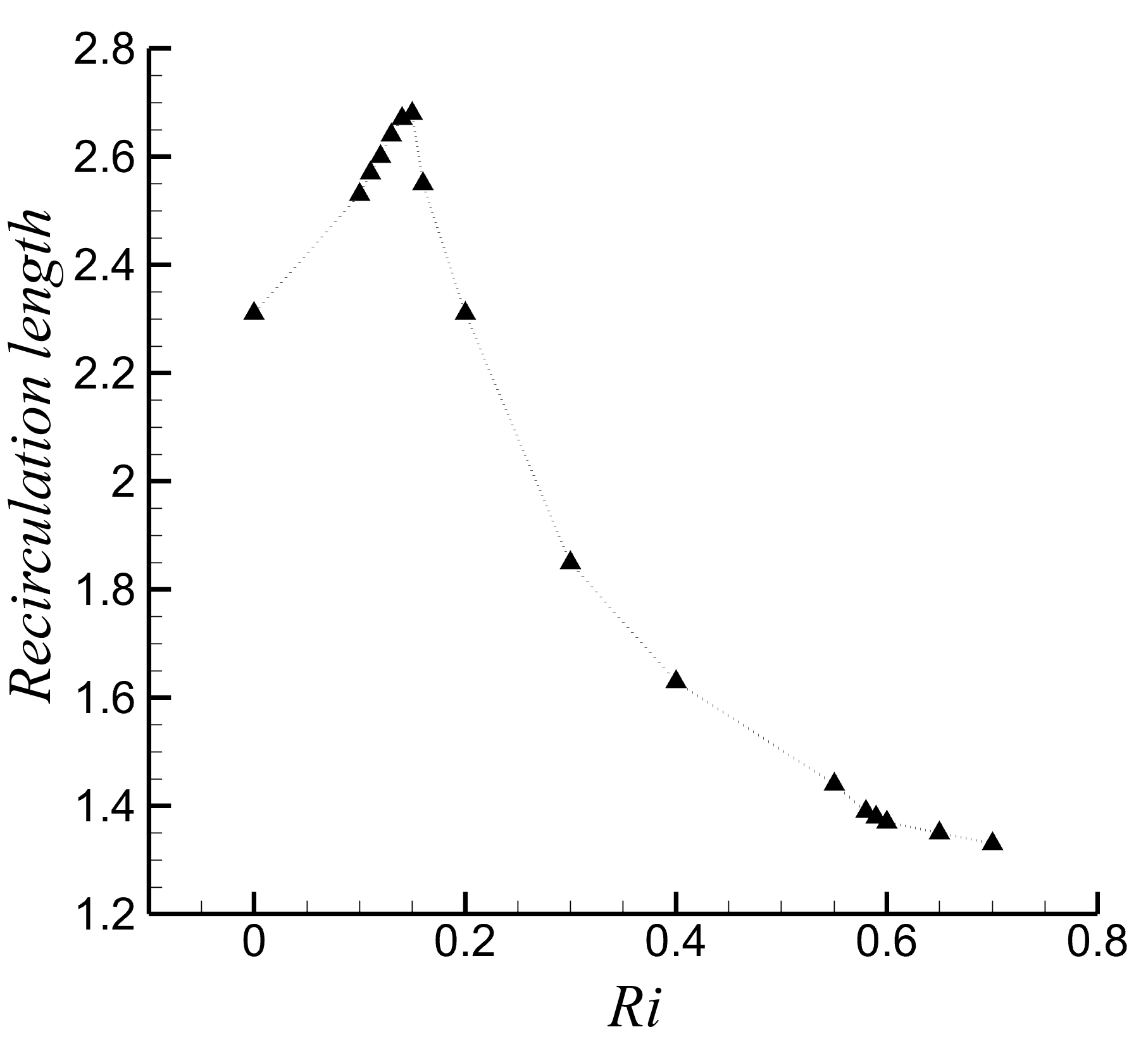}
\caption{Variation of recirculation region length with $\mathrm{Ri}$ for Re=100.}
\label{recirculation}
\end{minipage}
\end{figure}

The secondary flow arises from natural convection induced by the heated cylinder and acts alongside the externally imposed forced flow. Its presence has been shown to play a central role in suppressing vortex shedding~\citep{michaux1997near,kakade2010schlieren}. The influence of this secondary flow on flow acceleration is further confirmed in Figure~\ref{profile}, where a comparison among cases with varying Ri reveals a substantial increase in the vertical velocity downstream. \textcolor{\mycolorc}{The  centreline velocity variation at $Ri=0$ in Figure~\ref{profile} exhibits a non-monotonic downstream recovery. Immediately behind the cylinder, flow separation produces a recirculating region with a strong streamwise velocity deficit. As the influence of the near-wake recirculation diminishes downstream, the centreline velocity initially recovers from this deficit. In the intermediate wake, the organized K\'arm\'an vortex street induces cross-stream motion that results in a net outward redistribution of streamwise momentum away from the centreline region, thereby contributing to the decrease in the local centreline mean velocity. A similar non-monotonic variation of the centreline mean velocity has been reported for two-dimensional forced flow past a circular cylinder by \citet{qu2013quantitative}. Beyond approximately $y\approx32$, the mean-wake evolution changes: farther downstream, as the coherent vortical structures decay, viscous diffusion and continued cross-stream momentum redistribution progressively broaden the mean velocity-deficit profile and reduce the magnitude of the centreline deficit (Figure~\ref{deficit1}), leading to a gradual recovery of the centreline velocity toward its far-wake value. For $Ri>0$, buoyancy modifies this forced-wake evolution by adding upward momentum to the heated centreline region, leading to the sustained increase in centreline velocity seen in Figure~\ref{profile}. This downstream accumulation of buoyancy-induced streamwise momentum resembles the classical behaviour of buoyant plumes, wherein the vertical momentum flux increases with downstream distance due to the sustained action of the local buoyancy source~\citep{gebhart1970steady,puthenveettil2005plume}.}

The temperature and vorticity fields in the present configuration exhibit a similar vertically growing pattern, reinforcing the interpretation of the buoyancy-induced secondary flow as a two-dimensional plume superimposed on the base forced flow. As discussed by~\citet{hunt2022two}, such buoyant plumes are characterized by a downstream increase in momentum flux and a well-defined sense of vorticity, features that are consistent with the behaviour observed in the present simulations.

\begin{figure}
    \centering
    \begin{subfigure}{0.33\textwidth}
        \includegraphics[width=1\textwidth]{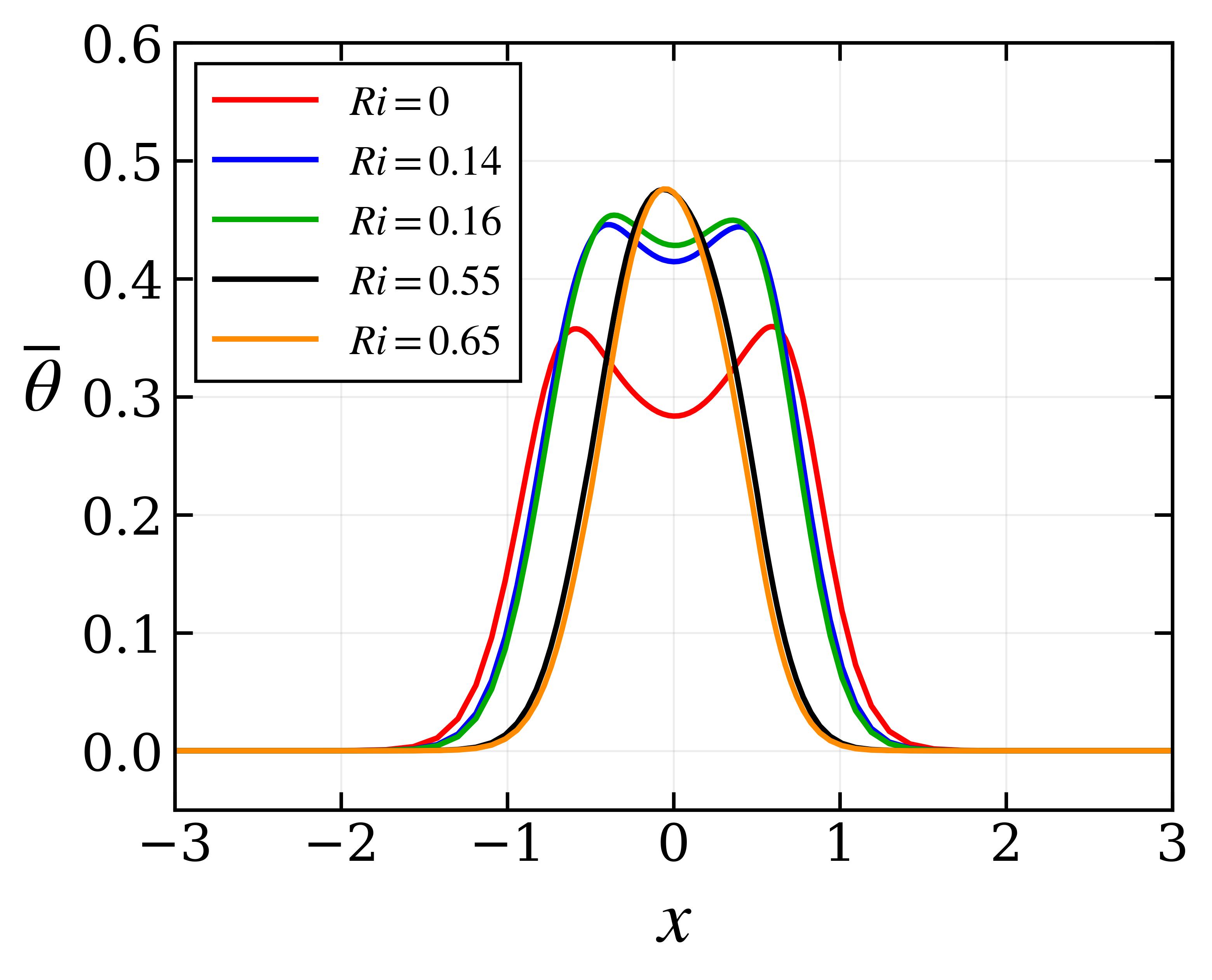}
        \subcaption{$y$ = 1.8}
    \end{subfigure}
    \hfill
    \begin{subfigure}{0.33\textwidth}
        \includegraphics[width=1\textwidth]{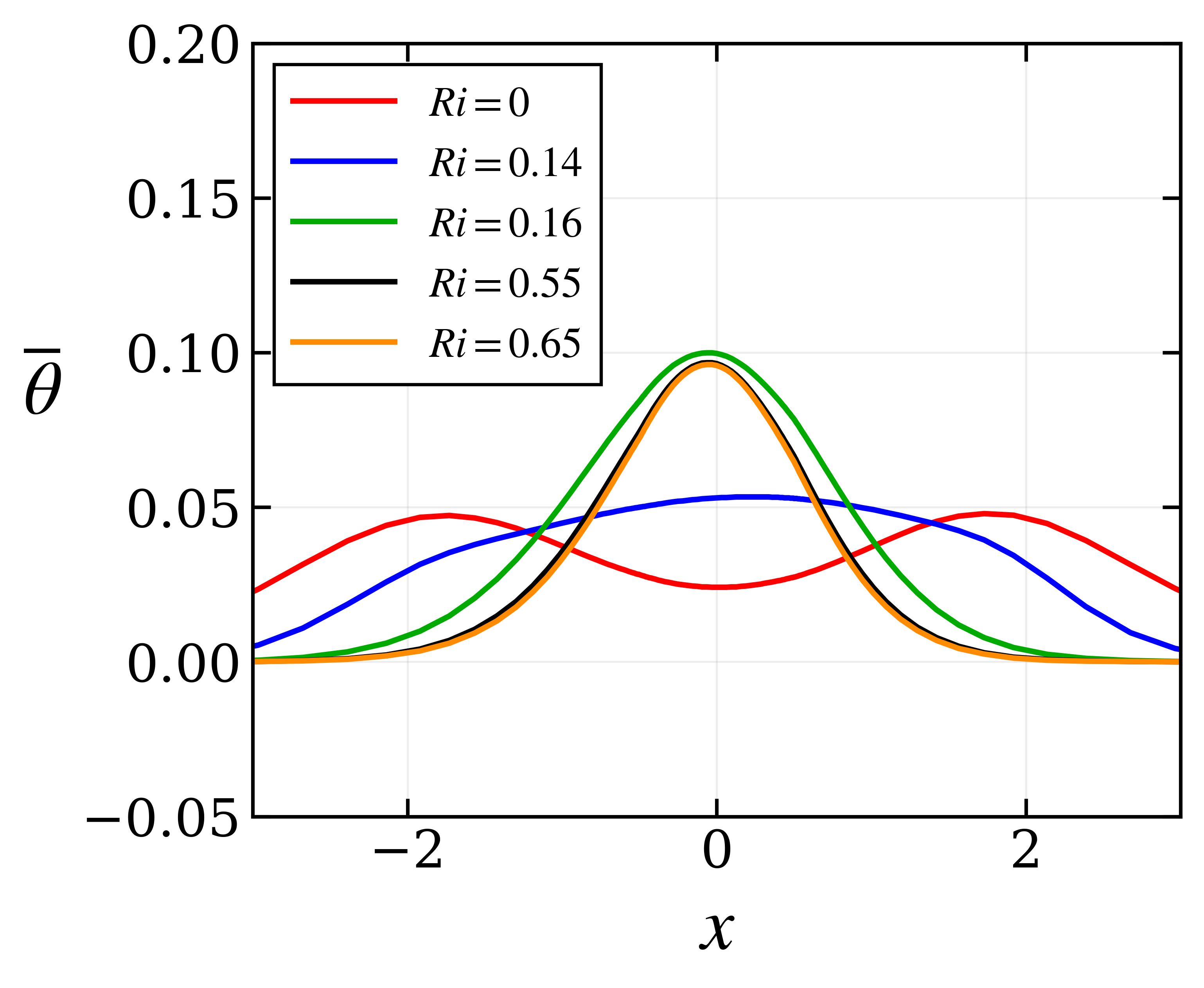}
        \subcaption{$y$ = 30}
    \end{subfigure}
    \hfill
    \begin{subfigure}{0.32\textwidth}
        \includegraphics[width=1\textwidth]{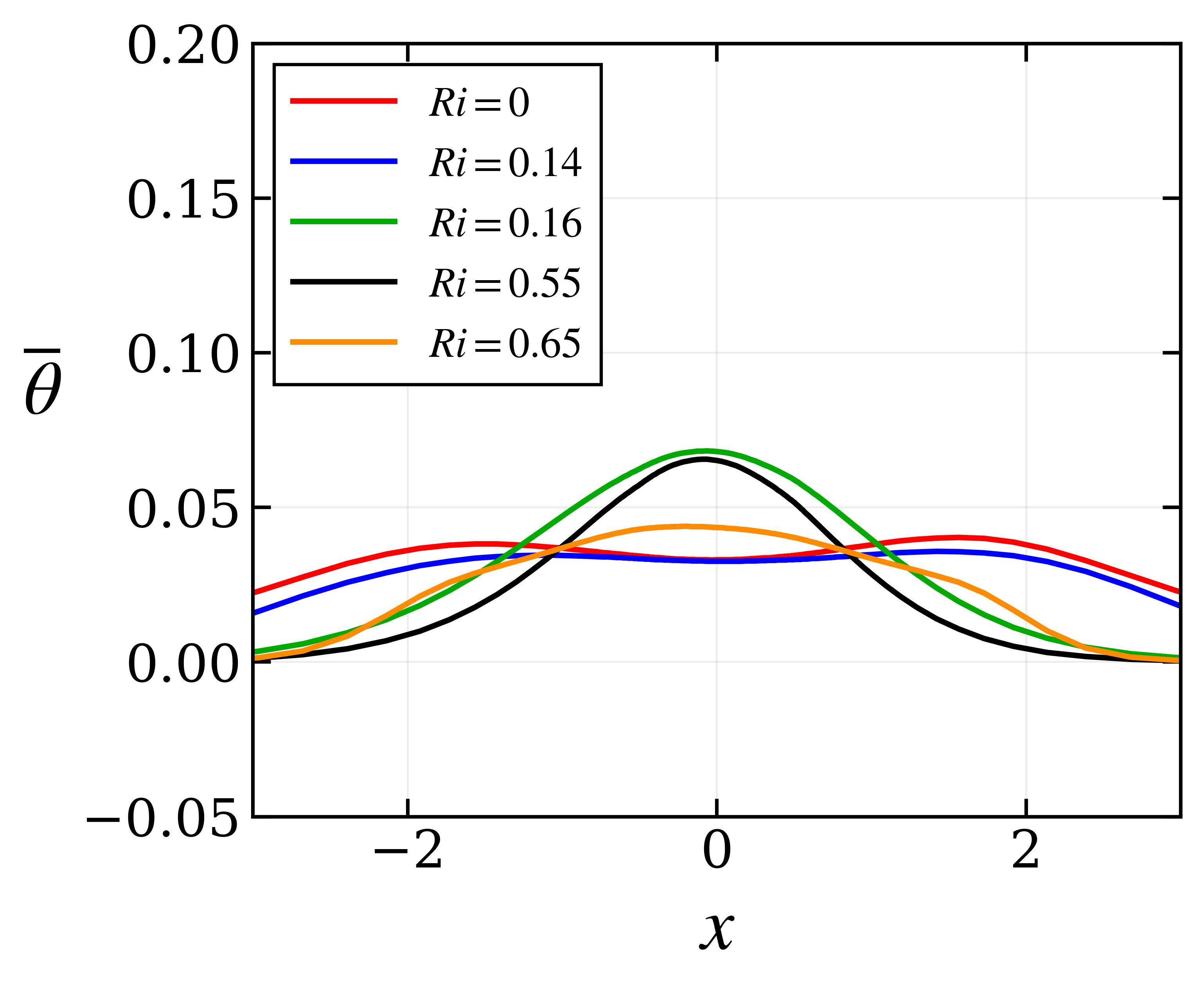}
        \subcaption{$y$ = 60}
    \end{subfigure}
    \caption{Time-averaged temperature profiles for different Ri at different $y$-coordinates (Re=100).}
    \label{deficit2}
\end{figure}

Complementary insights are obtained from the temperature profiles shown in Figure~\ref{deficit2}. The shape of these profiles can be effectively tied to the shape of the axial velocity profiles shown in Figure~\ref{deficit1}. \textcolor{\mycolorc}{A higher mean axial velocity enhances the downstream advective transport of thermal energy, leading to higher temperatures over greater downstream distances, as the heated fluid is convected downstream more rapidly before transverse thermal diffusion progressively smooths the temperature gradients across the shear layers. In other words, within the domain of the present flow field, regions of higher local streamwise velocity are associated with higher local temperatures in the shear layer at a given axial location.} Therefore, the ``double-peak'' like structure in the axial velocity profiles (which exists due to the presence of momentum deficit in the transverse central region of the shear layer) manifests as a corresponding ``double-peak'' structure in the temperature profiles as well. After sufficient mixing downstream, there is no momentum deficit but rather an added momentum due to buoyancy, caused by the vertical advection of hot fluid. This leads to the disappearance of the double-peaked structure in the temperature profiles and causes an extended thermal footprint of the heated cylinder further downstream. These trends are corroborated by the mean temperature contours in Figure~\ref{mean tempe}, which show an increasing vertical spread and a decreasing horizontal spread of the thermal field with increasing $\mathrm{Ri}$. Furthermore, a decrease in the overall temperature magnitude is observed downstream due to thermal diffusion.


The behaviour of the recirculation region is also elucidated from the mean velocity fields. The recirculation region is defined here as the zone along the cylinder centerline ($x = 0$) where the mean vertical velocity ($\overline{v}$) remains negative. Accordingly, the recirculation region length is measured as the vertical distance from the top of the cylinder to the point along $x = 0$ where $\overline{v}$ transitions from negative to positive. The variation of the recirculation length with Richardson number is shown in Figure~\ref{recirculation}.

The recirculation region length exhibits a non-monotonic dependence on $\mathrm{Ri}$: it initially increases with $\mathrm{Ri}$, reaching a maximum near the critical point for vortex shedding suppression ($\mathrm{Ri} = \mathrm{Ri}_{c}$), and subsequently decreases as $\mathrm{Ri}$ increases further. A possible explanation for this trend is provided by~\citet{michaux1997near}, who noted that the buoyancy-induced secondary flow adds momentum to the near-field momentum deficit region, thus reducing the strength of recirculation in the steady regime. This interpretation aligns with the mechanism described in section~\ref{sec:transport}, wherein the secondary flow acts to oppose the negative $v$ in the recirculating zone. In effect, the upward convective motion induced by buoyancy superimposes a positive vertical velocity around the flow centerline and helps in reducing the negative velocity within the recirculation region, thus resulting in a shorter recirculation length.

However, this mechanism does not explain the initial increase in recirculation length observed at lower $\mathrm{Ri}$ in the active vortex shedding regime. In this regime, the dominant effect is the buoyancy-induced acceleration of the shear layer, which enhances the mean axial velocity and delays flow reversal. This is evident from Figure~\ref{profile}, where increasing $\mathrm{Ri}$ corresponds to stronger upward momentum and a more energetic shear layer, consistent with the enhanced buoyancy source term ($\mathrm{Ri} \, \theta$) in the vertical momentum equation~\eqref{eq:momentum}. Under time-averaged conditions, this buoyancy term contributes to an upward mean flow that elongates the recirculation region. Therefore, the observed non-monotonicity results from the competition between two opposing effects: shear layer acceleration, which promotes recirculation growth, and secondary flow induced by buoyancy, which counteracts reverse flow and suppresses recirculation. The former dominates at low $\mathrm{Ri}$, leading to a lengthening of the recirculation region, while the latter becomes dominant beyond the critical point (vortex shedding suppression), resulting in a reduction in recirculation length with a further increase in $\mathrm{Ri}$.

\subsubsection{Drag characteristics}~\label{sec:drag}

\begin{figure}
\begin{minipage}{0.48\textwidth}
\centering
\captionsetup{width=0.95\textwidth}
\includegraphics[width=0.95\textwidth]{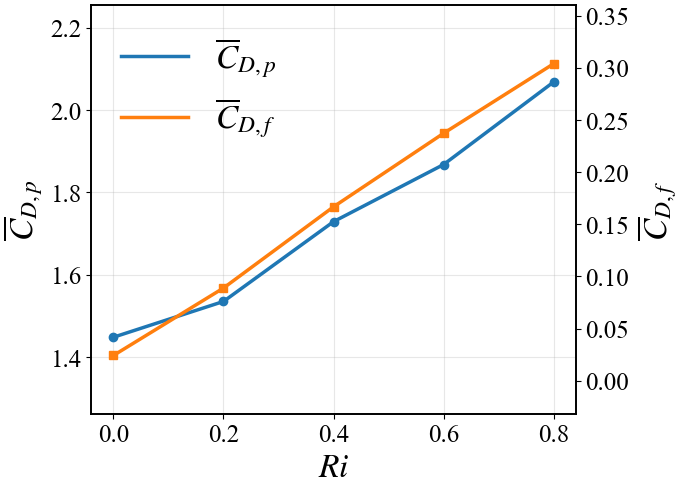}
\caption{Variation of the time-averaged pressure drag and friction drag with $Ri$ at Re=100.}
\label{drag_components}
\end{minipage}
\hfill
\begin{minipage}{0.48\textwidth}
\centering
\captionsetup{width=0.95\textwidth}
\includegraphics[width=0.95\textwidth]{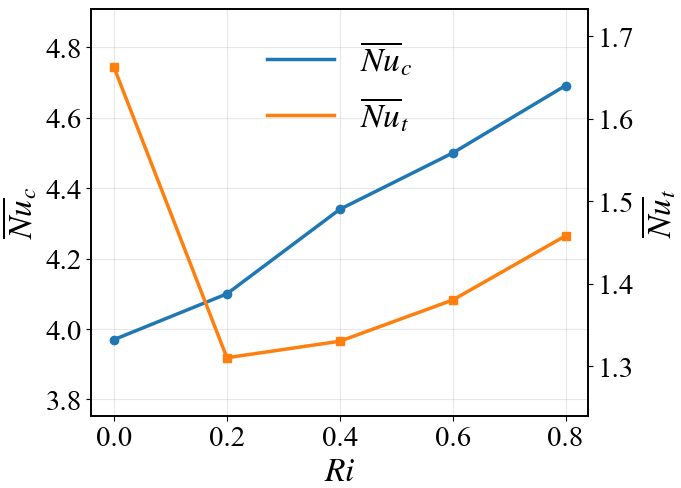}
\caption{Variation of the time-averaged Nusselt number with $Ri$ at $Re=100$. 
$\overline{Nu_t}$ and $\overline{Nu_c}$ denote the Nusselt numbers averaged over the top cylinder surface and the entire cylinder surface, respectively.}
\label{nusselt}
\end{minipage}
\end{figure}

The mean total drag coefficient for the cylinder is calculated by summing over the drag contributions from pressure and friction separately, i.e., $\overline{C}_D=\overline{C}_{D,p}+\overline{C}_{D,f}$, where $\overline{C}_{D,p}$ is the pressure drag caused by the normal force on the top and bottom surfaces of the cylinder induced by the flow pressure and $\overline{C}_{D,f}$ is the friction drag caused by the viscous shear (tangential) forces. The variation of the time-averaged total drag coefficient with Richardson number, shown in Figure~\ref{drag_validation}, reveals a clear monotonic increase with increasing $\mathrm{Ri}$. This behaviour reflects the growing influence of buoyant forces, which induce the secondary flow, thereby modifying the flow around the cylinder. This directly increases the pressure drag component. Furthermore, the acceleration of the shear layers near the cylinder wall due to added buoyant forces at higher Richardson numbers increases wall shear stress, thereby increasing friction drag. The increasing pressure drag and friction drag with $Ri$ is shown in Figure~\ref{drag_components}. Thus, both components contributing to the drag force increase monotonically with $\mathrm{Ri}$. Furthermore, the drag increases linearly both before and after the suppression of vortex shedding; however, the slope of the linear trend exhibits a distinct change near the first transition ($\mathrm{Ri} = \mathrm{Ri}_{c}$). In the vortex shedding regime in the near-field ($\mathrm{Ri} < \mathrm{Ri}_{c}$), the rate of increase is lower, suggesting that the unsteady wake dynamics partially mitigate the buoyancy-induced drag. Once vortex shedding is suppressed, the wake becomes steady and more symmetric, allowing the buoyant forces to contribute more directly to drag enhancement. In the present study, the major component of the drag force comes from the pressure drag, with the maximum contribution of friction drag being $\sim11.2\%$ to the total drag value for $Ri = 0.8$ (Figure~\ref{drag_components}). Note that the time-averaged lift coefficient is zero in every case due to the symmetric nature of the mean flow about the flow centreline ($x$=0).



\subsubsection{Heat transfer characteristics}~\label{sec:nusselt}

The variation of the time-averaged Nusselt number, both over the top surface and the entire surface of the cylinder, is shown in Figure~\ref{nusselt}. The surface-averaged Nusselt number increases monotonically with increasing Richardson number, indicating enhanced convective heat transfer with increasing buoyancy force. Similar trends are observed for the bottom, left, and right surfaces of the cylinder. This behaviour can be attributed to the acceleration of the near-wall shear layer at higher $\mathrm{Ri}$, which increases the mean flow velocity in the vicinity of the cylinder (see Figure~\ref{profile}), thereby enhancing convective transport. Interestingly, the Nusselt number on the top surface (Figure~\ref{nusselt}) exhibits a non-monotonic trend with $Ri$, showing a minimum near the onset of vortex shedding suppression. This behaviour reflects the influence of the evolving recirculation region above the cylinder, as discussed in Section~\ref{sec:transport}. As $Ri$ increases toward $Ri_{c}$, the recirculation region elongates, causing fluid to remain trapped over a larger region for longer residence times, which suppresses mixing with the ambient fluid near the top surface. This weakens convective heat transfer and lowers the Nusselt number. Beyond the transition point, the recirculation region begins to shrink with increasing $Ri$, reducing residence times and promoting more effective mixing adjacent to the surface, thereby improving convective heat transfer. Consequently, the Nusselt number on the top surface increases again with further increases in $Ri$, underscoring the non-monotonic behaviour.

\begin{figure}
\centering
\includegraphics[width=0.5\textwidth]{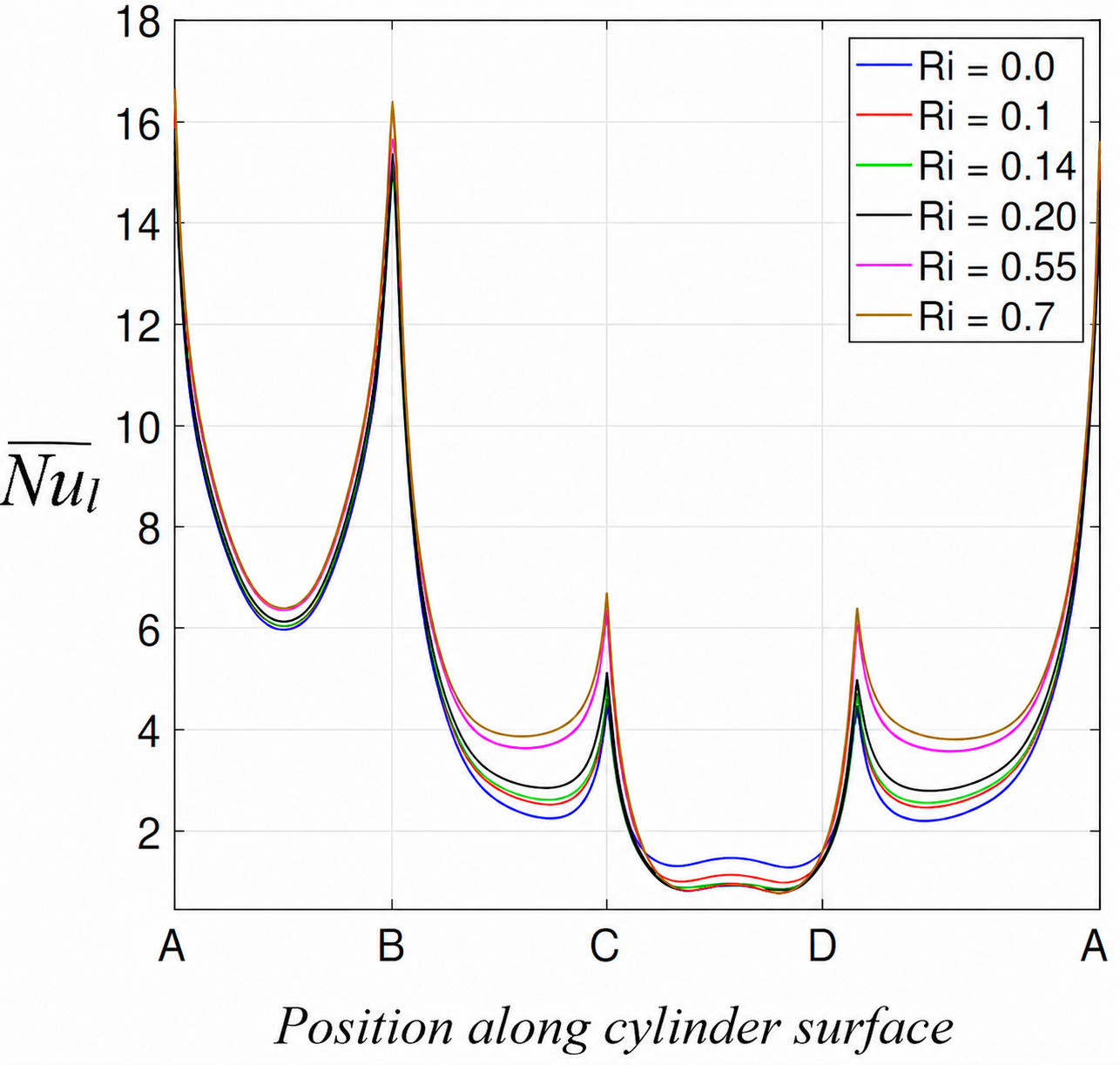}
\caption{Variation of time-averaged local Nusselt number along the cylinder surface for various $\mathrm{Ri}$ at Re=100.}
\label{mean_nu}
\end{figure}

The spatial variation of the local time-averaged Nusselt number ($Nu_l$) along the cylinder perimeter A–B–C–D–A (refer to Figure~\ref{domain}) is presented in Figure~\ref{mean_nu}. The majority of heat transfer occurs over the bottom surface (A–B), where the flow has the most effective contact with the cylinder having higher tangential velocity, resulting in the highest local Nusselt numbers. Over the side and top surfaces, flow separation limits direct interaction with the wall, reducing the heat transfer rate. Among these, the top surface is most affected by the wake dynamics due to its direct exposure to the recirculating region in the near-field.


\subsubsection{Vorticity inversion}~\label{sec:vortinversion}

Vorticity inversion refers to a change in sign of vorticity in the left and right sides of the wake, observed downstream in the mean vorticity contours. As shown in Figure~\ref{mean vorticity}, no vorticity inversion is present for $\mathrm{Ri} = 0$, where the shear layers downstream of the cylinder (wake structure) remain symmetric and are governed entirely by forced convection. For non-zero Richardson numbers, however, a distinct vorticity inversion is observed: the left vortex layer, initially possessing positive (clockwise) vorticity, eventually changes sign to negative (counter-clockwise), while the right vortex layer undergoes an opposite changeover. The streamwise distance from the cylinder at which this inversion occurs, also called inversion length, decreases with increasing $\mathrm{Ri}$.

This behaviour results from the interplay between two competing mechanisms: vorticity generated by flow separation at the upstream corners of the cylinder and oppositely signed vorticity induced by buoyancy-driven secondary flow. Under purely forced convection ($\mathrm{Ri} = 0$), separation at the upper corners leads to clockwise vorticity on the left and counter-clockwise vorticity on the right. At higher Re, the opposite-sign vortices detach from the bluff body alternately, resulting in the formation of a vortex street. This corresponds to the classical wake also called forward von K\'{a}rm\'{a}n vortex street. However, buoyancy acts to oppose this structure by inducing upward convective motion that generates counter-clockwise vorticity on the left and clockwise vorticity on the right. Essentially, the natural convective flow tends to push fluid vertically in the central region of the shear layer formed downstream of the cylinder, drawing in cooler fluid from the sides and inducing a counter-rotating circulation opposite in sign to the vorticity created by separation. This added vertical velocity creates an unsteady jet-like phenomenon. A similar jetting phenomenon, driven by added momentum, is observed behind a flapping foil, such as the undulating tail of a fish, which helps it move forward. This jet-like flow phenomenon is also termed reverse von K\'{a}rm\'{a}n vortex street. As the Richardson number increases, the vertical velocity increases, creating a jet-like phenomenon that helps limit the interaction of the two opposite-sign shear layers, thereby suppressing vortex shedding. The other reason may be that the increasing strength of the buoyant counter-vorticity acts on the vorticity of the separated shear layer, thereby inhibiting the formation of stronger separated vortices that would otherwise undergo shedding, effectively suppressing vortex shedding. 

The vorticity magnitude associated with flow separation decays with downstream distance, while the buoyancy-induced opposite-signed vorticity persists or strengthens. As a result, a point is eventually reached along each shear layer where the two contributions cancel, and the vorticity changes sign. This defines the onset of vorticity inversion, with its proximity to the cylinder decreasing as $\mathrm{Ri}$ increases. These trends, evident in the mean vorticity contours of Figure~\ref{mean vorticity}, underscore the role of natural convection in modulating wake dynamics. These trends are consistent with experimental observations of plume-induced cross-stream vorticity layers in heated bluff body flows, where the curvature and thinning of vortex sheets downstream have been attributed to the growing influence of buoyancy~\citep{puthenveettil2005plume,rogers2009natural}. As the buoyant force acts over a larger vertical extent, it promotes vorticity inversion by accelerating the upward motion of heated fluid and inducing entrainment from the sides. This entrainment creates a shear near the plume edge, thus contributing to the reversal of sign observed in the present simulations. When considered alongside the behaviour of the recirculation region, these observations reveal that buoyancy introduces a secondary flow with opposite-signed vorticity that both reduces the momentum deficit in the wake and suppresses vortex formation. Increasing $\mathrm{Ri}$ enhances this effect, promoting earlier vorticity cancellation and moving the inversion point closer to the cylinder. Given the increasing dominance of the secondary flow with $\mathrm{Ri}$ and its spatial growth with downstream distance, it can be inferred that this plume-like secondary flow plays a central role in the emergence of the far-field instability.

\subsection{Effect of Reynolds number}~\label{sec:Re_effect}

Although most of the mixed-convection analysis in this work is performed at fixed $\mathrm{Re = 100}$ with varying $\mathrm{Ri}$, the effect of Reynolds number on the observed transition scenario is briefly assessed using additional simulations at selected Reynolds numbers for which the flow is expected to be two-dimensional. Two trends were observed in the variation of flow stability with $\mathrm{Re}$, which are tabulated in Table~\ref{tab:critical_Ri_Re} (note that for all other Reynolds numbers except $\mathrm{Re = 100}$, the simulations were performed at larger intervals of the Richardson number because of limited computational resources, and hence the upper limit of the range within which the critical $\mathrm{Ri}$ is expected, has been reported). First, the critical $\mathrm{Ri}$ required to fully suppress near-field vortex shedding increases with increasing $\mathrm{Re}$. This behaviour is physically consistent with stronger inertial shear-layer instabilities and reduced viscous damping at higher $\mathrm{Re}$, implying that a larger buoyancy forcing (higher $\mathrm{Ri}$) is needed to stabilize the near-field dynamics. Second, the onset of far-field (plume-like) unsteadiness occurs at a lower critical $\mathrm{Ri}$ as $\mathrm{Re}$ is increased. As viscous dissipation (damping) weakens with increasing $\mathrm{Re}$, perturbations in the downstream buoyancy-driven shear layers are more readily amplified, favoring the far-field transition. Overall, it is observed that the effect of increasing Reynolds number is to destabilize the flow field, which is consistent with prior literature \citep{saha2000transition, yoon2010flow}. This destabilizing influence of increasing Reynolds number is also reflected in the shedding frequency, as quantified by the Strouhal number (shown for both $\mathrm{Ri}=0.0$ and $\mathrm{Ri}=0.8$ in Table~\ref{tab:St_vs_Re}). As $\mathrm{Re}$ increases, the reduced viscous damping and enhanced inertial amplification of shear-layer instabilities lead to faster temporal
oscillations of the wake, resulting in a monotonic increase in $\mathrm{St}$.

\begin{figure}
    \centering

\subfloat[]{
\includegraphics[width=0.46\textwidth]{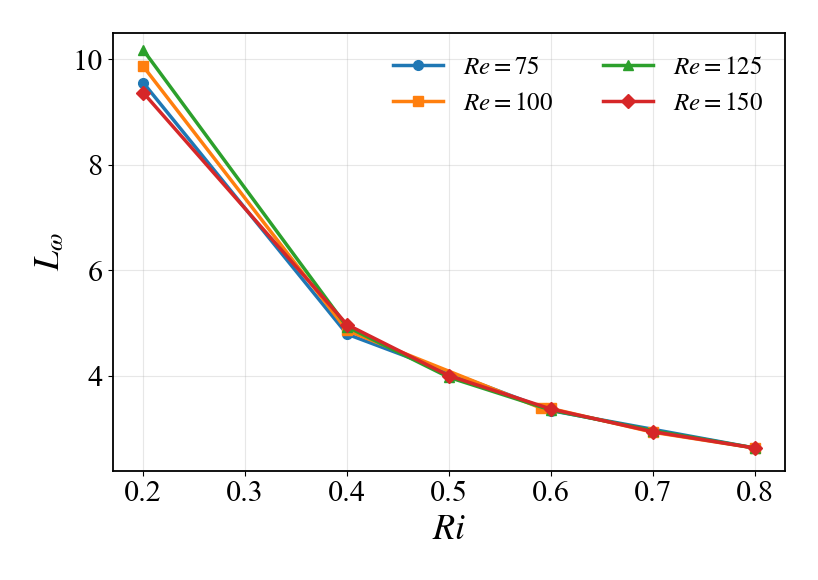}
\label{fig:inversion_length_Ri}
}
\hfill
\subfloat[]{
\includegraphics[width=0.48\textwidth]{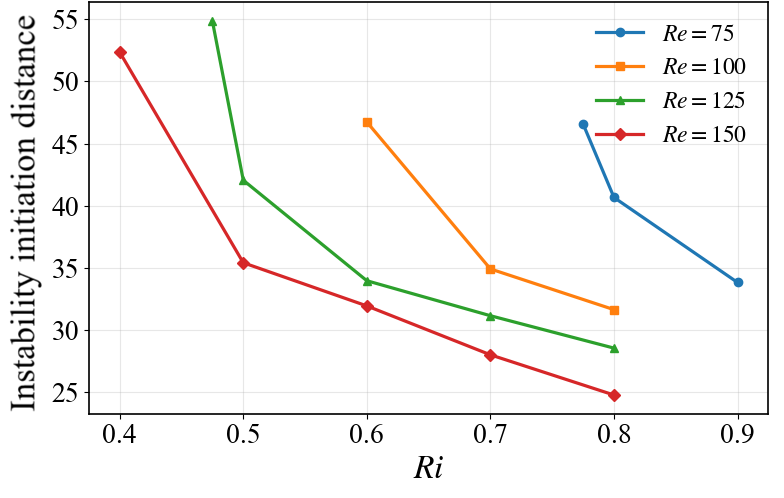}
\label{fig:local_farfield_analysis}
}

    \caption{\textcolor{\mycolorb}{Variation of (a) vorticity inversion length, $L_{\omega}$ and (b) far-field instability initiation distance with Richardson number for different Reynolds numbers.}}
\end{figure}

\textcolor{\mycolorb}{To investigate the effect of Re and Ri on the inversion length, we quantify the axial location at which the wake-like flow changes into a plume-like flow, using a vorticity-based inversion length, $L_{\omega}$. At each streamwise location, the mean vorticity is transversely averaged over the right half of the domain. The axial location at which this averaged vorticity changes sign is identified as $L_{\omega}$. This definition provides a measure of the changeover from the wake-dominated region to the buoyancy-dominated plume region. The physical interpretation is consistent with the corresponding centreline velocity field. In the wake-like region, the centreline velocity contains a momentum deficit relative to the surrounding shear-layer flow, which gives one sign of vorticity across half of the transverse width of the wake from the centreline. Farther downstream, buoyancy accelerates the heated fluid along the centreline, producing a momentum excess. The transverse velocity gradient therefore reverses, leading to the observed vorticity inversion. Thus, $L_{\omega}$ also represents the approximate axial location where the mean flow changes from a velocity-defect wake to a velocity-excess plume. Figure~\ref{fig:inversion_length_Ri} shows that $L_{\omega}$ decreases strongly with increasing $Ri$ for all Reynolds numbers considered. It was observed that the inversion lengths obtained from the centreline mean streamwise velocity closely follow the vorticity inversion lengths, confirming that both metrics identify the same downstream changeover from a velocity-defect wake to a velocity-excess plume. These observations confirm that stronger buoyancy causes the plume-like state to emerge progressively closer to the cylinder. Interestingly, the inversion length exhibits only a weak dependence on Reynolds number over the range considered in the present study, with noticeable Reynolds-number effects appearing only at lower Richardson numbers ($Ri \approx 0.2$). This behaviour can be interpreted by considering the physical meaning of the inversion location. The inversion length marks the downstream position at which the buoyancy-driven acceleration of the heated fluid becomes sufficiently strong to eliminate the wake momentum deficit and produce a velocity excess along the centreline. Since buoyancy acts directly through the Richardson number, increasing $Ri$ promotes earlier recovery of the centreline flow and shifts the inversion location upstream. In contrast, the influence of Reynolds number depends on the relative importance of buoyancy and inertia. At lower Richardson numbers, where buoyancy effects remain weak, the inversion location is strongly influenced by the wake structure and momentum transport characteristics, resulting in a more noticeable Reynolds-number dependence. As the Richardson number increases, buoyancy-driven acceleration becomes the dominant mechanism governing the flow evolution. Consequently, the inversion location becomes primarily controlled by the buoyancy forcing, and the Reynolds-number dependence weakens considerably, leading to a near collapse of the inversion length data at higher Richardson numbers.}

To further examine the role of local flow characteristics in the far-field instability, we identify the streamwise location at which the instability first becomes detectable. This location, denoted here as the instability initiation distance, is determined from the DNS-based fluctuating kinetic energy field. The fluctuating kinetic energy [$0.5(\tilde{u}^{2}+\tilde{v}^{2})$ with $\tilde{u}$ and $\tilde{v}$ being fluctuating velocities] along the centreline is extracted from the fluctuating kinetic energy field computed for the entire domain and the instability initiation distance is identified as the first downstream location from the cylinder centre at which this value exceeds $5\times10^{-4}$. This threshold corresponds approximately to velocity fluctuations of 2\% of \textcolor{\mycolorc}{$U_{\infty}$}, assuming comparable fluctuation amplitudes in the streamwise and transverse directions. While the exact instability initiation distance depends to some extent on the selected threshold value, all results reported herein are based on the same threshold of $5\times10^{-4}$, thereby enabling consistent comparisons across different Reynolds and Richardson numbers. The corresponding non-dimensional mean streamwise velocity at the instability initiation location (along the centreline) is denoted by $U_l = u_l/U_{\infty}$, which provides a local measure of the streamwise momentum available to the developing plume instability. Figure~\ref{fig:local_farfield_analysis} shows that the instability initiation distance decreases with increasing $Ri$ for all Reynolds numbers considered. Thus, stronger buoyancy makes the far-field instability detectable closer to the cylinder, consistent with the downstream amplification scenario discussed earlier. At fixed $Ri$, increasing $Re$ also generally shifts the instability initiation upstream, indicating that reduced viscous damping and stronger inertial transport promote earlier amplification of the plume-like disturbance. \textcolor{\mycolorc}{The variation of $U_l$ and the non-dimensional plume width, $\Delta_p$, at the initiation location is provided in Table~\ref{tab:critical_Ri_Re}. At each instability-initiation location, the local plume width $\delta_p$ is defined as the transverse width bounded by the two locations at which the magnitude of the transverse gradient of the mean streamwise velocity reaches the fixed threshold
$|\partial \overline{v}/\partial x|=0.01$. The same criterion is applied consistently to all cases. The corresponding non-dimensional plume width is denoted by $\Delta_p=\delta_p/B$. The local Reynolds numbers, $Re_{l,B}$ and $Re_{l,\delta_p}$ based on local velocity $U_l$ and cylinder width, and on local velocity $U_l$ and plume width, respectively, at the point of instability initiation, are defined here as}
\[ 
Re_{l,B}=\frac{u_l B}{\nu}=\frac{u_l}{U_\infty}\frac{U_\infty B}{\nu}=U_l Re ~~~~~\mbox{and} ~~~~~  Re_{l,\delta_p}=\frac{u_l \delta_p}{\nu}=\frac{u_l}{U_\infty} \frac{\delta_p}{B}\frac{U_\infty B}{\nu}=U_l \Delta_p Re 
\]
\textcolor{\mycolorc}{It is observed that a velocity scale $U_l$ and a length scale equal to cylinder width together do not collapse to a single critical value across all $Re$. Instead, the local mean streamwise velocity required for the onset of far-field instability decreases with increasing flow Reynolds number. This indicates that, at higher $Re$, the plume-like disturbance can become unstable at a lower local mean velocity, consistent with the reduced viscous damping of flow perturbations and the stronger inertial amplification reflected in the Strouhal-number trends reported in Table~\ref{tab:St_vs_Re}. The plume width at the instability-initiation location also decreases with increasing $Re$, further motivating its inclusion as the local length scale in the definition of $Re_{l,\delta_p}$. The values of $Re_{l,B}$ and $Re_{l,\delta_p}$ listed in Table~\ref{tab:critical_Ri_Re} show that replacing the cylinder width by the local plume width substantially improves the collapse across the four transition cases, reducing the coefficient of variation from approximately $15.7\%$ for $Re_{l,B}$ to approximately $7.5\%$ for $Re_{l,\delta_p}$. The remaining variation, which is largest for the $Re=75$ case, indicates that the collapse is approximate rather than exact. This improvement is physically consistent with the spatial character of the far-field instability. While the cylinder width $B$ provides the natural geometric length scale for the near-field bluff-body wake, the far-field instability develops within the downstream buoyant plume, whose local transverse dimension is represented more directly by $\delta_p$. The plume width therefore provides a more representative local length scale for characterizing the far-field instability. For a fixed Reynolds number, the local mean streamwise velocity $U_l$ at the instability-initiation location varies relatively weakly as $Ri$ is varied (Table~\ref{tab:local_St}).}

\textcolor{\mycolorc}{A similar local scaling can be introduced for the oscillation frequency. The conventional Strouhal number, $St=fB/U_\infty$, employs the cylinder width and freestream velocity and is a natural reference for the near-field bluff-body shedding process. In contrast, the far-field instability develops within the evolving buoyant plume, for which the local plume width and local mean streamwise velocity provide more representative scales. We therefore define a local plume-based Strouhal number at the instability-initiation location as}
\[
St_{l,\delta_p}
=
\frac{f\delta_p}{u_l}
=
\frac{St\,\Delta_p}{U_l}
\]
\textcolor{\mycolorc}{This quantity compares the observed oscillation frequency with the local convective frequency scale $u_l/\delta_p$. More broadly, the resulting local scaling quantities for all 13 simulated cases exhibiting far-field oscillations are summarized in Table~\ref{tab:local_St}. Across these cases, the plume-based local Reynolds number $Re_{l,\delta_p}$ exhibits an approximate collapse \textcolor{\mycolord}{about a mean value of 1319}, with a coefficient of variation of $6.26\%$. The corresponding locally scaled Strouhal number $St_{l,\delta_p}$ similarly exhibits an approximate collapse \textcolor{\mycolord}{about a mean value of 0.542} across the investigated $(Re,Ri)$ combinations, with a coefficient of variation of $6.87\%$. Together, these results indicate that the local plume width and local mean velocity provide representative scales for the far-field instability, with its characteristic oscillation timescale scaling approximately with the local convective timescale $\delta_p/u_l$ at the instability-initiation location.}

\begin{table}
\color{black}
\centering
\small
\setlength{\tabcolsep}{8pt}
\caption{\textcolor{\mycolorc}{Critical Richardson number for near-field vortex-shedding suppression and far-field instability at different Reynolds numbers. Also shown are the non-dimensional local mean streamwise velocity, $U_l$, at the far-field instability initiation location, non-dimensional plume width, $\Delta_p$ and the corresponding local Reynolds numbers. All data is reported for the original domain (Figure~\ref{domain}).}}
\label{tab:critical_Ri_Re}
\begin{tabular}{ccccccc}
\toprule
$\mathrm{Re}$ &
Critical $\mathrm{Ri}$ &
Critical $\mathrm{Ri}$ &
$U_l$ &
$Re_{l,B}$ & $\Delta_p$ & $Re_{l,\delta_p}$ \\
&
(Near-field transition) & 
(Far-field transition) &
& & &
\\
\midrule
75  & 0.10  & 0.775 & 2.49 & 187 & 7.16 & 1337\\
100 & 0.15  & 0.590 & 2.28 & 228 & 6.64 & 1514\\
125 & 0.20  & 0.475 & 2.09 & 261 & 6.04 & 1578\\
150 & 0.25  & 0.400 & 1.92 & 288 & 5.70 & 1642\\
\bottomrule
\end{tabular}
\end{table}


\begin{table}
\centering
\small
\setlength{\tabcolsep}{8pt}
\caption{Variation of the Strouhal number with Reynolds number for selected Richardson numbers. For $Ri = 0.0$, $St$ is associated with near-field unsteadiness, whereas for $Ri = 0.8$, it is associated with far-field unsteadiness.}
\label{tab:St_vs_Re}
\begin{tabular}{ccc}
\toprule
$\mathrm{Re}$ & $\mathrm{St}$ ($Ri=0.0$) & $\mathrm{St}$ ($Ri=0.8$) \\
\midrule
75  & 0.1389 & 0.2098 \\
100 & 0.1512 & 0.2400 \\
125 & 0.1577 & 0.2458 \\
150 & 0.1621 & 0.2613 \\
\bottomrule
\end{tabular}
\end{table}

\begin{figure}
\centering
\subfloat[]{
\includegraphics[width=0.44\textwidth]{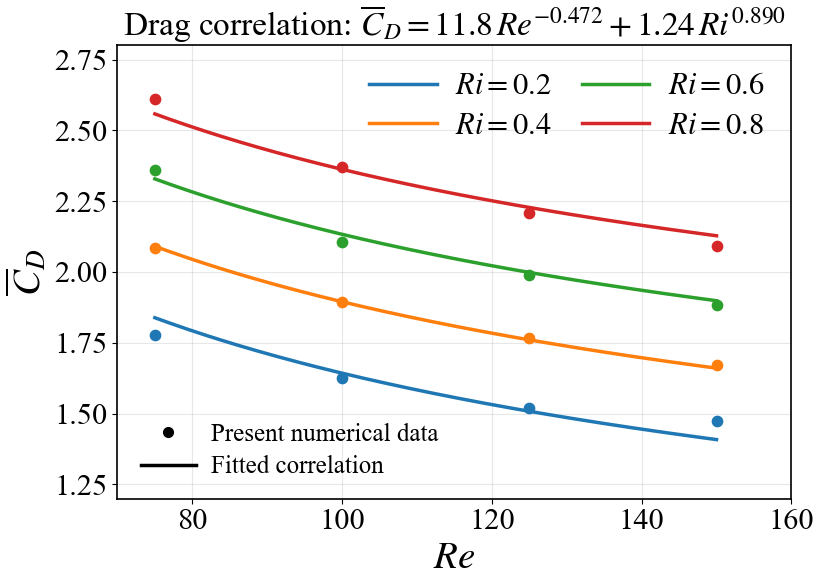}
}
\hfill
\subfloat[]{
\includegraphics[width=0.48\textwidth]{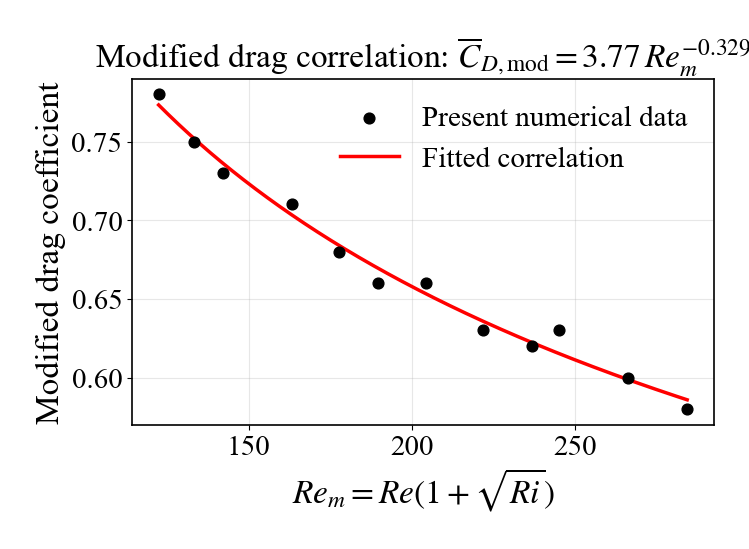}
}

\caption{
(a) Mean drag coefficient variation with Reynolds number for varying Ri. (b) Collapse of modified drag coefficient,
$\overline{C}_{D,\mathrm{mod}}
=
\overline{C}_D/(1+\sqrt{Ri})^2$, using the modified Reynolds number,
$Re_m=Re(1+\sqrt{Ri})$, together with the fitted correlation.
}
\label{WithRe}
\end{figure}

\begin{figure}
    \centering
    \begin{subfigure}{0.49\textwidth}
        \includegraphics[width=1\textwidth]{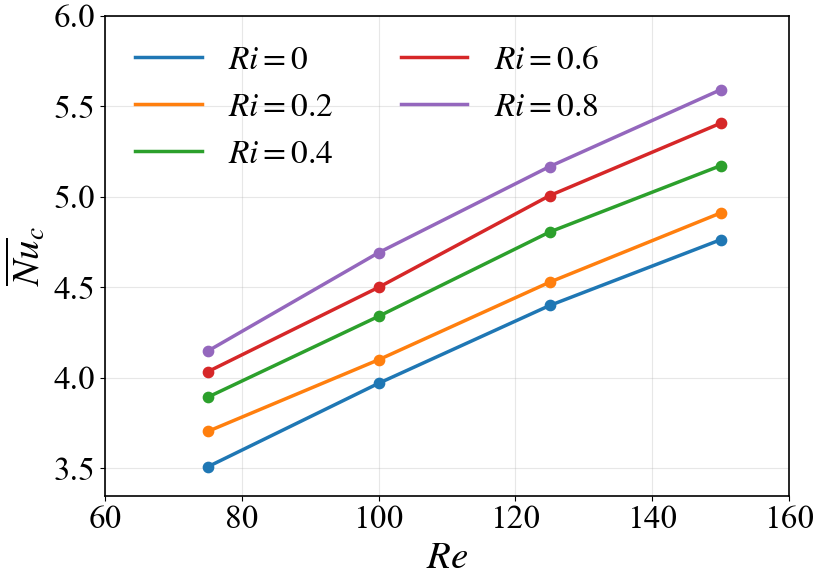}
        \subcaption{}
        \label{nure}
    \end{subfigure}
    \hfill
    \begin{subfigure}{0.49\textwidth}
        \includegraphics[width=1\textwidth]{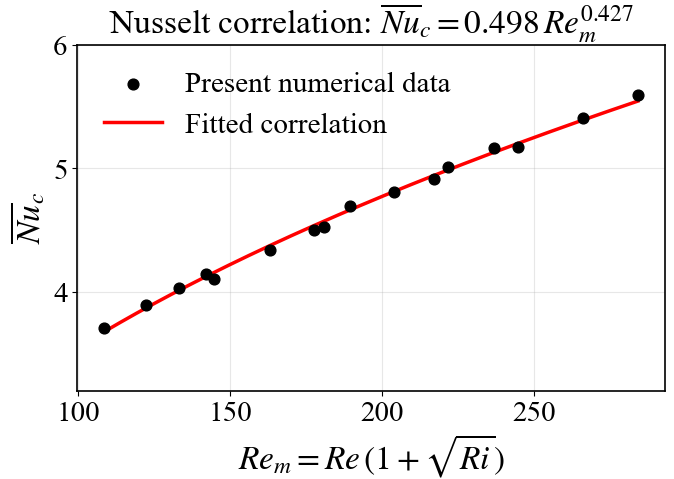}
         \subcaption{}
       \label{Nu_ReM}
    \end{subfigure}
    \caption{Variation of (a) $\overline{Nu}_c$ with Re for varying Ri and (b) $\overline{Nu}_c$ with $\mathrm{Re_m}$.}
    \label{correlations}
\end{figure}

In addition to the transition behaviour, Figs~\ref{WithRe}(a) and \ref{correlations}(a) show the variation of time-averaged drag coefficient and Nusselt number with $\mathrm{Re}$ for different $\mathrm{Ri}$ values, respectively. As before, the total mean drag coefficient $\overline{C}_D$ (sum of mean pressure and friction contributions) is plotted, where the contribution due to the friction component is higher for cases with higher Richardson numbers (with maximum contribution up to $\sim16\%$) and becomes negligible for cases with lower $\mathrm{Ri}$ values. For each $\mathrm{Ri}$, the mean cylinder Nusselt number $\overline{Nu}_c$ increases with increasing $\mathrm{Re}$, consistent with enhanced convective heat transport and steeper near-wall temperature gradients at higher Reynolds number. In contrast, the total mean drag coefficient $\overline{C}_D$ decreases with increasing $\mathrm{Re}$. This is consistent with observations for laminar flow past square cylinders~\citep{sen2011flow}, where increasing $\mathrm{Re}$ leads to enhanced pressure recovery at the cylinder base and a corresponding reduction in pressure drag due to systematic reorganization of the steady wake structure (longer recirculation bubble).

Here, an attempt has been made to obtain compact \textcolor{\mycolord}{empirical} trends for the time-averaged drag and heat-transfer response under buoyancy-aided mixed convection. For this purpose, we fit empirical correlations to the parametric simulation database for mixed-convective cases (i.e., for $Ri>0$).

For the mean drag coefficient, an $Re$--$Ri$ correlation of the form
\begin{equation}
\overline{C}_D \;=\; C_1 \;+\; C_2\,Re^{\,m} \;+\; C_3\,Ri^{\,n}
\label{eq:Cd_corr}
\end{equation}
is adopted, where $C_1$, $C_2$, $C_3$, $m$, and $n$ are determined from a least-squares fit to the simulation data by minimizing the sum of squared deviations between the correlation and the computed mean drag coefficients. The above structure is motivated by a simple superposition of forced and buoyancy-driven contributions: the term $C_2Re^{m}$ represents the inertia-dominated (forced-flow) component that varies primarily with $Re$, whereas $C_3Ri^{n}$ captures the additional mixed-/free-convective contribution introduced by buoyancy (quantified by $Ri$). This formulation is consistent with the observations presented later in the paper (see Sec.~\ref{modes_analysis}), where the total flow dynamics are shown to be effectively interpretable as a superposition of forced inertial and free-convective components. The resulting model captures the variation of $\overline{C}_D$ with both $Re$ and $Ri$ exceedingly well over the investigated parameter range (see Figure~\ref{WithRe}(a)). The maximum discrepancy between the correlation and the simulation values is $4.4\%$ (occurring at \textbf{$Re=150,\,Ri=0.2$}).

\textcolor{\mycolorb}{To provide a more physically motivated interpretation of the mean drag coefficient and the mean Nusselt number averaged over the entire cylinder surface in a mixed convective flow, we employ an effective mixed-convection velocity scale based on an empirical combination of forced and buoyancy-driven velocity scales. In forced convection, the characteristic velocity entering the Reynolds number is naturally taken as the freestream velocity $U_\infty$. However, in buoyancy-aided mixed convection, thermal buoyancy induces an additional flow whose characteristic velocity scale is given by $U_b \sim \sqrt{g\,\beta\,\Delta T\,B}$~\citep{saha2000unsteady,nirmalkar2013mixed,daniel2013aiding}, where $B$ is the cylinder width and $\Delta T = T_s - T_{\infty}$. This buoyancy-driven motion augments the inertial freestream transport and effectively enhances convective heat transfer. Accordingly, the modified characteristic velocity for buoyancy-aided mixed convection is defined empirically as the sum of the forced and buoyant contributions, $U_{ch,m}=U_\infty+U_b$. This scale represents the Newtonian-fluid limit of the mixed-convection velocity scaling employed previously for related aiding-buoyancy flows in power-law fluids~\citep{nirmalkar2013mixed,daniel2013aiding}.}

Using $Re=U_\infty B/\nu$ and $Gr=g\beta\Delta T\,B^3/\nu^2$, the modified Reynolds number based on the modified characteristic velocity scale for mixed convection is now defined as:
\begin{equation}
Re_m \;=\; \frac{U_{ch,m}B}{\nu}
\;=\; \frac{(U_\infty + \sqrt{g\beta\Delta T\,B})B}{\nu}
\;=\; Re \;+\; \sqrt{Gr}
\;=\; Re\left(1+\sqrt{Ri}\right)
\label{eq:Rem_def}
\end{equation}
The modified Reynolds number captures the combined influence of inertial forced flow and free-convective flow components. Hence, this construction also inherently adopts the forced-buoyancy superposition, consistent with findings reported in Sec.~\ref{modes_analysis}.

{\color{black}Since the drag coefficient is defined based on the dynamic pressure associated with the chosen characteristic velocity scale, the introduction of the modified mixed-convective velocity scale also modifies the corresponding drag coefficient. The conventional mean drag coefficient is defined as (Appendix A):

\begin{equation}
\overline{C}_D
=
\frac{\overline{F}_y}
{\frac{1}{2}\rho U_\infty^2 B},
\end{equation}

whereas using the mixed-convective velocity scale gives

\begin{equation}
\overline{C}_{D,\mathrm{mod}}
=
\frac{\overline{F}_y}
{\frac{1}{2}\rho U_{ch,m}^{2} B}.
\end{equation}

Substituting $U_{ch,m}=U_\infty(1+\sqrt{Ri})$ gives

\begin{equation}
\overline{C}_{D,\mathrm{mod}}
=
\frac{\overline{C}_D}
{(1+\sqrt{Ri})^2}
\label{eq:Cd_mod}
\end{equation}

The modified drag coefficient hence represents the drag force normalized using the effective mixed-convective velocity scale, thereby allowing comparison across different mixed-convection conditions. The modified mixed-convective velocity scaling may also be used to interpret the variation of the flow oscillation frequency under buoyancy-aided conditions. Using the modified characteristic velocity scale, the corresponding modified Strouhal number (corresponding to the far-field oscillations) is obtained as (see Appendix A for reference)
\begin{equation}
St_m
=
\frac{fB}{U_{ch,m}}
=
\frac{fB}{U_\infty(1+\sqrt{Ri})}
=
\frac{St}{1+\sqrt{Ri}} .
\label{eq:Stm_def}
\end{equation}

Using the modified drag coefficient and modified Strouhal number defined in Eqs.~\eqref{eq:Cd_mod} and ~\eqref{eq:Stm_def}, the drag, heat-transfer, and frequency responses are correlated using a single-parameter power-law,
\begin{subequations}
\begin{align}
\overline{C}_{D,\mathrm{mod}}
&=
a_d\,Re_m^{\,b_d}
\label{eq:Cd_corr_2}
\\
\overline{Nu}_c
&=
a_n\,Re_m^{\,b_n}
\label{eq:Nu_corr}
\\
St_m &= a_s\,Re_m^{\,b_s}
\label{eq:St_corr}
\end{align}
\end{subequations}
}
{\color{black}The constants $a_d,\ a_n, \ a_s$ and $b_d,\ b_n, \ b_s$ are obtained from least-squares fitting in log--log space, which effectively minimizes relative deviations between the correlation and the numerical data. \textcolor{\mycolord}{Given the relatively limited range of $Re_m$ covered by the present simulations, these power-law relations are intended as compact empirical representations of the investigated parameter range rather than as universal predictive correlations.}

Figure~\ref{WithRe}(b) shows that the modified Reynolds number, together with the corresponding drag normalization provides an excellent collapse of the drag data once mixed convection becomes sufficiently developed. Since buoyancy-induced modifications of drag remain relatively weak at lower Richardson numbers, the drag scaling is constructed using only cases with $Ri>0.2$. In this regime, Eq.~\eqref{eq:Cd_corr_2} accurately captures the numerical data with $R^2=0.986$ and a maximum relative deviation of only $2.34\%$. The same modified Reynolds number is then employed to correlate the heat-transfer response through Eq.~\eqref{eq:Nu_corr}. Unlike drag, the collapse for $\overline{Nu}_c$ remains effective even at lower Richardson numbers, including $Ri=0.2$ (Figure~\ref{Nu_ReM}). This difference likely arises because heat transfer responds directly to buoyancy-induced modifications of the thermal boundary layer, whereas the drag force remains dominated by inertial pressure and viscous contributions until buoyancy becomes sufficiently strong. Equation~\eqref{eq:Nu_corr} predicts the numerical data with a maximum relative deviation of $1.4\%$. Since $Re_m = Re+\sqrt{Gr}$, at sufficiently large Grashof numbers the scaling asymptotically approaches the natural-convection limit. However, under the present conditions, both inertial and buoyancy-driven transport remain important, making the combined scaling appropriate.}

\begin{figure}
\centering
\includegraphics[width=0.6\textwidth]{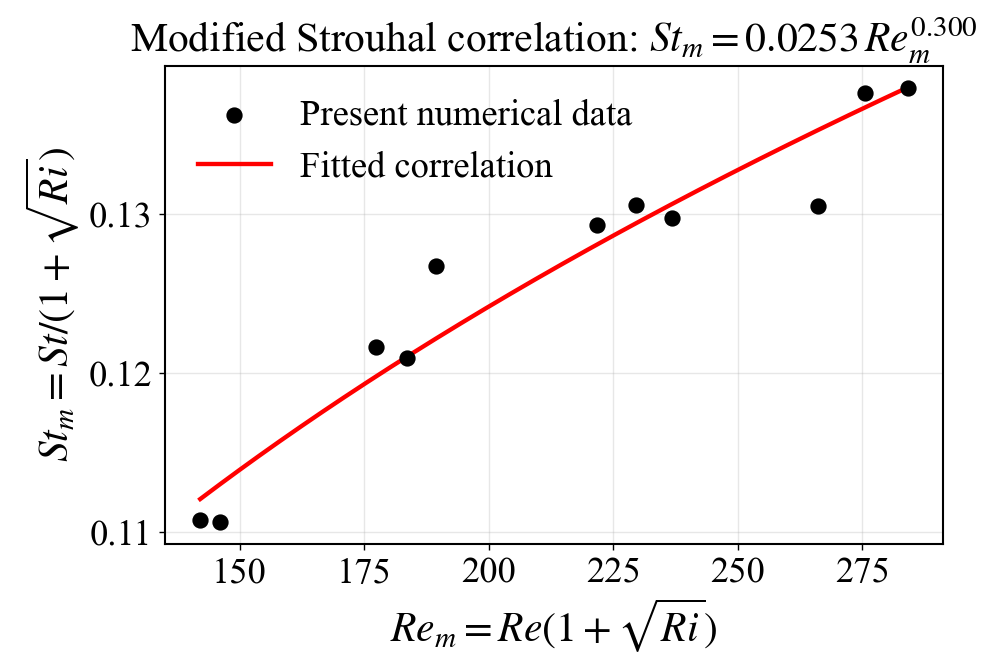}
\caption{Variation of the modified Strouhal number, $St_m$, with the modified Reynolds number, $Re_m$.}
\label{fig:Stm_Rem}
\end{figure}

Figure~\ref{fig:Stm_Rem} shows the variation of the modified Strouhal number, $St_m$ (for far-field unsteadiness) with the modified Reynolds number ($Re_m$). It is observed that the modified scaling produces a substantially improved collapse of the frequency data compared to the conventional Strouhal number ($St$), for which no comparable single-parameter correlation could be obtained. This suggests that a substantial part of the increase in $St$ is associated with the increase in the effective mixed-convective velocity scale produced by the combined action of forced and buoyancy-driven transport. The DNS data are reasonably represented by Eq.~\eqref{eq:St_corr}, which yields an $R^2$ value of $0.93$ and a maximum relative error of $3.7\%$. The remaining scatter may also be partially attributable to uncertainty in the FFT-based frequency estimation. Since a fixed number of samples ($2^{21}$) was used for all cases, while the temporal sampling interval varied across simulations, the resulting frequency resolution differed slightly from case to case due to the explicit method-based simulations, introducing a modest uncertainty in the extracted Strouhal numbers. \textcolor{\mycolorc}{For a fixed Reynolds number, the modified Strouhal number ($St_m$) also substantially reduces the Richardson-number dependence of the far-field oscillation frequency compared with $St$. Independently, the local mean streamwise velocity $U_l$ at the instability-initiation location varies relatively weakly with $Ri$ for a given $Re$, as also evident from Table~\ref{tab:local_St}.}

\textcolor{\mycolorc}{The modified Strouhal number ($St_m$) and the local plume-based Strouhal number ($St_{l,\delta_p}$) serve complementary purposes. $St_m$ together with $Re_m$ provides a compact empirical correlation in terms of the externally imposed Reynolds and Richardson numbers, and the systematic dependence observed in Figure~\ref{fig:Stm_Rem} is therefore represented by a power-law fit. In contrast, when the dimensional oscillation frequency ($f$) is scaled using the local mean-flow velocity ($u_l$) and plume width ($\delta_p$) at the instability-initiation location, the non-dimensional frequency $St_{l,\delta_p}$ reveals an approximate collapse across the investigated cases, with a coefficient of variation of 6.87\%.}

\begin{table}
\color{black}
\centering
\small
\setlength{\tabcolsep}{5pt}
\caption{\textcolor{\mycolorc}{Local flow scales evaluated at the far-field instability-initiation location for the investigated unstable cases. Here, $\Delta_p=\delta_p/B$, $Re_{l,\delta_p}=U_l\Delta_p Re$, and $St_{l,\delta_p}=St\Delta_p/U_l$.}}
\label{tab:local_St}
\begin{tabular}{ccccccc}
\toprule
$\mathrm{Re}$ & $\mathrm{Ri}$ & $U_l$ & $\Delta_p$ & $Re_{l,\delta_p}$ & $St$ & $St_{l,\delta_p}$ \\
\midrule
75  & 0.80 & 2.45 & 6.65 & 1222 & 0.2098 & 0.5693 \\
75  & 0.90 & 2.48 & 6.14 & 1142 & 0.2155 & 0.5333 \\
100 & 0.60 & 2.23 & 6.22 & 1387 & 0.2158 & 0.6019 \\
100 & 0.70 & 2.25 & 5.62 & 1265 & 0.2221 & 0.5551 \\
100 & 0.80 & 2.31 & 5.31 & 1227 & 0.2400 & 0.5519 \\
125 & 0.50 & 2.06 & 5.59 & 1439 & 0.2248 & 0.6097 \\
125 & 0.60 & 2.09 & 5.11 & 1335 & 0.2294 & 0.5609 \\
125 & 0.70 & 2.19 & 4.82 & 1320 & 0.2398 & 0.5279 \\
125 & 0.80 & 2.27 & 4.64 & 1316 & 0.2458 & 0.5024 \\
150 & 0.50 & 1.95 & 4.78 & 1398 & 0.2222 & 0.5447 \\
150 & 0.60 & 2.06 & 4.58 & 1415 & 0.2289 & 0.5089 \\
150 & 0.70 & 2.14 & 4.24 & 1361 & 0.2546 & 0.5045 \\
150 & 0.80 & 2.20 & 4.02 & 1325 & 0.2613 & 0.4775 \\
\bottomrule
\end{tabular}
\end{table}

\subsection{Linear stability analysis}

To gain further insight into the mechanisms driving the two observed transitions, namely the suppression of near-field vortex shedding and the onset of far-field unsteadiness, a linear stability analysis (LSA) in a domain size of $78.5B \times 50B$ is performed. The analysis is conducted at a fixed Reynolds number of $\mathrm{Re=100}$, allowing the role of buoyancy (through variations in $\mathrm{Ri}$) in governing the instability mechanisms to be isolated. This section presents an investigation of the leading eigenvalues and corresponding perturbation modes, with the aim of characterizing the nature and spatial structures of the underlying instabilities.

\subsubsection{Eigenvalue analysis}
\label{sec:eigenvalue_analysis}
\begin{figure}
    \centering
    \begin{subfigure}{0.49\textwidth}
        \includegraphics[width=1\textwidth]{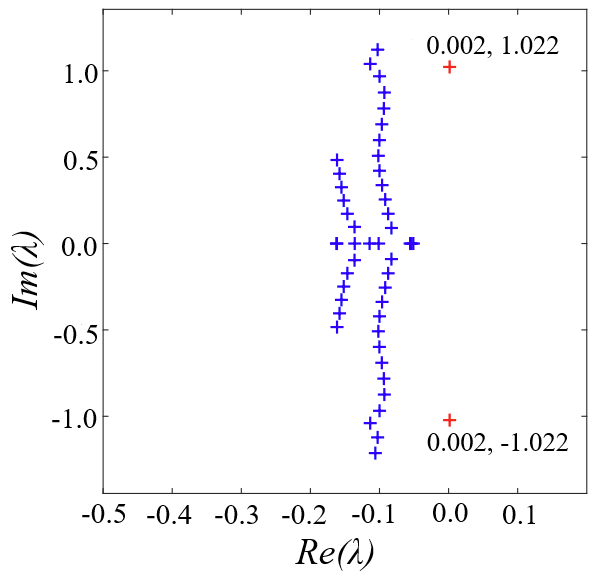}
        \subcaption{Ri = 0.14}
    \end{subfigure}
    \hfill
    \begin{subfigure}{0.49\textwidth}
        \includegraphics[width=1\textwidth]{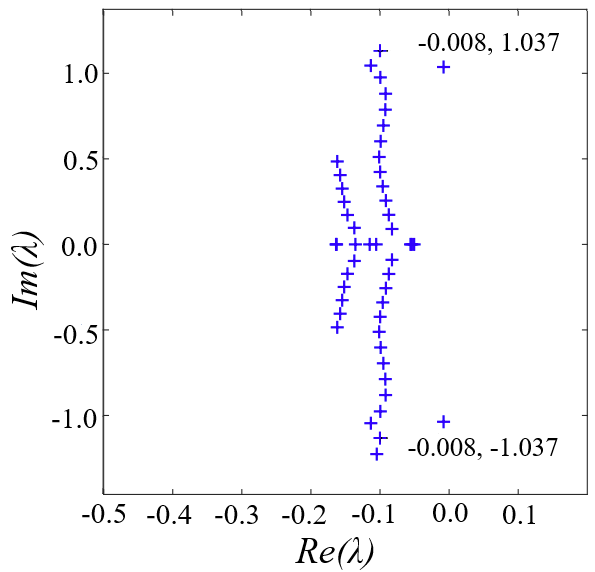}
        \subcaption{Ri = 0.15}
    \end{subfigure}
    \begin{subfigure}{0.49\textwidth}
        \includegraphics[width=1\textwidth]{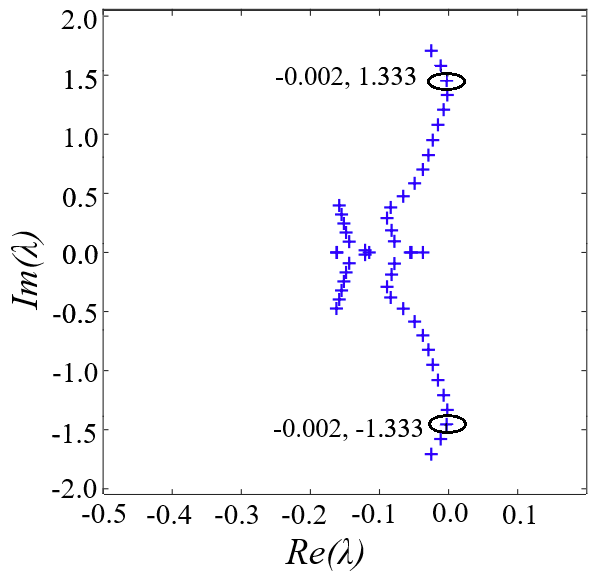}
        \subcaption{Ri = 0.61}
    \end{subfigure}
    \hfill
    \begin{subfigure}{0.49\textwidth}
        \includegraphics[width=1\textwidth]{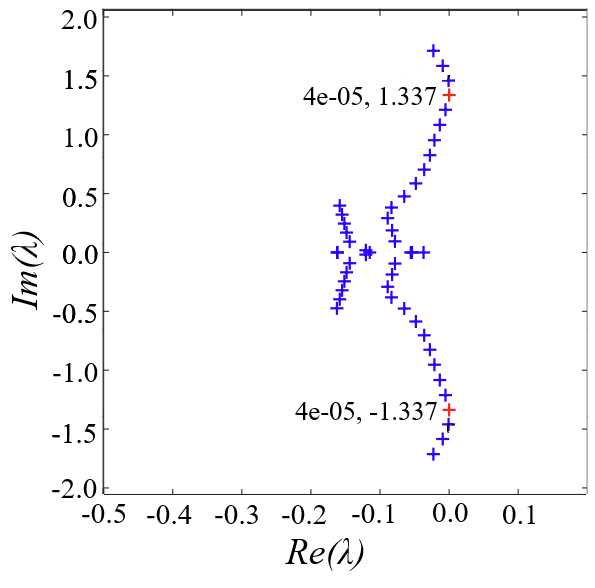}
        \subcaption{Ri = 0.62}
    \end{subfigure}    
    \caption{Eigenvalue plots at various Richardson numbers, (a) Ri=0.14, (b) Ri=0.15, (c) Ri=0.61 and (d) Ri=0.62 for Re=100.}
    \label{EV}
    \end{figure}

The eigenvalue spectra obtained from linear stability analysis for various Richardson numbers are presented in Figure~\ref{EV}. For $\mathrm{Ri} = 0.15$, all eigenvalues possess negative real parts, indicating a stable flow. However, for $\mathrm{Ri} = 0.14$, a pair of complex-conjugate eigenvalues with positive real parts emerges, marking the onset of instability. This transition behaviour implies that the first critical Richardson number corresponding to the suppression of vortex shedding lies between $\mathrm{Ri} = 0.14$ and $0.15$, consistent with previous observations~\citep{dushe2022linear}. This result is in exact agreement with the DNS findings.

As the Richardson number is increased beyond this point, the flow remains stable until $\mathrm{Ri} = 0.62$, where LSA again identifies a complex-conjugate eigenvalue pair with a positive real part. This indicates a second transition to instability, consistent with the DNS-predicted far-field instability. The second critical Richardson number, therefore, lies between $\mathrm{Ri} = 0.61$ and $0.62$ in contrast to the DNS study where the corresponding value is between 0.58 and 0.59. This slight discrepancy can be explained based on the fact that the amplitude of the flow oscillations due to far-field instability is much larger compared to the oscillation amplitudes caused by vortex shedding (see Figure~\ref{signal1_far}), making the linear approximation a little questionable for the far-field instability. Nonetheless, the overall agreement between DNS and LSA remains excellent.

In both transitions, the critical Richardson number corresponds to the condition where the real part of the most unstable eigenvalue crosses zero. At these points, the eigenvalues are purely imaginary, pointing to a Hopf bifurcation. \textcolor{\mycolorb}{Thus, within the finite original domain considered for the LSA, both the near-field vortex-shedding suppression and the onset of far-field instability are associated with Hopf bifurcations. For the far-field mode, this should be interpreted together with the DNS domain-dependence results discussed earlier: the apparent critical Richardson number depends on whether the domain is long enough for the downstream buoyancy-driven disturbance to amplify to observable levels.}

As noted earlier in Table~\ref{table:st_validation}, the imaginary part of the leading eigenvalues obtained from LSA captures the monotonic increase in Strouhal number with the Richardson number, and determines the temporal frequency of the most unstable mode. The consistency between LSA and DNS results highlights the reliability of the LSA framework not only in capturing the onset of instability via the real part of the eigenvalues, but also in accurately predicting the frequency of flow oscillations through their imaginary components.

\subsubsection{Perturbation mode structures}~\label{modes_analysis}

\begin{figure}
    \centering
        \includegraphics[width=0.9\textwidth]{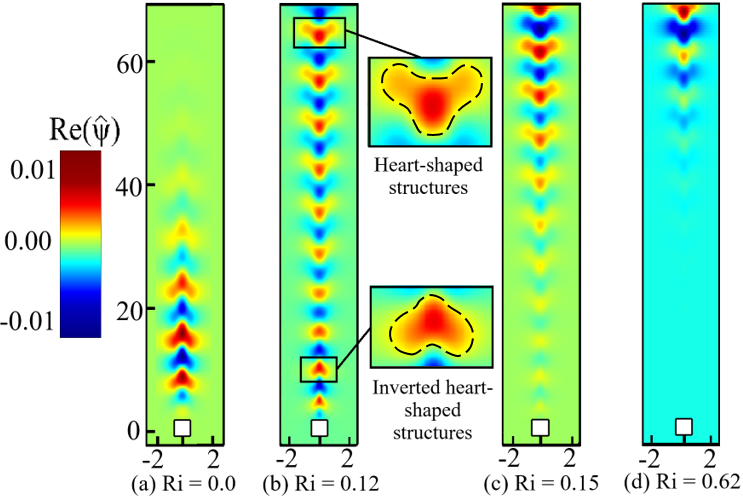}
    \caption{Stream function contours of the real part of the most unstable eigenmode at various Ri, (a) Ri=0.0, (b) Ri=0.12, (c) Ri=0.15 and (d) Ri=0.62 for Re=100.}
    \label{REM}
\end{figure}

To analyze the structure of the perturbation modes, the complex stream function $\hat{\psi}$ was reconstructed at each grid point using the perturbation velocity components $\hat{u}$ and $\hat{v}$ associated with the most unstable eigenmode—identified as the mode corresponding to the eigenvalue with the largest real part. Contour plots of the real part of $\hat{\psi}$ were then generated to visualize the spatial structure of the instability (Figure~\ref{REM}). Similar contours were also obtained using the imaginary part of $\hat{\psi}$ and are omitted here for brevity. Two distinct classes of perturbation structures were observed: inverted heart-shaped contours, prevalent in the near-field region, and heart-shaped contours emerging farther downstream. These are shown as zoomed insets for Ri = 0.12 in Figure~\ref{REM}. At $\mathrm{Ri} = 0.0$, the flow exhibits prominent inverted heart-shaped structures that decay downstream (Figure~\ref{REM}a), consistent with the presence of shear-layer instabilities corresponding to vortex shedding confined to the near-field region. These structures vanish at $\mathrm{Ri} = 0.15$, which corresponds to the critical point of vortex shedding suppression, suggesting that they are characteristic signatures of the near-field instability mechanism.

In contrast, as the Richardson number increases beyond the suppression threshold, the perturbation field begins to exhibit heart-shaped structures located in the far-field region. These structures become more pronounced at higher $\mathrm{Ri}$, indicating that the perturbations increasingly become stronger downstream. This spatial redistribution of instability is consistent with observations from DNS, where the flow transitions from a near-field shedding regime to a far-field instability. The diminishing near-field perturbation amplitude and the growing far-field structures with increasing $\mathrm{Ri}$ confirm that the dominant instability mechanism shifts downstream as buoyant effects intensify. Thus, the two classes of mode shapes, namely the inverted heart-shaped and heart-shaped modes, can be associated with the near-field and far-field instability mechanisms, respectively.


The spatial growth of perturbations in the far-field for high $\mathrm{Ri}$ bears a strong resemblance to the instability characteristics of buoyant plumes. In fact, DNS studies of horizontal plate heat source~\citep{hernandez2015natural} also show that thermal plumes can develop spatially growing instabilities associated with the progressive amplification of buoyancy-driven disturbances even in the absence of external forcing. The heart-shaped perturbation contours observed in the present LSA results may therefore be viewed as manifestations of these intrinsic plume instabilities, arising from the continued downstream acceleration and lateral entrainment of the buoyant flow. The observed downstream intensification of perturbations thus reflects the sustained action of the buoyant source term and the lateral entrainment that defines plume dynamics. These results strengthen the interpretation that the far-field instability is not merely a wake-like phenomenon but emerges due to the superposition of a growing, unsteady plume instability on the base flow.

\begin{figure}
    \centering
    \begin{subfigure}{0.49\textwidth}
        \includegraphics[width=1\textwidth]{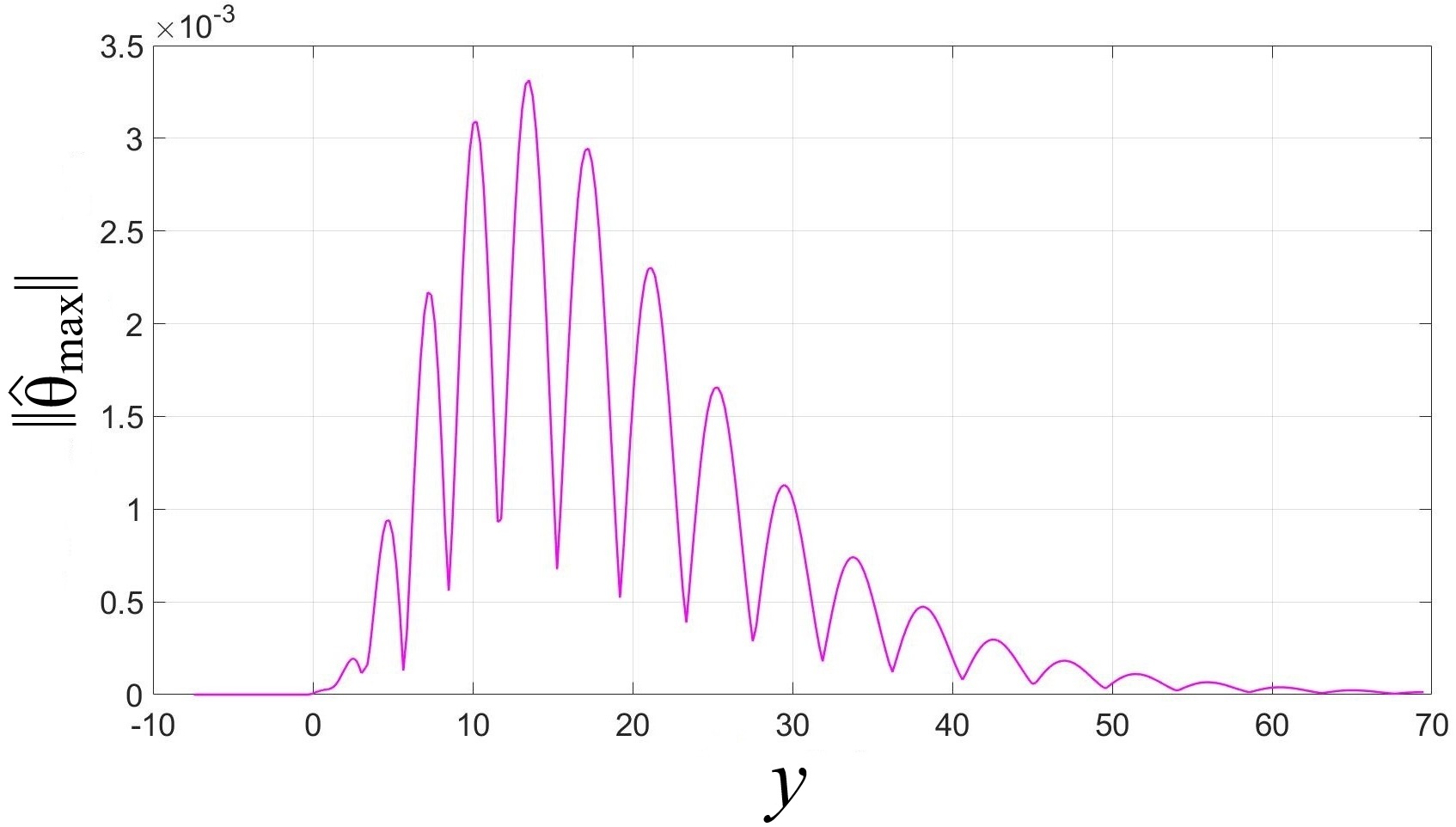}
        \subcaption{Ri = 0.0}
    \end{subfigure}
    \hfill
    \begin{subfigure}{0.49\textwidth}
        \includegraphics[width=1\textwidth]{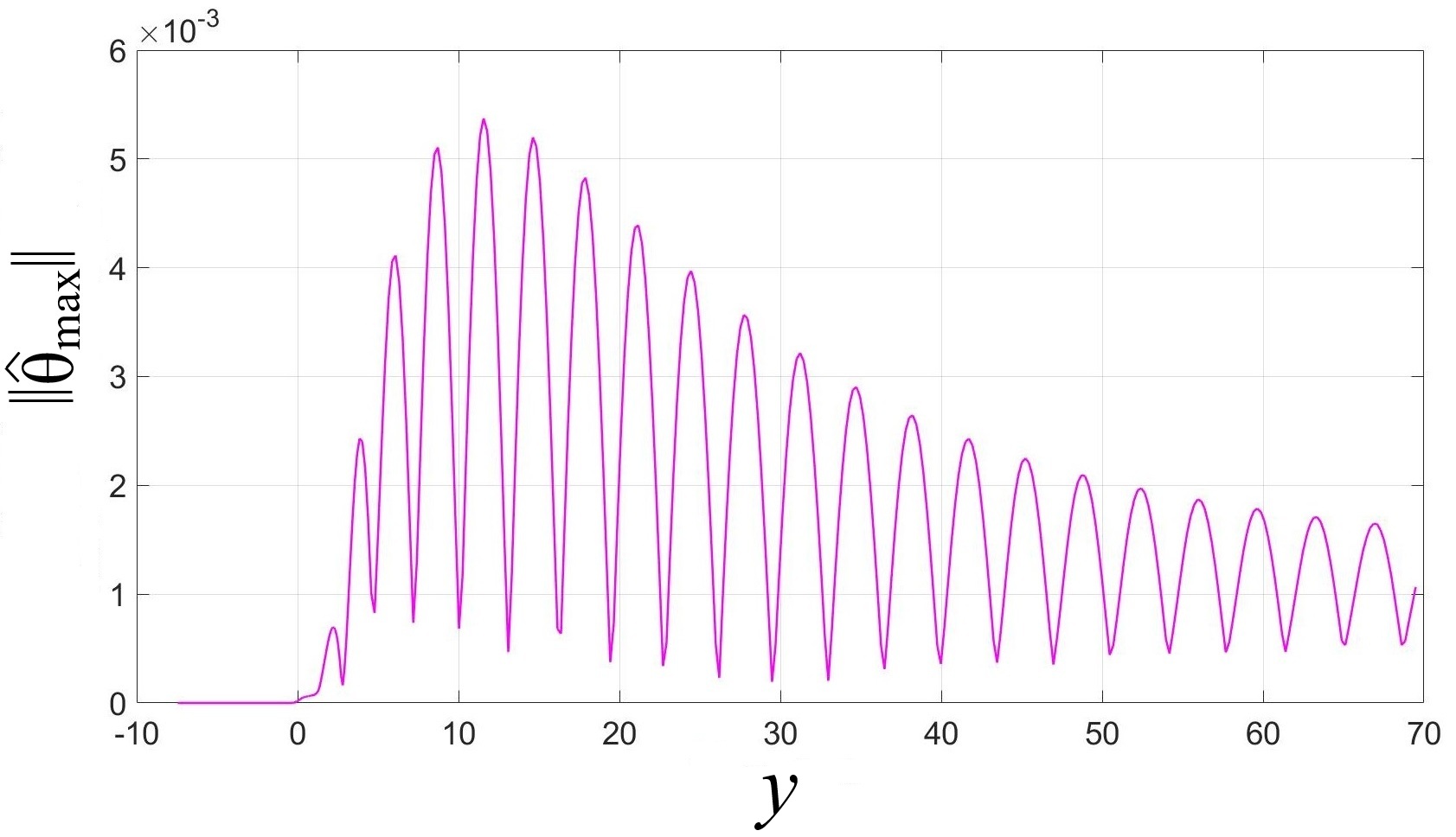}
        \subcaption{Ri = 0.12}
    \end{subfigure} 

    \begin{subfigure}{0.49\textwidth}
        \includegraphics[width=1\textwidth]{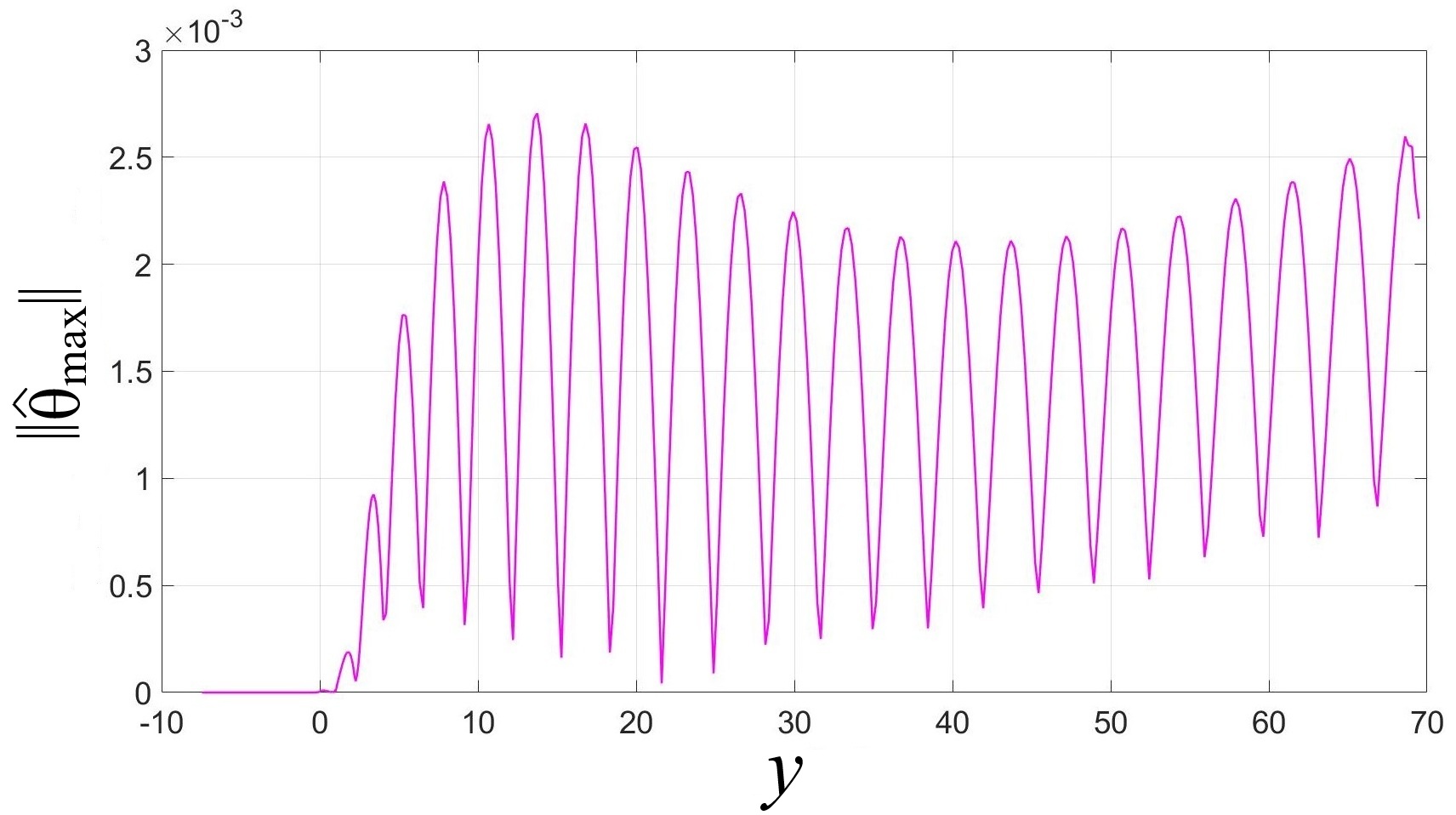}
        \subcaption{Ri = 0.14}
    \end{subfigure} 
    \hfill
        \begin{subfigure}{0.49\textwidth}
        \includegraphics[width=1\textwidth]{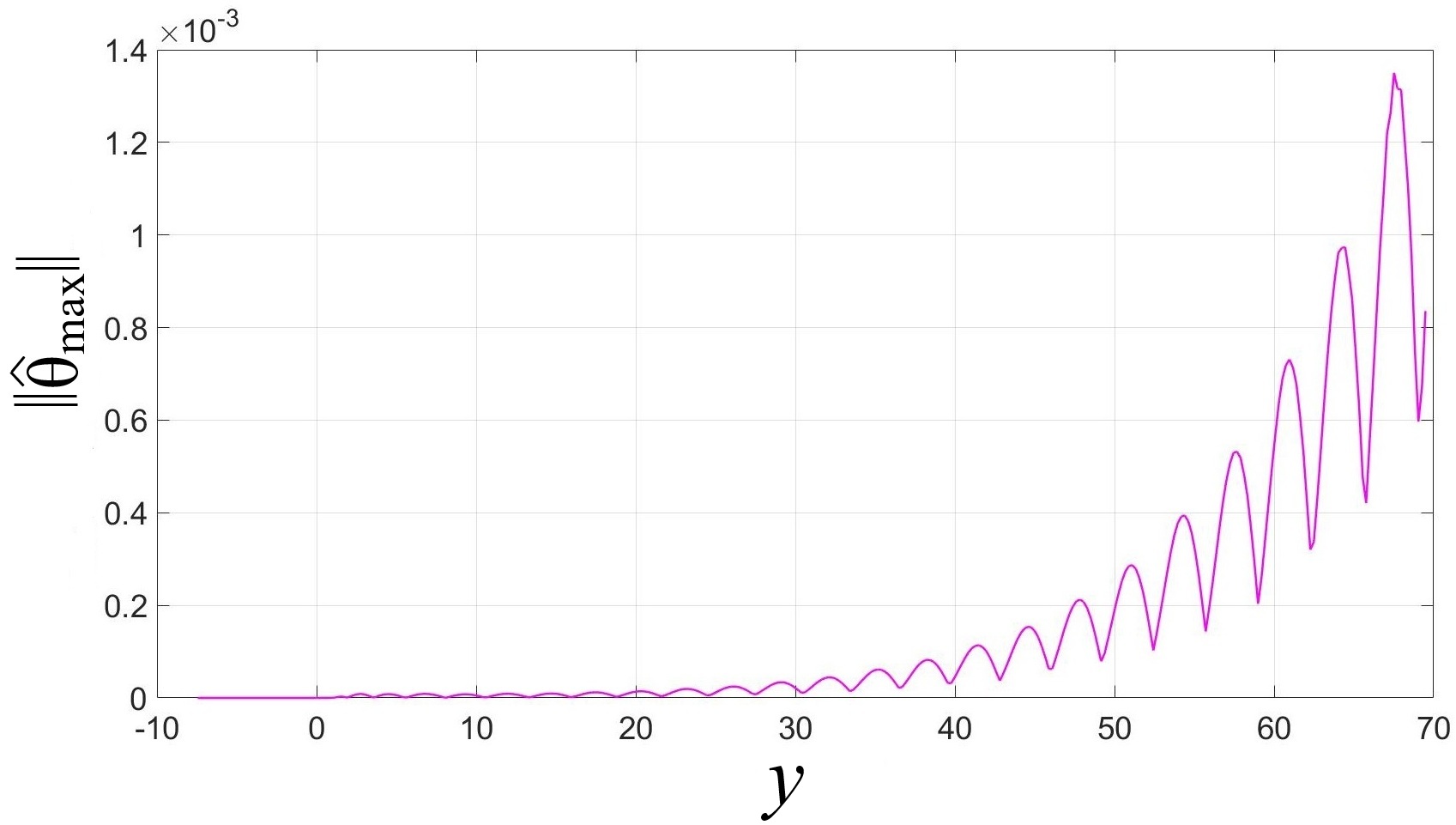}
        \subcaption{Ri = 0.65}
    \end{subfigure}
    \caption{Variation of peak temperature perturbation magnitude with $y$-coordinate for varying Ri at Re=100.}
    \label{thermal}
\end{figure}

The variation in flow perturbations is further corroborated by the behaviour of the thermal perturbation magnitude plotted in Figure~\ref{thermal}, where the maximum amplitude at each $y$-coordinate is reported. In cases of Ri=0.0 and Ri=0.12, the perturbation magnitude initially decays with increasing vertical distance from the cylinder, consistent with the spatially decaying nature of the near-field instability. However, for Ri=0.14 and 0.65, a qualitative shift is observed. For $\mathrm{Ri} = 0.14$, the perturbation magnitude initially decreases and subsequently increases in the far-field region, whereas for $\mathrm{Ri} = 0.65$, the perturbation magnitude is nearly zero near the cylinder and begins to grow beyond a certain distance. Similar variations were observed for perturbations of other flow variables. These trends suggest a transition in the dominant instability mechanism from near-field (spatially decaying) to far-field (spatially growing) with increasing $\mathrm{Ri}$. 

This behaviour can be explained by recognizing that the overall flow is a superposition of two distinct components: a forced convective flow associated with $\mathrm{Ri} = 0$, and a natural convective flow arising purely from buoyancy (i.e., the secondary flow). The forced convective component is characterized by perturbations that originate near the cylinder due to separation but decay downstream. In contrast, the natural convective component behaves akin to a buoyant plume, exhibiting a downstream growth in perturbation amplitude due to the sustained action of the buoyancy force. This analogy to buoyant plumes is supported by previous studies, such as that by \citet{hunt2022two}, which describe a downstream increase in momentum flux and perturbation intensity within such flows.

Accordingly, the net perturbation field in the present problem can be viewed as a composite of a decaying (forced) and a growing (natural) component. At low $\mathrm{Ri}$, the decaying forced-convective perturbation component dominates the observable perturbation field, resulting in a monotonic decline in perturbation amplitude with distance from the cylinder (Figure~\ref{thermal}a). For intermediate $\mathrm{Ri}$ (Figure~\ref{thermal}b), the domain height is insufficient for the growing component to surpass the decaying one, and thus only the decay is visible. However, at higher $\mathrm{Ri}$ (Figure~\ref{thermal}c), the growing natural convective component eventually dominates, leading to a reversal in the spatial trend of the perturbation magnitude. Finally, for $\mathrm{Ri} = 0.65$ (Figure~\ref{thermal}d), the near-field forced-convective perturbation component is strongly suppressed, and the perturbation field is dominated by the growing natural-convective mode.

These findings are consistent with the spatial patterns observed in the perturbation stream function contours, and they reinforce the interpretation of the far-field instability as a buoyancy-driven phenomenon. Moreover, the peak amplitudes of the thermal and velocity perturbations were found to be of comparable magnitude. This equivalence can be attributed to the Prandtl number ($\mathrm{Pr} = 0.7$), which is of order unity, leading to similarly weighted thermal and hydrodynamic effects in the instability mechanism \citep{su2000linear}.

If one assumes that the spatial growth of buoyancy-induced perturbations associated with the natural convection component continues downstream, it follows that such perturbations may eventually trigger far-field unsteadiness beyond a sufficiently large downstream distance from the cylinder, even at relatively low Richardson numbers, for the fixed Reynolds number considered here. From this perspective, the onset of far-field instability may not be associated with a sharply defined critical Richardson number at $\mathrm{Re=100}$, but instead depends on the cumulative downstream amplification of buoyancy-driven disturbances as they evolve away from the near wake.

This interpretation provides a refinement of the earlier conclusion regarding the existence of a second Richardson-number threshold for the far-field transition. Rather than representing an abrupt bifurcation at a unique $\mathrm{Ri}$, the far-field unsteadiness observed in the present simulations may be viewed as emerging gradually as buoyancy-induced perturbations grow and reorganize over increasing downstream distance. A definitive assessment of this behaviour would require simulations extending further downstream than those considered here.
 \newline

\section{Conclusions}
\label{sec:conclusions}

This study examined buoyancy-influenced mixed convection past a heated square-section cylinder using two-dimensional direct numerical simulations (DNS) and global linear stability analysis (LSA) at $\mathrm{Re}=100$ to investigate how variations in Richardson number govern transitions between distinct wake regimes. Three characteristic flow behaviours were identified: (i) active vortex shedding, (ii) a steady wake with suppressed shedding, and (iii) a far-field unsteadiness associated with buoyant plume dynamics.

At low $\mathrm{Ri}$, the wake is dominated by classical vortex shedding driven by shear-layer separation at the upstream corners of the cylinder. Increasing $\mathrm{Ri}$ in this regime leads to a longer recirculation zone and a rise in mean drag, heat transfer and Strouhal number. LSA captures this instability through the leading global mode, whose near-field perturbation structure manifests as inverted heart-shaped patterns in the stream function eigenmodes.

As $\mathrm{Ri}$ increases further, vortex shedding is suppressed beyond a critical value ($\mathrm{Ri}_c \approx 0.15$). This transition is driven by a buoyancy-induced secondary flow that partially fills the wake momentum deficit and introduces opposite signed vorticity, thereby damping the shear-layer instability and rendering the near wake steady. Although the mean drag coefficient
and Nusselt number continue to increase with $\mathrm{Ri}$, the mean recirculation length decreases gradually beyond $\mathrm{Ri}_c$. The counter-vorticity introduced by the secondary flow also gives rise to vorticity inversion, with its inception occurring closer to the cylinder as $\mathrm{Ri}$ is increased. Consistent with DNS, LSA confirms the stabilization of the near-field shedding mode and reveals the emergence of additional downstream modes that reflect the increasing influence of the natural-convective component of the flow.

At sufficiently high $\mathrm{Ri}$, a second unsteadiness appears in the far-field region. Rather than emerging at a sharply defined Richardson number, the far-field instability is interpreted as arising beyond a sufficiently large downstream distance for a given $\mathrm{Ri}$, as buoyancy-driven perturbations continue to grow and reorganize downstream. LSA identifies this transition through heart-shaped perturbation structures localized in the far field, which strengthen with increasing $\mathrm{Ri}$, while the original near-field shedding mode progressively weakens. Together, the DNS and eigenmode structures support the interpretation that the far-field unsteadiness originates from buoyant plume dynamics developing on top of the decaying forced-convective wake.

Taken together, the results suggest that the flow may be viewed as a superposition of a decaying forced-convective field and a growing buoyant plume, providing a unified framework for interpreting mixed convection wake transitions. Buoyancy plays a dual role: it suppresses near-field vortex shedding in the vicinity of the cylinder, while promoting plume-driven instability further downstream once the natural-convective component becomes sufficiently strong. Additional DNS at selected Reynolds numbers indicate that increasing $\mathrm{Re}$ has an overall destabilizing influence, modifying wake-transition characteristics and increasing the shedding frequency, in qualitative agreement with prior observations. Specifically, increasing Reynolds number shifts the two buoyancy-induced transitions in opposite directions: stronger inertia delays the suppression of near-field vortex shedding to higher Richardson numbers, while simultaneously promoting the earlier onset of far-field unsteadiness by reducing viscous damping in the buoyancy-driven shear layers. This asymmetric influence of Re highlights the fundamentally different physical origins of the two instabilities: one governed by near-field shear-layer dynamics and the other by downstream plume amplification. \textcolor{\mycolorc}{Analysis of the local flow at the onset of far-field instability further shows that the choice of local length scale is important. When the cylinder width is retained as the length scale, the local Reynolds number exhibits appreciable variation across Reynolds numbers. Replacing the cylinder width with the local plume width substantially improves the collapse, demonstrating that the evolving plume width provides a more representative length scale for the far-field instability. Across the 13 investigated far-field cases, the plume-based local Reynolds number $Re_{l,\delta_p}$ \textcolor{\mycolord}{ranges from 1142 to 1439, with a} coefficient of variation of approximately $6.26\%$, while the corresponding plume-based Strouhal number, $St_{l,\delta_p}=St\Delta_p/U_l$, \textcolor{\mycolord}{ranges from 0.4775 to 0.6097, with a} coefficient of variation of $6.87\%$. For a fixed Reynolds number, the local mean streamwise velocity $U_l$ at the instability-initiation location also varies relatively weakly with $Ri$. Together, these observations indicate that the local plume width and local mean velocity provide physically relevant scales for the far-field instability, with its characteristic oscillation timescale scaling approximately with the local convective timescale $\delta_p/u_l$.} In addition to these transition effects, for each $Ri$, the mean drag coefficient and mean Nusselt number are found to decrease and increase with Re, respectively. The combined influence of Reynolds and Richardson numbers on integral transport quantities is captured through simple empirical correlations for the mean drag and Nusselt numbers; in particular, the introduction of a mixed-convective characteristic velocity scale that superposes forced and buoyancy-driven contributions enables the surface-averaged Nusselt number data to collapse onto a single scaling relation across the explored parameter space.


\section*{Acknowledgements}
The authors would like to thank Dr. Basheer Ahmad Khan for his valuable feedback and insightful discussions that contributed significantly to this work. The authors acknowledge the computational support provided by the High Performance Computing (HPC) facility at the Indian Institute of Technology Kanpur, Kanpur.

\section*{Declaration of interests}
The authors report no conflict of interest.

\section*{Data availability statement}
The data supporting the findings of this study are available from the corresponding author upon reasonable request.
\appendix
\renewcommand{\thetable}{B.\arabic{table}}
\setcounter{table}{0}
\section{Calculation of global flow parameters}
This section presents the mathematical expressions for calculating the parameters used in the present study. For ease of expression, the four sides of the square cylinder have been labelled as: $S_{b}, S_{t}, S_{l}, S_{r}$ corresponding to the bottom, top, left and right sides of the cylinder, respectively. From Figure \ref{domain}, $S_{b}$ = side AB, $S_t$ = side CD, $S_l$ = side BC and $S_r$ = side DA. \newline

\begin{enumerate}
\item \textbf{Drag coefficient ($C_D$):} The drag coefficient is a non-dimensional number which gives the magnitude of the force experienced by the cylinder in the direction of incoming flow. The mathematical expression for the drag force can be written as:

\begin{equation}
    C_{D} = \frac{F_{y}}{(1/2)\rho U_{\infty}^2B}
\end{equation} \newline

Where, $F_{y}$ is the force on the cylinder along the Y-direction per unit thickness of the cylinder (as in the current work, a two-dimensional flow has been considered). \textcolor{\mycolorb}{The denominator uses the dynamic pressure multiplied by the projected frontal area per unit thickness. For the present two-dimensional square cylinder, this reference area per unit thickness reduces to the cylinder width, $B$.} The drag force experienced by the square cylinder comes from the action of two forces: the force due to pressure and the force due to viscous stresses. Thus, $F_{y}$ can be broken down into two components as:
\begin{equation}
   F_{y} = F_{yp} + F_{yv} 
\end{equation}

Where, $F_{yp}$ = form drag (pressure drag) and $F_{yv}$ = drag due to viscous stresses. These are calculated as follows:

\begin{equation}
    F_{yp} = \int_{S_b}^{} p \,ds - \int_{S_t}^{} p \,ds
\end{equation}

\begin{equation}
    F_{yv} = \int_{S_r}^{} \rho\nu\left(\frac{\partial v}{\partial x} + \frac{\partial u}{\partial y} \right)\,ds - \int_{S_l}^{} \rho\nu\left(\frac{\partial v}{\partial x} + \frac{\partial u}{\partial y} \right)\,ds
\end{equation} \newline

In these dimensional force expressions, $p, u, v, x, y$ and $s$ denote dimensional quantities. Dividing these forces by $(1/2)\rho U_{\infty}^2B$ converts the dimensional integrands into dimensionless quantities used in the present study, which facilitates their calculation. The first order derivatives in the above expression have been discretized using first order schemes and the values of the integrals have been calculated by finding the values of their integrands on the cylinder wall grid cells and summing over these values over the desired cylinder boundaries. \newline

\item \textbf{Lift coefficient ($C_L$):} The lift coefficient in the present work gives the magnitude of the sideways force experienced by the cylinder due to the surrounding fluid. The mathematical expression for the lift force can be written as:
\begin{equation}
    C_{L} = \frac{F_{x}}{(1/2)\rho U_{\infty}^2B}
\end{equation}

Where $F_{x}$ is the force on the cylinder along the X-direction per unit thickness of the cylinder. The procedure for evaluating this force is identical to the one used for evaluating the drag force. Thus, the total lift force is decomposed into two components, the pressure force and the viscous force:
\begin{equation}
   F_{x} = F_{xp} + F_{xv} 
\end{equation}

Where, $F_{xp}$ = pressure force and $F_{xv}$ = viscous force. These are calculated as follows:

\begin{equation}
    F_{xp} = \int_{S_l}^{} p \,ds - \int_{S_r}^{} p \,ds
\end{equation}

\begin{equation}
    F_{xv} = \int_{S_t}^{} \rho\nu\left(\frac{\partial u}{\partial y} + \frac{\partial v}{\partial x} \right)\,ds - \int_{S_b}^{} \rho\nu\left(\frac{\partial u}{\partial y} + \frac{\partial v}{\partial x} \right)\,ds
\end{equation} \newline

The numerical scheme for evaluating these forces and calculating the lift coefficient is identical to that for the drag coefficient. \newline

\item \textbf{Nusselt number ($Nu$):} The Nusselt number is a measure of the heat transfer occurring between the cylinder and the surrounding fluid. The local Nusselt number is defined for the present problem as: 

\begin{equation}
    Nu_l = \frac{hB}{k}
\end{equation}

Where $h$ is the convective heat transfer coefficient. For a cylinder with constant temperature, the heat transfer relation at the liquid-solid interface is given by:

\begin{equation}
  \left. -k\frac{\partial T}{\partial n} \right \vert_{n=0} = h(T_s - T_{\infty})
\end{equation}

Here $n$ is the direction normal to the cylinder surface. When the above equation is non-dimensionalized, the local Nusselt number ($Nu_{l}$) is obtained as the normal derivative of the non-dimensional temperature at the cylinder walls:

\begin{equation}
    Nu_{l} = -\frac{\partial \theta}{\partial n}
\end{equation}
 The local Nusselt number, $Nu_l$, is averaged in time and over the surface of the cylinder to calculate the average Nusselt numbers.
\newline
\item \textbf{Strouhal number ($St$):} The Strouhal number is the non-dimensional form of the most dominant frequency in the time-series data of the flow variables, which is obtained by performing a Fast Fourier Transform (FFT) of the data. The mathematical expression for the Strouhal number can be written as:

\begin{equation}
   St = \frac{fB}{U_{\infty}} 
\end{equation}

Where $f$ is the dominant oscillation frequency calculated from the transverse velocity signal.
\end{enumerate}



\bibliographystyle{AAA}
\bibliography{AAA}

\end{document}